\documentclass[pdflatex,sn-vancouver-num]{sn-jnl}% Vancouver Numbered Reference Style
\usepackage{graphicx}
\usepackage{amsmath,amssymb,amsfonts}
\usepackage{amsthm}
\usepackage{mathrsfs}
\usepackage[title]{appendix}
\usepackage{array}
\usepackage{subfigure}
\usepackage{comment}
\usepackage{multirow}
\usepackage{booktabs}
\usepackage{longtable}
\usepackage{pifont}
\usepackage[dvipsnames]{xcolor}
\usepackage[table]{xcolor}
\usepackage{xfp}
\usepackage{anyfontsize}
\usepackage{url}
\usepackage{rotating}
\usepackage{float}
\usepackage{enumitem}
\setlist[description]{leftmargin=\parindent,labelindent=\parindent}

\newcommand\myMatha[1]{
  \fpeval{(100-#1)*2}}
\newcommand\myMathb[1]{
  \fpeval{2*#1}}
  
\newcommand{\scaleaa}[1]{\cellcolor{Maroon!\myMatha{#1}}#1\%}
\newcommand{\scalebb}[1]{\cellcolor{Maroon!\myMathb{#1}}#1\%}
\newcommand{\scalecc}[1]{\cellcolor{Maroon!\myMathb{#1}}\noindent {\color{white}#1\%}}
\newcommand{\scalea}[1]{\cellcolor{Maroon!#1}\noindent {\color{white}#1\%}}
\newcommand{\scaleb}[1]{\cellcolor{Maroon!#1}#1\%}

\newcommand{\capivaraicon}[1]{\includegraphics[width=#1]{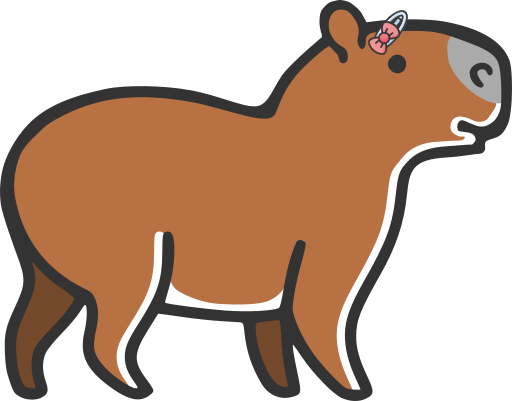}}

\newcommand{\afirmacaoum}{``The AIET is easy to use/answer''}
\newcommand{\afirmacaodois}{``The AIET is useful for identifying the risks of the language model''}
\newcommand{\afirmacaotres}{``The AIET is useful for generating responsible documentation about the language model''}
\newcommand{\afirmacaoquatro}{``It is complicated to identify the ethical considerations of the language model using the AIET''}
\newcommand{\cpl}{``Agree''}
\newcommand{\cep}{``Weakly agree''}
\newcommand{\dep}{``Weakly disagree''}
\newcommand{\dpl}{``Disagree''}
\newcommand{\neu}{``Neutral''}

\begin{document}

\tolerance=999
\sloppy

\title[Article Title]{Evaluation of AI Ethics Tools in Language Models: A Developers' Perspective Case Study}

\author*[1]{\fnm{Jhessica} \sur{Silva}}\email{jhessica.silva@ic.unicamp.br}

\author[1]{\fnm{Diego} \sur{A. B. Moreira}}

\author[1]{\fnm{Gabriel} \sur{O. dos Santos}}

\author[2]{\fnm{Alef} \sur{Ferreira}}

\author[1]{\fnm{Helena} \sur{Maia}}

\author*[1]{\fnm{Sandra} \sur{Avila}}\email{sandra@ic.unicamp.br}

\author*[1]{\fnm{Helio} \sur{Pedrini}}\email{helio@ic.unicamp.br}

\affil[1]{\orgdiv{Instituto de Computação}, \orgname{Universidade Estadual de Campinas (UNICAMP)}, \orgaddress{\
%street{Av. Albert Einstein, 1251}, 
\city{Campinas}, %\postcode{13083-852}, 
\state{São~Paulo}, \country{Brazil}}}

\affil[2]{\orgdiv{Instituto de Informática}, \orgname{Universidade Federal de Goiás (UFG)},\orgaddress{
%\street{Alameda Palmeiras}, 
\city{Goiânia}, %\postcode{74690-900}, 
\state{Goiás}, \country{Brazil}}}

% Abstract
\abstract{
In Artificial Intelligence (AI), language models have gained significant importance due to the widespread adoption of systems capable of simulating realistic conversations with humans through text generation. Because of their impact on society, developing and deploying these language models must be done responsibly, with attention to their negative impacts and possible harms. In this scenario, the number of AI Ethics Tools (AIETs) publications has recently increased. These AIETs are designed to help developers, companies, governments, and other stakeholders establish trust, transparency, and responsibility with their technologies by bringing accepted values to guide AI's design, development, and use stages. However, many AIETs lack good documentation, examples of use, and proof of their effectiveness in practice. This paper presents a methodology for evaluating AIETs in language models. Our approach involved an extensive literature survey on 213~AIETs, and after applying inclusion and exclusion criteria, we selected four AIETs: Model Cards, ALTAI, FactSheets, and Harms Modeling. For evaluation, we applied AIETs to language models developed for the Portuguese language, conducting 35 hours of interviews with their developers. The evaluation considered the developers' perspective on the AIETs' use and quality in helping to identify ethical considerations about their model. The results suggest that the applied AIETs serve as a guide for formulating general ethical considerations about language models. However, we note that they do not address unique aspects of these models, such as idiomatic expressions. Additionally, these AIETs did not help to identify potential negative impacts of models for the Portuguese language.}

%%\pacs[JEL Classification]{D8, H51}
%%\pacs[MSC Classification]{35A01, 65L10, 65L12, 65L20, 65L70}

\maketitle

%%==================================%%
%%           Introduction           %%
%%==================================%%
\section{Introduction}

Language models have been reshaping how we interact with machines thanks to popularizing systems capable of simulating realistic conversations. This popularization has occurred mainly with the launch of ChatGPT~\citep{chatgpt}, which has a simple interface for user interaction. These models are trained \textit{``on the task of string prediction, whether it operates over characters, words or sentences, and sequentially or not''}~\citep{bender2020}, given a previous context. Typically, language models take text as input to predict new strings. 

Researchers from various fields have warned of the ethical threats these models pose. \citet{hovy2016} show that the social impacts of language models range from the exclusion of people due to demographic biases present in the data used for training to the generation of erroneous or false answers due to overgeneralization of the models during training. \citet{bender2021} warn that large language models trained with large datasets have a high environmental and financial cost and---yet---do not guarantee diversity, propagating worrying biases such as stereotypical associations or negative feelings towards specific groups. \citet{weidinger2021} conducted significant work investigating the ethical and social risks of harm caused by language models and pointed out 21 potential risks.

In the context of the Portuguese, a language with medium-computational resources\footnote{Despite Portuguese being ranked among the top ten languages globally based on the number of native speakers, it is considered a medium-resource language within the context of AI~\citep{nicholas2023losttranslationlargelanguage}.}, as it has little data available for use in AI technologies, the most multilingual language models while lacking competitiveness~\citep{santos2023capivara}, also fail to adequately represent the cultural, historical, and social nuances of the population represented by the language~\citep{bender2021}. This deficiency is largely due to the predominant use of English-language data during training, with minimal data from other languages incorporated~\citep{johnson2022}. Particularly, when it comes to Portuguese expert models, although they perform better, the vast majority are trained with textual data translated from English. Consequently, while these models target Portuguese speakers, they tend to replicate all aspects of the global north. They may perpetuate errors and biases arising from automatic translation processes~\citep{santos2023capivara,hovy2021five}. 

These points show the importance of ethically evaluating language models and providing end users with comprehensive documentation that explains the associated challenges of the technology. However, several current models are only being advertised accompanied by ``technical reports'' that lack substantial information about the model's training processes, architecture, or ethical implications. One of the reasons for this is that many of these models are being proposed by computer professionals without the support of a multidisciplinary team. Unfortunately, many of these computer professionals finish their basic training without a solid knowledge of subjects related to ethics and technologies~\citep{brown2024teaching,goetze2023integrating}. Consequently, they usually do not have a basis for how to carry out ethical formulations about the technologies they propose. On the other hand, specific conferences such as Conference on Fairness, Accountability, and Transparency (FAccT), Empirical Methods in Natural Language Processing (EMNLP), and Conference on Neural Information Processing Systems (NeurIPS) have begun to include dedicated sections for ethical considerations in their calls for papers, showing the importance of reflecting on ethics in technology research. Due to the impacts and potential risks that language models can generate for society, these models must be developed and deployed responsibly to mitigate these threats. 

In light of the presented challenges, this paper aims to contribute to the urgent open discussions in our society about ethics and AI technologies. To this end, we conducted a case study on the use of AI Ethics Tools (AIETs)~\citep{mitchell2019,altai,harmsmodeling,arnold2019factsheets} to assess the ethical considerations of language models developed in Portuguese.  This study was based on interviews with developers from four language models developed for the Portuguese language using AIETs as a guide. The AIETs are practical methods that can be applied during the design, development, and use of AI to identify potential ethical problems that the technology may generate and propose ways to mitigate these problems. AIETs can be a starting point for designing and developing more ethical AI technologies, especially language models, that aim to mitigate biases, avoid spreading inequalities, and prevent the malicious use of their numerous capabilities. 

The main contributions of our paper are as follows:
\begin{itemize}
    \item We introduced a methodology to select and evaluate AIETs in language models carefully. For this, we conducted an extensive literature survey of AIETs. We applied inclusion and exclusion criteria to filter all 213 found AIETs, resulting in four selected AIETs; 
    \item We conducted a case study where we applied the selected AIETs through interviews with the developers of language models. To our knowledge, this is the first study to evaluate the use of AIETs in language models;
    \item We introduced a process to assess AIETs from the developers' perspective. We designed a questionnaire to extract developers' perceptions of the analyzed AIETs to achieve this. We aimed to evaluate the usability and effectiveness of AIETs in helping developers of language models to identify ethical considerations about the developed model by comparing the results of each AIET with the others.    
\end{itemize}

The remainder of this paper is structured as follows. Section~\ref{sec:backArelated} overviews AIETs and the works most closely related to this paper. Section~\ref{sec:methodology} presents the methodology adopted in this study. Section~\ref{sec:results} presents the results and discusses the developers' perspectives and opinions on the AIETs applied in an interview format. Section~\ref{sec:limitations} presents the limitations and future work of this study. Finally, Section~\ref{sec:conclusion} presents the conclusions and the ethical considerations of this study.

%%==================================%%
%%   Background and Related Work    %%
%%==================================%%
\section{Background and Related Work}
\label{sec:backArelated}

This section presents a brief overview of AIETs and the most closely works related to this paper. Section~\ref{sec:AIETbackground} provides the related concepts about AIETs, whereas Section~\ref{sec:AIETs_overview} presents an overview of the works that analyze AIETs as a whole. We adopt the term AI Ethics Tools (AIETs), as~\citet{ayling2022}, to refer to all tools, toolkits, methods, frameworks, and research that aim to promote AI ethics throughout the AI life cycle. In some works, other terms are also used to refer to AIETs, such as AI ethics toolkits~\citep{wong2023}, Ethical AI frameworks~\citep{prem2023}, and AI ethics frameworks~\citep{qiang2023}. 

\subsection{AIETs background}
\label{sec:AIETbackground}

The AIETs are practical methods that can be utilized during the design, development, and use stages of AI technologies to identify potential ethical issues that the technology may generate and, in some cases, propose strategies for mitigating these problems. In recent years, researchers and companies in the public and private sectors have proposed several AIETs to evaluate ethical principles. Several papers analyze these AIETs, presenting an overview of their applications and impacts and the implications associated with translating ethical principles into practical tools for use in AI technologies~\citep{morley2020,ayling2022,palladino2022,wong2023,prem2023, ortega2024applying}.

The AIETs currently available are diverse, with different proposals and approaches, differing according to the stage of the AI technology life-cycle where they can be applied (problem specification, design, testing, and use stages) and according to their (i) categories, (ii) final objectives, (iii) topics and principles covered by each one, and (iv) target audience~\citep{ayling2022, palladino2022, prem2023}. 
In addition, AIETs can be applied to datasets and/or to the developed models. Table~\ref{tab:IAcycleAIET} shows examples of Ethical Considerations and AIETs by AI life-cycle stages and Table~\ref{tab:egAIETs} shows examples of AIETs by each type of classification mentioned.

\begin{table}[!htb]
    \renewcommand{\arraystretch}{1.8}
    \footnotesize
    \centering
    \caption{Examples of ethical considerations and AIETs by stages in the AI life-cycle}
    \begin{tabular}{>{\raggedright\arraybackslash}p{1.8cm}>{\raggedright\arraybackslash}p{5.5cm}>{\raggedright\arraybackslash}p{4.5cm}}
        \toprule
        \textbf{Stage} & \textbf{Examples of Ethical Considerations} & \textbf{Examples of AIETs}\\
        \midrule
        
        \textbf{Problem specification} & {\tiny$\bullet$} Responsibility (Have the requirements been clearly documented?)\newline {\tiny$\bullet$} Non-maleficence (Who are the future users and how will they be affected?)\newline {\tiny$\bullet$} Sustainability (How might this AI impact the environment?)
        & {\tiny$\bullet$} \textit{IDEO’s AI Ethics Cards}~\citep{ideo2019}\newline {\tiny$\bullet$} \textit{Corporate Digital Responsibility}~\citep{LOBSCHAT2021875}\newline {\tiny$\bullet$} \textit{Ethics for Designers}~\citep{ethicsfordesigners} \\

        \textbf{Design} & {\tiny$\bullet$} Privacy (Has personal data been anonymized?)\newline {\tiny$\bullet$} Freedom and Autonomy (Was the data collected with consent?)\newline {\tiny$\bullet$} Beneficence (Is the AI development process aimed at the well-being of users?) & {\tiny$\bullet$} \textit{Datasheets for Datasets}~\citep{gebru2021}\newline {\tiny$\bullet$} \textit{Data statements for NLP}~\citep{bender2018data}\newline {\tiny$\bullet$} \textit{The Dataset Nutrition Label}~\citep{holland2018} \\

        \textbf{Testing} & {\tiny$\bullet$} Transparency (Are the results explainable or interpretable?) \newline {\tiny$\bullet$} Privacy (Can AI generate confidential information?) \newline {\tiny$\bullet$} Justice and Fairness (Did AI perform differently for any groups?) & {\tiny$\bullet$} \textit{Aequitas}~\citep{saleiro2019aequitas}\newline {\tiny$\bullet$} \textit{AI Explainability 360 Toolkit}~\citep{10.1145/3430984.3430987}\newline {\tiny$\bullet$} \textit{What-if Tool}~\citep{whatiftool} \\

        \textbf{Use} & {\tiny$\bullet$} Security and Safety (Is the AI robust against adversary attacks?)\newline {\tiny$\bullet$} Trust (Can users trust AI results?) \newline {\tiny$\bullet$} Dignity (Does AI respect users' values?) & {\tiny$\bullet$} \textit{Model Cards for Model Reporting}~\citep{mitchell2019}\newline {\tiny$\bullet$} \textit{Harms Modeling}~\citep{harmsmodeling}\newline {\tiny$\bullet$} \textit{ALTAI}~\citep{altai}\\\bottomrule
    \end{tabular}
    \label{tab:IAcycleAIET}
\end{table}

\begin{table}[!htb]
    \renewcommand{\arraystretch}{1.6}    
    \footnotesize
    \caption{Examples of AIETs by categories, final objectives, principles and target audience}
    
    \begin{tabular}{p{0.1cm}>{\raggedright\arraybackslash}p{2.1cm}>{\raggedright\arraybackslash}p{0.1cm}>{\raggedright\arraybackslash}p{9cm}}
        \toprule
        & \textbf{AIET Type} && \textbf{Examples of AIETs}\\
        \midrule

        \parbox[t]{1mm}{\multirow{8}{*}{\rotatebox[origin=c]{90}{\textbf{Categories}}}} & \textbf{Checklists or Questionnaires} && 
        {\tiny$\bullet$} \textit{Datasheets for Datasets}~\citep{gebru2021}\newline 
        {\tiny$\bullet$} \textit{FactSheets: Increasing Trust in AI Services through Supplier’s Declarations of Conformity}~\citep{arnold2019factsheets}\\
        
        & \textbf{Frameworks} && 
        {\tiny$\bullet$} \textit{AI-RFX Procurement Framework}~\citep{airfx}\newline 
        {\tiny$\bullet$} \textit{Zeno: An Interactive Framework for Behavioral Evaluation of Machine Learning}~\citep{10.1145/3544548.3581268}\\
        
        & \textbf{Metrics} && 
        {\tiny$\bullet$} \textit{LIME -- Local Interpretable Model-Agnostic Explanations}~\citep{10.1145/2939672.2939778}\newline 
        {\tiny$\bullet$} \textit{SHAP -- SHapley Additive exPlanations}~\citep{shap2017}\\
        
        & \textbf{Codes and Algorithm} && 
        {\tiny$\bullet$} \textit{BOLD: Dataset and Metrics for Measuring Biases in Open-Ended Language Generation}~\citep{dhamala2021bold}\newline 
        {\tiny$\bullet$} \textit{XAI -- An eXplainability toolbox for machine learning}~\citep{EthicalMLxai} \\\hline

        \parbox[t]{1mm}{\multirow{12}{*}{\rotatebox[origin=c]{90}{\textbf{Objectives}}}} & \textbf{Audit} && 
        {\tiny$\bullet$} \textit{Auditing LLMs: A Three-Layered Approach}~\citep{mokander2024auditing}\newline 
        {\tiny$\bullet$} \textit{Closing the AI Accountability Gap: Defining an end-to-end framework for internal algorithmic auditing}~\citep{raji2020closing}\\
        
        & \textbf{Requirements assessment and/or Supervision} && 
        {\tiny$\bullet$} \textit{ALTAI -- The Assessment List for Trustworthy Artificial Intelligence}~\citep{altai}\newline 
        {\tiny$\bullet$} \textit{A seven-layer model with checklists for standardising fairness assessment throughout the AI lifecycle}~\citep{agarwal2024seven} \\
        
        & \textbf{Risk assessment} && 
        {\tiny$\bullet$} \textit{Harms Modeling}~\citep{harmsmodeling}\newline 
        {\tiny$\bullet$} \textit{PROBAST: A Tool to Assess the Risk of Bias and Applicability of Prediction Model Studies}~\citep{wolff2019probast} \\
        
        & \textbf{Documentations} && 
        {\tiny$\bullet$} \textit{Model Cards for Model Reporting}~\citep{mitchell2019}\newline  
        {\tiny$\bullet$} \textit{The Dataset Nutrition Label}~\citep{holland2018}\\

        & \textbf{Case study or Tutorial} && 
        {\tiny$\bullet$} \textit{Judgment Call the Game: Using Value Sensitive Design and Design Fiction to Surface Ethical Concerns Related to Technology}~\citep{ballard2019judgmentcall}\newline  
        {\tiny$\bullet$} \textit{Community jury}~\citep{communityjury}\\
        
        & \textbf{Technical} && 
        {\tiny$\bullet$} \textit{TensorFlow Privacy}~\citep{tensorflowprivacy}\newline  
        {\tiny$\bullet$} \textit{Adversarial Robustness Toolbox}~\citep{nicolae2019adversarialrobustnesstoolboxv100}\\\hline

        \parbox[t]{1mm}{\multirow{6}{*}{\rotatebox[origin=c]{90}{\textbf{Principles}}}} & \textbf{Transparency} && {\tiny$\bullet$} \textit{AI Explainability 360 Toolkit}~\citep{10.1145/3430984.3430987}\\
        & \textbf{Responsibility} && {\tiny$\bullet$} \textit{Closing the AI accountability gap: Defining an end-to-end framework for internal algorithmic auditing}~\citep{raji2020closing}\\
        & \textbf{Beneficence} && {\tiny$\bullet$} \textit{Consequence Scanning: An Agile event for Responsible Innovators}~\citep{doteveryone}    \\
        & \textbf{Dignity} && {\tiny$\bullet$} \textit{Equity Evaluation Corpus}~\citep{kiritchenko2018examining}\\
        & \textbf{Privacy} && {\tiny$\bullet$} \textit{Should I disclose my dataset? Caveats between reproducibility and individual data rights}~\citep{m-benatti-etal-2022-disclose}\\
        \hline
        
        \parbox[t]{1mm}{\multirow{4}{*}{\rotatebox[origin=c]{90}{\textbf{Audience}}}} & \textbf{Data Scientist} && {\tiny$\bullet$} \textit{\textit{Aequitas: A Bias and Fairness Audit Toolkit}}~\citep{saleiro2019aequitas}\\ 
        & \textbf{Examiners} && {\tiny$\bullet$} \textit{TuringBox: An Experimental Platform for the Evaluation of AI Systems}~\citep{ijcai2018p851}\\ 
        & \textbf{Students} && {\tiny$\bullet$} \textit{AI Audit: A Card Game to Reflect on Everyday AI Systems}~\citep{Ali_Kumar_Breazeal_2024}\\
        \bottomrule
    \end{tabular}
    \label{tab:egAIETs}
    \vspace{0.1cm}
    {\footnotesize Observations: Some AIETs can take on more than one classification; AIETs can have more than one objective, ethical principle or target audience, although the table shows only one; Examples have not been provided for all the existing principles and possible target audiences.}
\end{table}

As mentioned earlier, AIETs aim to help developers, researchers, companies, governments, and other stakeholders create more ethical AIs. However, although promising, many AIETs lack good documentation, examples of use, evidence of their effectiveness, and are not yet well adopted by their target audience, as well as being proposed for AIs in general~\citep{morley2020,qiang2023}. This paper practically investigates these issues, applying AIETs to language models. It seeks to understand the effectiveness of AIETs in mapping ethical issues already identified by the literature on these models. We assess interviews with the developers of language models proposed for the Portuguese language, the usability of AIETs, their readability, whether they can guide ethical thinking about technology, and the possible future adoption by those interviewed.

\begin{figure}
    \centering
    \includegraphics[width=0.9\linewidth]{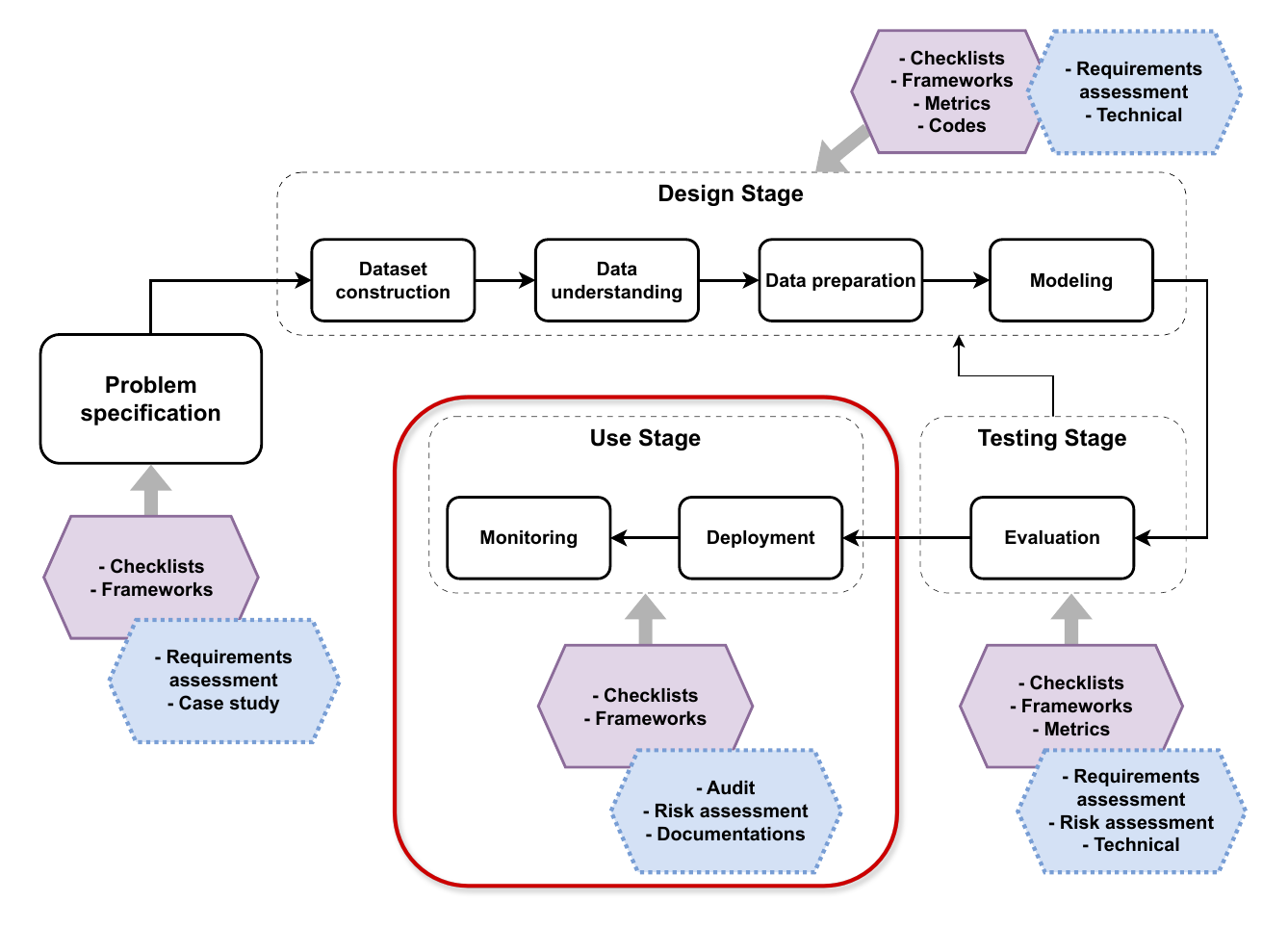}
    \caption{AIETs can be applied throughout the AI life-cycle, from problem specification to the use stage. Colored hexagons indicate examples of AIETs types that can be applied in each of the stages. In purple, full border, the categories of AIETs. In blue, dotted border, the objectives. The AIETs analyzed in this study are exclusively applied in the use stage of the AI life-cycle, indicated by the solid red border}
    \label{fig:iacycle}
\end{figure}

All AIETs analyzed in this research---Model Cards by~\citet{mitchell2019}, ALTAI by AI HLEG~\citep{altai}, FactSheets by~\citet{arnold2019factsheets}, and Harms Modeling by~\text{\citet{harmsmodeling}}---fall into the Checklists and Questionnaires category and are applied exclusively in the use stage of the AI life-cycle, as shown in Fig.~\ref{fig:iacycle}. In other words, these AIETs consist of lists of criteria, checks, or questions that guide the ethical analysis and decision-making related to AI during the deployment and use stages. The primary objective of AIETs in this category is to offer guidance on various ethical considerations associated with AI, including details about datasets, the components of AI systems,  requirements, system behaviors, potential biases, the impacts of the technology, and raising ethical considerations. 

\subsection{Overview of AIETs}
\label{sec:AIETs_overview}

This section presents the most closely works related to this paper, providing an overview of the works that analyze AIETs as a whole.
Several AIETs have been proposed to evaluate ethical principles in practice. These principles have been extensively discussed in the literature~\citep{zhou2020,schiff2020s,floridi2019translating,khan2022ethics,correa2023}. \citet{jobin2019} carried out a significant work with a comprehensive overview of 84 AI ethics guidelines\footnote{Here, as a way of presenting concepts related to this paper, one of the main works in the area was presented, carried out by~\citet{jobin2019}. However, the work by~\citeauthor{jobin2019} has already been updated by other authors, such as~\citet{correa2023}, who have added more guidelines to the review (200--until 2023). However, the results are close and the discussions pointed out by~\citeauthor{jobin2019} are still current.}. The results indicate that the guidelines converge on five ethical principles: transparency, justice and fairness, non-maleficence, responsibility, and privacy. Other less recurrent principles are beneficence, freedom and autonomy, trust, sustainability, dignity, and solidarity. \citet{jobin2019} also discussed that although there is convergence around some principles, the analysis revealed significant semantic and conceptual divergences in how these ethical principles are interpreted and in the specific recommendations or areas of concern derived from each one. \citet{ryan2020} provided a compilation of the content of the normative requirements arising from these guidelines for a better understanding of the codes of ethics included in each principle.
 
In recent years, some papers have significantly discussed about the implications of translating ethical principles into viable AIETs for use in AI technologies and their impacts, and the following are the most related to this paper. 

\citet{morley2020} were among the first to present a typology of over 100 AIETs for applying ethics in AI, discussing ways that ethical principles could be implemented in AI technologies. The authors categorized the AIETs into five ethical AI principles: beneficence, non-maleficence, autonomy, justice, and explainability, and by application stages in the AI life-cycle. They discuss ``\textit{how efforts to date have been too focused on the `what' of ethical AI }(\textit{i.e., debates about principles and codes of conduct\emph{)} and not enough on the `how' of applied ethics}'' and point out that AIETs offer little help on how to use them in practice, have limited documentation and require a high level of skill to use.  

\citet{prem2023} extended the work of~\citet{morley2020} by systematically analyzing AIETs from an operational and implementation perspective. The author classified AIETs in terms of their categories (\emph{e.g.}, checklists, declarations, audits, and metrics), their implementations (\emph{e.g.}, algorithms, libraries), and the ethical problems they aim to solve (general ethical aspects, privacy, fairness and bias, explainability, accountability, transparency, correctness and accuracy, diversity, robustness, and reproducibility). \citeauthor{prem2023} also discussed the complexity involved in putting the principles into practice by developers since the principles are complex and challenging to implement (``\textit{For example, it remains unclear what `interpretability' means and how it should be implemented.''\emph{)}.}

\citet{ayling2022} presented a typology with 39 AIETs to apply ethics in AI, as does~\citet{morley2020}, aiming to translate ethical principles into practice. What sets this work apart is the presentation of AIETs' analyses based on the sectors that develop and use them, the support for various stakeholders, and the categories and stages of the AI life-cycle in which they can be applied. The authors also contrasted that AIETs emerged in a scenario with no regulations for AI technologies, which could minimize their adoption and application.

\citet{wong2023} qualitatively analyzed 27 AIETs to critically examine how the work of ethics is imagined and how these AIETs support it. The authors' analysis focused on the discourses that AIETs rely on when talking about ethical issues and their implications, and finally, they presented three recommendations to improve the way AIETs support AI ethics: \textit{``\emph{(1)} provide support for the non-technical dimensions of AI ethics work; \emph{(2)} support the work of engaging with stakeholders from non-technical backgrounds; \emph{(3)} structure the work of AI ethics as a problem for collective action.''}.

The four papers presented previously discuss the gaps in putting principles into practice through the use of AIETs and carry out a general assessment of these AIETs and their specifications, providing insights into how ethics currently being applied in AI technologies. However, analyses were conducted only theoretically and did not present practical research and comparisons. 

\citet{qiang2023}, as far as we know, were the first authors to assess AIETs practically in a comparative case study. The authors compared four AIETs---Foresight into AI Ethics Toolkit~\citep{fttool}, Guidelines for Trustworthy AI~\citep{egttool}, Ethics and Algorithms Toolkit~\citep{eattool}, and Algorithmic Impact Assessment~\citep{aiatool}---from the perspectives of an auditor conducting an AI ethics risk assessment for a company and that of the company receiving the AIETs' final documentation. The authors discussed the lack of clarity and explanations of AIETs outcomes. In addition, the authors pointed out the need for more standardization in the results of AIETs, which makes their use more complicated. The authors applied the AIETs to an AI that was still being developed, and the results suggest that each analyzed AIET provides a different benefit and that there is no unique solution to address all AI ethical issues adequately. The authors also pointed out that, to better adopt AIETs, it is necessary to specify the suitability and expected benefits of existing AIETs. 

This paper is related to the work developed by~\citet{qiang2023}. However, their studies focused on AIETs that could be used in the development stage of the analyzed AI. In contrast, our work addresses the scenario of AIETs applied to an already-developed AI. We focused on evaluating the effectiveness of existing AIETs in helping developers of language models identify the risks and harms of their model while comparing the AIETs analyzed. We chose language models as a case study because the literature has already mapped several risks of these models, so we can compare whether the ethical considerations identified through AIETs align with the literature. 

%%==================================%%
%%             Method               %%
%%==================================%%
\section{Method}
\label{sec:methodology}

\begin{figure}[!htb]
\centering
    \subfigure[][AIET Selection]{\includegraphics[width=\textwidth,height=\textheight,keepaspectratio, trim={0, 11.5cm, 0, 0cm}, clip]{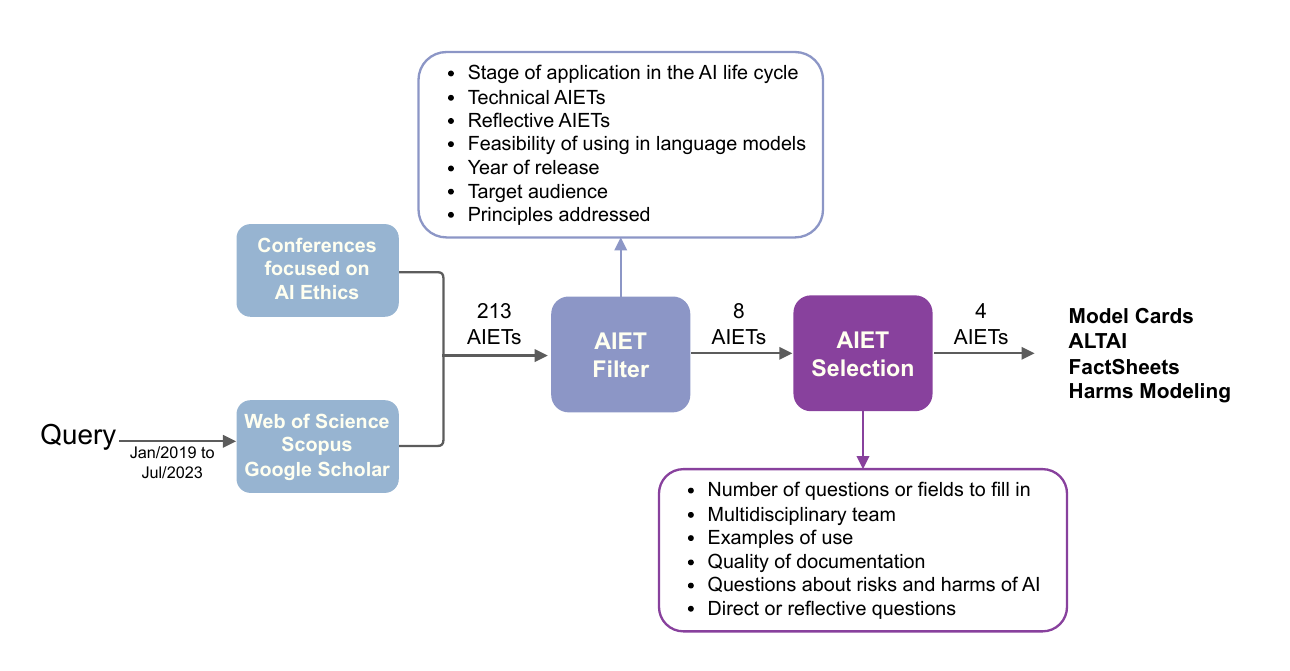}\label{fig:m1}}\\
    \subfigure[][Language Model Selection]{\includegraphics[scale=0.55, ,keepaspectratio, trim={0, 19.5cm, 0, 0cm},clip]{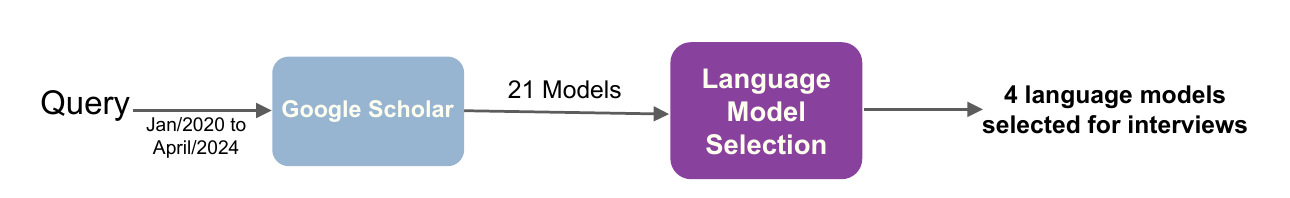}\label{fig:m2}}\\
    \subfigure[][Interviews: AIET Application and Evaluation]{\includegraphics[width=\textwidth,height=\textheight,keepaspectratio, trim={0, 19.7cm, 0, 0cm},clip]{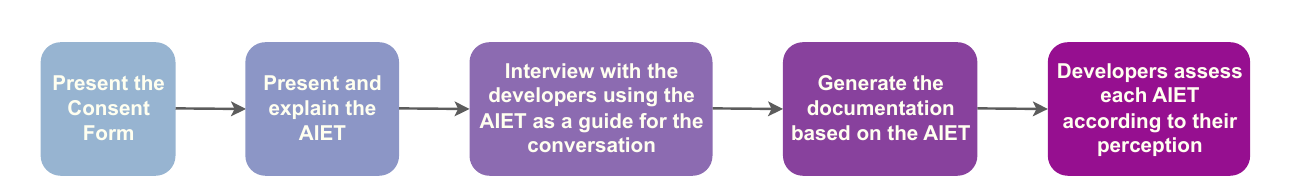}\label{fig:m3}}\\
    \caption{Methodology overview. The initial stage of this project involved conducting a bibliographic survey of AIETs and selecting the specific AIETs for evaluation (Fig.~\ref{fig:m1}). Following the selection, we assembled the interview scripts from the AIETs questions and carried out an initial test. Once we defined the protocol, we submitted a project to the Research Ethics Committee to obtain authorization to conduct the study, given that it involved interviews with human participants. After securing the Ethics Committee's approval, we carried out a bibliographic survey to select language models in Portuguese and  invite developers to participate in the study (Fig.~\ref{fig:m2}). Finally, we conducted the interviews and the AIETs were assessed by the developers (Fig.~\ref{fig:m3})}
    \label{fig:method}
\end{figure}

Our main goal is to understand how an AIET can help developers of language models to raise ethical issues. In this section, we describe the methodology for achieving this goal, detailing the AIETs selected, the language model selection, and how the AIETs were applied and evaluated. Fig.~\ref{fig:method} shows an overview of the methodology adopted in this paper. 

\subsection{AIET Selection}
\label{sec:AIETs_filtered}

To evaluate AIETs in language models, we initially conducted an extensive literature survey. This survey was carried out from January 2019 to July 2023\footnote{To keep this paper up to date, we provide the list of AIETs released until March 2025. In this way, we provide both the original list of AIETs surveyed up to July 2023 (Appendix~\ref{sec:aietslist}) and the updated list from August 2023 to March 2025 (Appendix~\ref{sec:aietslistupdate}). The selection of the analyzed AIETs considered only the ones released up to July 2023.} and aimed to identify the available AIETs up to that point. We performed the searches using the search engines 
Web of Science, %\footnote{\url{https://webofscience.help.clarivate.com/en-us/Content/home.htm}}, 
Scopus, %\footnote{\url{https://www.elsevier.com/pt-br/products/scopus}}, 
and Google Scholar %\footnote{\url{https://scholar.google.com/}} 
with the following query:

\begin{quote}
    \centering\texttt{(tool* OR framework* OR method*) AND ("artificial intelligence" OR "machine learning" OR "natural language processing" OR "language model*") AND ethic*}
\end{quote}

We also searched the AIETs directly in journals and conferences focused on AI Ethics:
FAccT (Conference on Fairness, Accountability, and Transparency), %\footnote{\url{https://dl.acm.org/conference/facct}}, 
AIES (Conference on AI, Ethics, and Society), %\footnote{\url{https://dl.acm.org/conference/aies}}, 
AI and Ethics, %\footnote{\url{https://link.springer.com/journal/43681}}, 
AI \& Society, %\footnote{\url{https://link.springer.com/journal/146}}, 
and Philosophy \& Technology. %\footnote{\url{https://link.springer.com/journal/13347}}.

During the literature review, we found five articles that mapped other AIETs (more than 130 AIETs)~\citep{morley2020,ayling2022,palladino2022,wong2023,prem2023}. At the end of the search, we collected 213~AIETs\footnote{288 AIETs with the update made until March 2025.}---including the AIETs indicated by the five articles cited above. Due to the abundance of AIETs, we selected those that best suited the purpose of this research using the following inclusion and exclusion criteria:

\begin{itemize}

    \item \textit{Stage of application in the AI life-cycle}: AIETs can be applied at different stages of the AI life-cycle. In this paper, we evaluate the ethical considerations of a ready-for-use language model. Therefore, the selected AIET must be applied in this final stage of the AI life-cycle, the use stage. We excluded all AIETs that can only be applied to the previous stages (problem specification, design, and testing) such as~\citep{turing2021turing, gebru2021, reisman2018algorithmic}.
        
    \item \textit{Technical AIETs}: Some AIETs were proposed for unique and specific objectives such as privacy~\citep{tensorflowprivacy}, explainability~\citep{EthicalMLxai}, and robustness~\citep{nicolae2019adversarialrobustnesstoolboxv100}, and required their application in the target model codes, thus being technical AIETs. These AIETs did not meet the objective of this work and, therefore, were excluded.
    
    \item \textit{Reflective AIETs}: We kept all the AIETs that generated ethical considerations about the developed model. Usually, these AIETs are questionnaires or checklists, where developers can produce documentation of their AI developed focusing on the ethical considerations raised. In this way, we excluded all the AIETs that did not help developers raise ethical issues, such as AIETs that were published to serve as a guide on how to carry out an ethical audit on an AI without presenting questionnaires or checklists~\citep{mokander2024auditing, raji2020closing}.

    \item \textit{Feasibility of using the AIET in language models}: Some AIETs have specific focuses, such as only image models~\citep{dhurandhar2018explanations,goodman2020advboxtoolboxgenerateadversarial}. Since we focused on analyzing a language model, we excluded all AIETs that were not feasible to use in our work.
    
    \item \textit{Year of release}: We kept AIETs published from 2019 onwards to focus on tools developed during the recent proliferation of language models and associated ethical guidelines.
    
    \item \textit{Target audience}: Some AIETs were developed to be answered by companies through group tutorials and company-wide workshops~\citep{communityjury}. In this work, we selected AIETs that could be applied exclusively to the model's developers, and that could be applied in an interview format. 

    \item \textit{Principle addressed}: Some AIETs have been developed with a focus on just one ethical principle, such as only on fairness~\citep{saleiro2019aequitas,10091496} or only on privacy~\citep{GOLDSTEEN2023101352,guidetouk}. We kept the AIETs that had a combination of at least three ethical principles. This analysis was carried out by checking the items contained in the tools (\textit{e.g.}, questions, checklists) to identify the presence of the principles already discussed in the literature~\citep{jobin2019}.
\end{itemize}

Following these inclusion and exclusion criteria, we end up with eight AIETs~\citep{mitchell2019,harmsmodeling,altai,oecd,agarwal2024seven,arnold2019factsheets,aipb,mlmm}. Due to the time required to apply all AIETs found through interviews with the developers, we decided to evaluate four AIETs. 

To select the final AIETs, we classified the AIETs found  (see Table~\ref{tab:8-aiets}) by (i)~the number of questions or fields the AIET contained, (ii)~whether the team proposing the AIET was multidisciplinary, (iii)~whether the AIET presented examples of use in its documentation, (iv)~the quality of the documentation---whether it was possible to understand how to use the AIET, (v)~whether the AIET contained questions or fields to fill in about risks and harms of the developed AI, and (vi)~whether the AIET contained direct questions (answered with yes or no) or had reflective questions, since the way the question is presented changes the way developers can answer the question, (\textit{e.g.}, ``During the development of the model, were there any concerns about biases? [The answer to this question can be more direct, and can even be answered with just `yes' or `no']'' and ``What concerns related to bias were taken into account during the development of the model? [The answer to this question will not be as direct as the previous question, because if the participant's answer was `yes' in the previous question, in this one they will describe in more detail the concerns taken into account]'').

The selection aimed to use different AIETs to assess developers' perspectives of the different AIETs and find their preferences for future use. The resulting collection of eight works appears in Table~\ref{tab:8-aiets}. The AIETs selected are Model Cards~\citep{mitchell2019}, ALTAI~\citep{altai}, FactSheets~\citep{arnold2019factsheets}, and Harms Modeling~\citep{harmsmodeling}. In the following, we briefly describe the four AIETs selected.

\textbf{\textit{Model Cards for Model Reporting}}~\citep{mitchell2019}, referred hereafter as \textit{Model Cards}, are documents that accompany an AI to provide transparent and accountable documentation about the technology, allowing interested parties to be informed about all aspects of the AI in question, both in technical details and ethical positions. To this end, \citeauthor{mitchell2019} proposed a structure with nine topics to comprise a Model Cards that must be filled in by those responsible for the proposed AI: (1) model details, (2) intended use, (3) factors, (4) metrics, (5)~evaluation data, (6) training data, (7) quantitative analyses, (8) ethical considerations, and (9) caveats and recommendations.

\textbf{\textit{The Assessment List for Trustworthy Artificial Intelligence} (ALTAI)}~\citep{altai} aims to help developers, companies, and/or organizations (self-)assess an AI through a questionnaire to create more trustworthy technologies by identifying how AI can generate risks for society and the environment and what actions to take to minimize these risks. ALTAI contains questions categorized into seven topics: (1)~human agency and oversight, (2) technical robustness and safety, (3) privacy and data governance, (4) transparency, (5) diversity, non-discrimination, and fairness, (6)~societal and environmental well-being, and (7) accountability.

\textbf{\textit{FactSheets: Increasing Trust in AI Services through Supplier's Declarations of Conformity}}~\citep{arnold2019factsheets}, referred hereafter as \textit{FactSheets}, aims to create a documentation about an AI through a questionnaire, which is designed to give stakeholders more trust in the technology. FactSheets helps to create documentation on the developed AI to inform interested parties about how AI works, its applications, limitations, and other information. FactSheets contains questions categorized into five topics: (1) statement of purpose, (2) basic performance, (3) safety, (4) security, and (5) lineage.

\textbf{\textit{Harms Modeling}}~\citep{harmsmodeling} aims to anticipate a technology's potential for risks and identify deficiencies in the technology that could put society and the environment at risk. Harms Modeling comprises stages where those responsible for the technology define the purpose and use cases, analyze the impacts and their stakeholders, and carry out a harm assessment of the technology. Harms Modeling addresses several types of harm that should be assessed according to their severity, scale, probability, and frequency of occurring and affecting personal and social well-being, and are grouped within the following types of risks: (1) risk of injury (physical injury and emotional or psychological injury), (2) denial of consequential services (opportunity loss and economic loss), (3) infringement on human rights (dignity loss, liberty loss, privacy loss, and environmental impact), and (4) erosion of social and democratic structures (manipulation and social detriment).

\begin{sidewaystable}[p]
    \setlength{\tabcolsep}{3mm}
    \renewcommand{\arraystretch}{1.7}
    \footnotesize
    \centering
    \caption{AIETs overview. Here, we compare the eight AIETs filtered according to the criteria (i)~\textbf{\#Q}: Number of questions or fields to fill,  
   (ii) \textbf{IA}: if it provides information about the authors, like name and affiliations, (iii) \textbf{EG}: if it provides examples of use, (iv) \textbf{DQ}: Documentation quality (good, regular, and bad),  (v) \textbf{RH}: if it evaluates risks and harms, (vi) \textbf{D/R}: if it has direct or reflective questions, and (vii) the \textbf{Differential} about the AIETs. \textbf{T}:~Application time (hours) in the testing interviews}
    \begin{tabular}{
    >{\raggedright\arraybackslash}p{2.3cm}
    >{\raggedright\arraybackslash}p{1.3cm}
    >{\centering\arraybackslash}p{0.5cm}
    >{\centering\arraybackslash}p{0.6cm}
    >{\centering\arraybackslash}p{0.5cm}
    >{\centering\arraybackslash}p{0.6cm}
    >{\raggedright\arraybackslash}p{1cm}
    >{\centering\arraybackslash}p{0.6cm}
    >{\raggedright\arraybackslash}p{1.3cm}
    >{\raggedright\arraybackslash}p{4.5cm}}
    \hline
    \textbf{AIET} & \textbf{Reference} & \textbf{T} & \textbf{\#Q} & \textbf{IA} & \textbf{EG} & \textbf{DQ}  & \textbf{RH} & \textbf{D/R}    & \textbf{Differential}
    \\\hline
    \arrayrulecolor{black!15}
    \textbf{Model Cards} & \citep{mitchell2019} & 2h & 25 & \checkmark & \checkmark & good  & \checkmark & --   & Only AIET for responsible documentation in the format of a fill-in form.\\\hline
    
    \textbf{FactSheets} & \citep{arnold2019factsheets} & 2h & 119 & \checkmark & \checkmark & good  & \checkmark & both   & Best example of use. It has questions that make references to other AIETs~\citep{gebru2021, bender2018data}.\\\hline
    
    Machine Learning Maturity Model & \citep{mlmm} & -- & 64 & $\times$ & $\times$ & regular  & $\times$ & direct   & All topic sections aim to assess the quality of the AI developed team.\\\hline
    
    AI Procurement in a Box & \citep{aipb} & -- & 119 & \checkmark  & $\times$ & bad & \checkmark  & both  & Provides case studies, but does not show the use of the AIET.\\\hline
    
    \textbf{ALTAI} & \citep{altai} & 3h & 142 & \checkmark & $\times$ & regular & \checkmark & direct     & Provides a glossary of important terms.\\\hline
    
    \textbf{Harms Modeling }& \citep{harmsmodeling} & 4h & 94 & $\times$ & \checkmark & regular  & \checkmark & reflective   & Focuses exclusively at assessing risk and harms and evaluate the potential magnitude of them.\\\hline
    
    OECD framework for the classification of AI systems  & \citep{oecd} & -- & 37 & \checkmark & \checkmark & good & $\times$  & direct   & Provides extensive documentation about the AIET and AI topics.\\\hline
    
    Seven-Layer Model with Checklists  & \citep{agarwal2024seven} & -- & 60 & \checkmark & \checkmark & good & \checkmark & both    & Focuses exclusively on fairness assessment.\\\arrayrulecolor{black}\hline
    \end{tabular}
    \label{tab:8-aiets}
\end{sidewaystable}

\subsection{Scripting and Testing the Interviews}

All AIETs selected for this work fall within the Checklists and Questionnaires category, where the AIET presents a set of questions that the participants must consider. In this way, the AIETs were used as a script for each interview, using the AIET questions as a guide for the conversation. 

To test the methodology proposed for the interviews, we conducted a pilot study with the developers of the \capivaraicon{1.2em}CAPIVARA language model~\citep{santos2023capivara}, a model developed by members of the Natural Language Processing group of the Artificial Intelligence and Cognitive Architectures Hub (H.IAAC), the research group in which this work was done. \capivaraicon{1.2em}CAPIVARA is an open source model proposed for low-resource languages focusing on Portuguese. For the test, we invited and interviewed the three main developers of the model to fill in the selected AIETs. Each interview session had a maximum duration of one hour, resulting in 11 interviews, all of which every developer attended. We randomly selected the AIETs' order and applied them as follows: Model Cards, FactSheets, ALTAI, and Harms Modeling. The total time spent interviewing the developers for each AIET was approximately two hours for Model Cards, two hours for FactSheets, three hours for ALTAI, and four hours for Harms Modeling.

The AIETs best evaluated by the developers were Model Cards and Harms Modeling, as will be presented in Section~\ref{sec:res_capivara}. Considering the duration necessary for conducting the interviews and the overall feasibility of the study, the subsequent evaluations within this research will be exclusively focused on the AIETs' Model Cards and Harms Modeling.

\subsection{Language Model Selection}
\label{sec:sel_modelos_linguagem}

Following the final AIET selection, we sought authorization from the Research Ethics Committee to conduct the study, as it involves collecting data from human participants. Once the project received approval, we invited the language model developers to participate in the study.

To select the language models developed with textual data in Portuguese\footnote{In this paper, we focus only on language models developed exclusively for the Portuguese language, thus not considering multilingual models.}, a literature search was carried out in February 2024 (updated in March 2025\footnote{As with the AIETs, to keep this paper up to date, we provide the list of language models released until March 2025. However, the selection of the interviewed models considered only the ones released up to March 2024.}) with the following query in the search engine Google Scholar, where we visited the first 20~pages, filtering the search from 2020\footnote{As far as we know, the first language models developed for the Portuguese language were published in 2020.} onwards: 

\begin{quote}
    \centering\texttt{"language model"~AND~"portuguese"}
\end{quote}

\begin{figure}[!htb]
    \centering
    \includegraphics[width=\linewidth, trim={1.2cm, 1.6cm, 0.7cm, 0cm}, clip]{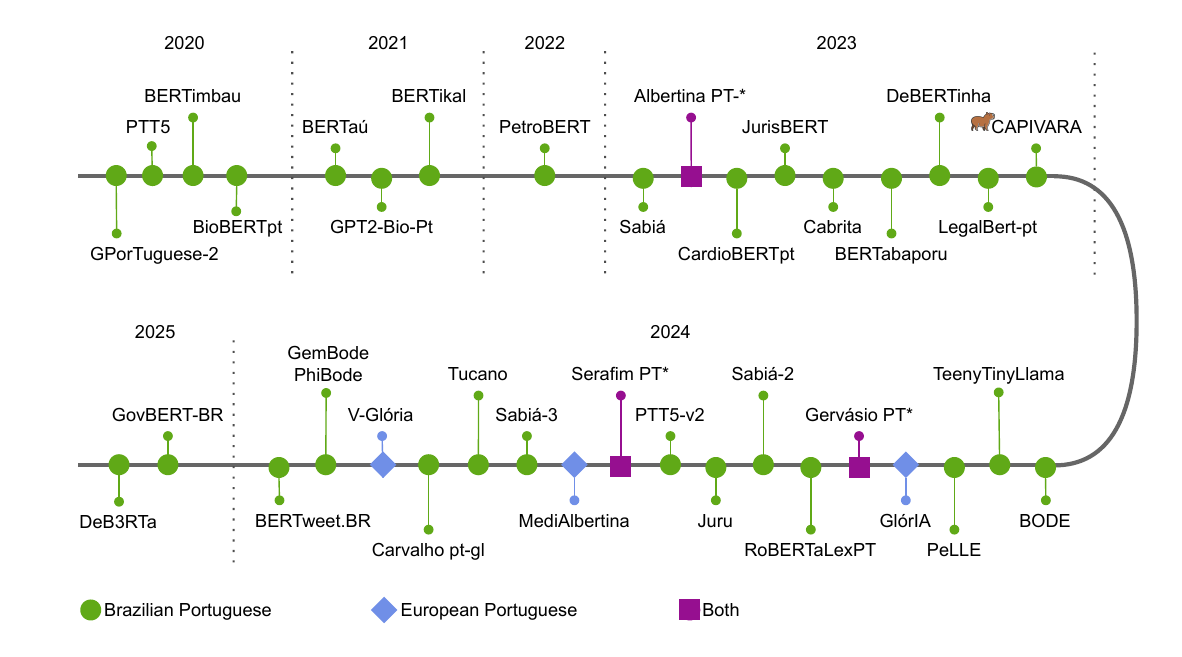}
    \caption{Language models developed for the Portuguese language releases from 2020 to March 2025: GPorTuguese-2~\citep{pierre2020gpt2smallportuguese}, PTT5~\citep{ptt5}, BERTimbau~\citep{bertimbau}, BioBERTpt~\citep{biobertpt}, BERTáu~\citep{bertau}, GPT2-Bio-Pt~\citep{gpt-2bio}, BERTikal~\citep{polo2021legalnlpnaturallanguage}, PetroBERT~\citep{petrobert}, Sabiá~\citep{sabia}, Albertina PT-*~\citep{albertina}, CardioBERTpt~\citep{cardiobertpt}, JurisBERT~\citep{JurisBERT}, Cabrita~\citep{cabrita}, BERTabaporu~\citep{bertabaporu}, DeBERTinha~\citep{debertinha}, CAPIVARA~\citep{santos2023capivara}, LegalBert-pt~\citep{legalbert-pt}, BODE~\citep{bode}, TeenyTinyLlama~\citep{teenytinyllama}, PeLLE~\citep{demello2024pelle}, GlórIA~\citep{gloria}, Gervásio PT*~\citep{santos_advancing_2024}, RoBERTaLexPT~\citep{robertalexpt}, Sabiá-2~\citep{almeida2024sabi}, Juru~\citep{junior2024juru}, PTT5-v2~\citep{piau_ptt5-v2_2025}, Serafim PT*~\citep{gomes_open_2025}, MediAlbertina~\citep{nunes_medialbertina_2024}, Sabiá-3\citep{abonizio_sabia-3_2025}, Tucano~\citep{correa_tucano_2024}, Carvalho pt-gl~\citep{gamallo_galician-portuguese_2025}, V-Glória~\citep{simplicio_v-gloria_2024}, GemBode e PhiBode~\citep{garcia_gembode_2025}, BERTweet.BR~\citep{carneiro_bertweetbr_2025}, GovBERT-BR~\citep{silva_govbert-br_2025}, and DeB3RTa~\citep{pires_deb3rta_2025}. \includegraphics[width=0.25cm, trim={0 0 8.5cm 0}, clip]{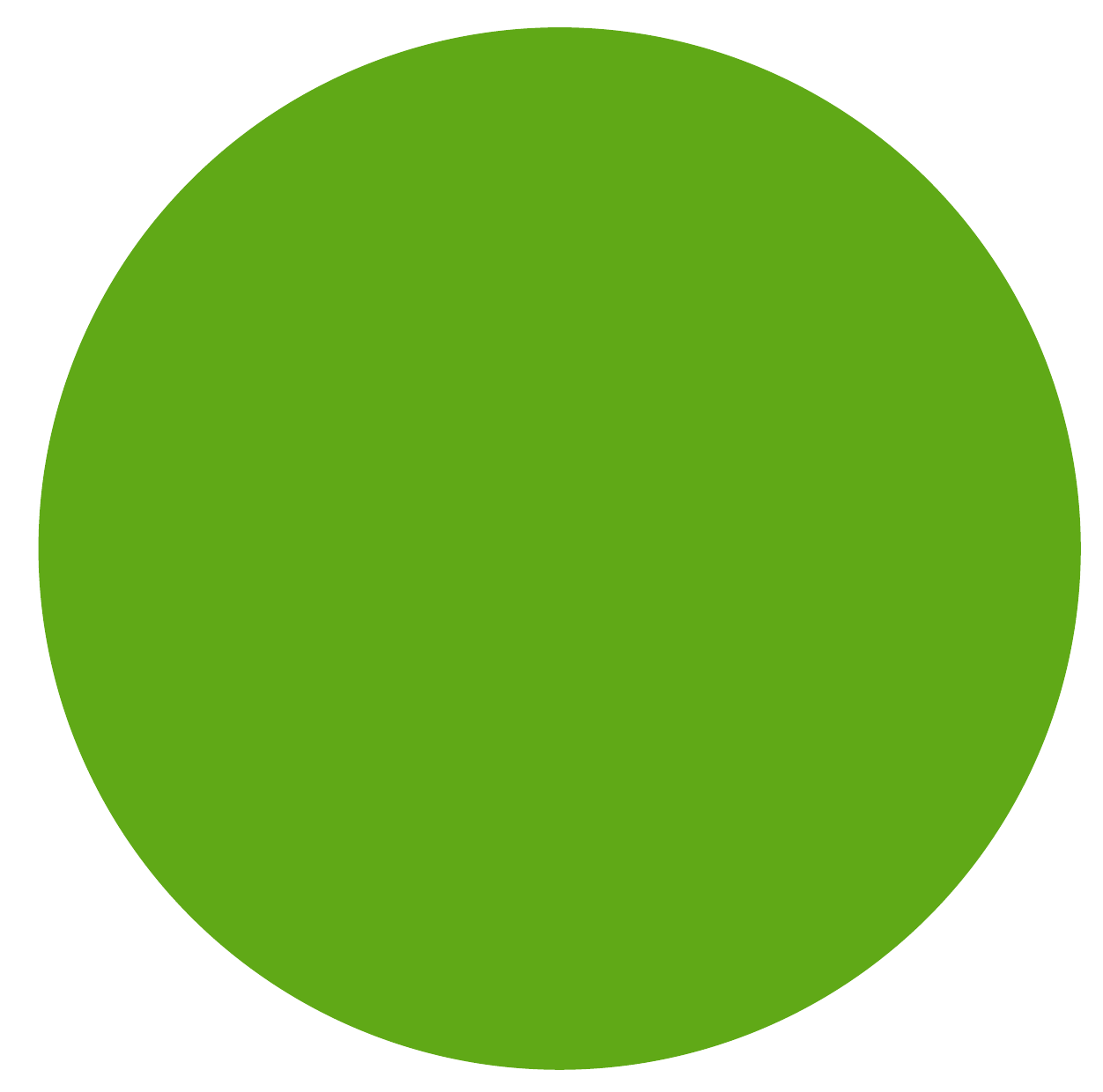}~Brazilian Portuguese, \includegraphics[width=0.3cm, trim={0cm 0cm 0.5cm 0}, clip]{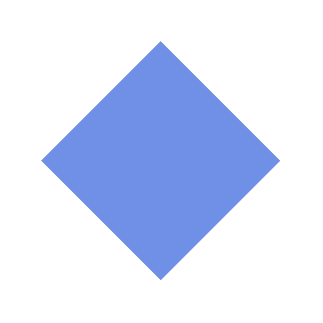}~European Portuguese, \includegraphics[width=0.25cm, trim={0 0 0.5cm 0}, clip]{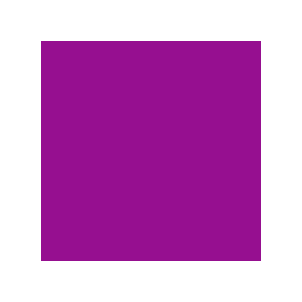}~Both, Brazilian and European Portuguese}
    \label{fig:LMs_time}
\end{figure}

Fig.~\ref{fig:LMs_time} presents the surveyed language models. Among these, we randomly selected four models to participate in the interviews\footnote{We selected only four models due to the time required to conduct the study. As detailed in Section~\ref{sec:limitations}, carrying out a study on an AIET requires a significant amount of time from both the researchers and the participants. In total, we conducted 35 interview sessions, each lasting roughly one hour. Additionally, documenting each conversation required about two hours. For each group of participants, the interviews were spread over 2 to 3 months.}. Volunteer participants were recruited by the researcher responsible for this study via e-mail.  To preserve the confidentiality of the participants, we will not disclose the specific models interviewed in this paper. In addition, aspects of ethnicity, sexual orientation, gender identity, class, social group, and physical size were not relevant to the selection of participants. Furthermore, we did not collect such information because the study population was small, making the participants easily identifiable, and thus breaking their confidentiality.
A total of 11~developers participated in the interviews, which were organized into four groups of 2, 2, 3, and 4 participants, corresponding to the group of developers from each model. Developers from different models had no interaction during the interviews and their participation are anonymous\footnote{Note that CAPIVARA model was a pilot case outside this anonymous group.}.

\subsection{AIET Application and Evaluation}

\subsubsection{Interview and AIET Documentation}

After selecting the AIETs and having the study approved by the Research Ethics Committee, we remotely interviewed the language model developers who agreed to participate using the online meeting platform \textit{Google Meet}. We recorded all the interviews and later transcribed them with the appropriate authorizations and consent from all the participants. Each interview session lasted a maximum of one hour and included the participation of all developers from the same model. 

At the first meeting, we presented to the participants a Consent Term setting out the objectives and justifications of the study, the importance of the study, the procedures, and methodology that would be adopted, how the data collected would be handled, the possible discomforts and risks of the study, the benefits and the ethical guarantees for the participants. In addition, at the beginning of each interview, we took a moment to answer any questions the participants might have and to help with whatever was necessary for the progress of the research.

At the beginning of each interview with a new AIET, we took a moment to introduce the AIET to the developers and answer any questions they had about it. The AIETs were answered verbally by the developers during the interviews, and the interviews were conducted in Portuguese---we read all the questions in Portuguese---we presented the English version for better understanding when a direct translation was not possible---and the developers answered the questions in Portuguese.

After the interviews, we used audio recordings and transcriptions to create the final documentation for each AIET. This documentation was prepared following the requirements of each AIET, that is, it is the filled-in AIET. This documentation is confidential and only the developers had access to the final document for their model.

\subsubsection{Developers' Feedback}

Based on the final documentation---the filled-in AIETs---the participants were invited to individually complete a questionnaire to identify their perspective of the AIETs and the ethical considerations raised by each applied AIET.

For each applied AIET, we asked each developer their agreement with statements such as ``The AIET is easy to use/respond'', ``The AIET is useful for identifying the risks of the language model'', ``The AIET is useful for generating responsible documentation about the language model'', and ``It is complicated to identify the ethical considerations of the language model using the AIET''. In addition, we asked whether they would use a given AIET again in other projects, whether the developer had already thought about the ethical considerations of their model before the interviews, which AIET raised the most ethical concerns about their model, the documentation they consider most informative about ethical considerations, and whether they already knew and used AIETs. We asked them to rate each AIET used between 1 and 5, with 5 being the most positive assessment of the AIET. Table~\ref{tab:forms1} presents the complete questionnaire used for this stage. 

Following the risks listed by~\citet{weidinger2021}, and adding the following risks to the list: ``Lack of representation of cultural and social aspects of the Brazilian population'', ``Lack of representation of cultural and social aspects of the Portuguese-speaking population'', and ``Low performance in the Portuguese language'', we asked developers if they identified that risk as appearing in the language model they developed, and if so, which AIET helped them to identify that risk. Table~\ref{tab:forms2} presents the complete questionnaire used for this stage.

This questionnaire took each developer around 30 minutes to complete, and they had 15 days after receiving the final documentation to individually complete the questionnaire. We completed the interviews  in July 2024. Section~\ref{sec:results} presents and discusses the results obtained in our study.

\begin{table}[!htp]
    \centering
    \renewcommand{\arraystretch}{1.48}
    \footnotesize
    \caption{Questionnaire to extract developers' perceptions about the analyzed AIETs [Part 1]}
    
    \newcommand{\rc}[3]{\multicolumn{#1}{>{\raggedright\arraybackslash}p{#2\textwidth}}{#3}}
    \newcommand{\bc}{$\bigcirc$}
    \newcommand{\ebc}[1]{\end{tabular}\\\hline\begin{tabular}{#1}}
    
    \begin{tabular}{|l|}\hline\arrayrulecolor{black!15}
    
    \begin{tabular}{lllll}
    \rc{5}{0.93}{\textbf{Considering the AIET Harms Modeling, indicate your opinion on the following statements:}}\\
    
    \rc{5}{0.93}{\afirmacaoum}\\
    \hspace{0.5cm}\bc~Agree & \bc~Weakly agree & \bc~Neutral & \bc~Weakly disagree & \bc~Disagree \\

    \rc{5}{0.93}{\afirmacaodois}\\
    \hspace{0.5cm}\bc~Agree & \bc~Weakly agree & \bc~Neutral & \bc~Weakly disagree & \bc~Disagree \\

    \rc{5}{0.93}{\afirmacaotres}\\
    \hspace{0.5cm}\bc~Agree & \bc~Weakly agree & \bc~Neutral & \bc~Weakly disagree & \bc~Disagree \\

    \rc{5}{0.93}{\afirmacaoquatro}\\
    \hspace{0.5cm}\bc~Agree & \bc~Weakly agree & \bc~Neutral & \bc~Weakly disagree & \bc~Disagree \\
     
    \ebc{lllll}

    \rc{5}{0.93}{\textbf{Considering the AIET Model Cards, indicate your opinion on the following statements:}}\\
    
    \rc{5}{0.93}{\afirmacaoum}\\
    \hspace{0.5cm}\bc~Agree & \bc~Weakly agree & \bc~Neutral & \bc~Weakly disagree & \bc~Disagree \\

    \rc{5}{0.93}{\afirmacaodois}\\
    \hspace{0.5cm}\bc~Agree & \bc~Weakly agree & \bc~Neutral & \bc~Weakly disagree & \bc~Disagree \\

    \rc{5}{0.93}{\afirmacaotres}\\
    \hspace{0.5cm}\bc~Agree & \bc~Weakly agree & \bc~Neutral & \bc~Weakly disagree & \bc~Disagree \\

    \rc{5}{0.93}{\afirmacaoquatro}\\
    \hspace{0.5cm}\bc~Agree & \bc~Weakly agree & \bc~Neutral & \bc~Weakly disagree & \bc~Disagree \\
     
    \ebc{lllll}
    
    \rc{5}{0.93}{\textbf{Which AIET would you prefer to use again if you were to propose/develop a new model?}}\\
    \hspace{0.5cm}\bc~Harms Modeling & \bc~Model Cards &&&\\

    \ebc{lllll}
    
    \rc{5}{0.93}{\textbf{Although you indicated a preference in the previous question, would you actually use the AIET again?}}\\
    \hspace{0.5cm}\bc~Yes & \bc~No & \bc~Maybe &&\\
    
    \ebc{lllll}
    
    \rc{5}{0.93}{\textbf{Which AIET helped you the most to identify the ethical considerations of your model?}}\\
    \hspace{0.5cm}\bc~Harms Modeling & \bc~Model Cards &&&\\

    \ebc{lllll}
    
    \rc{5}{0.93}{\textbf{Which AIET's final documentation revealed the most ethical considerations about your model?}}\\
    \hspace{0.5cm}\bc~Harms Modeling documentation & \bc~Model Cards documentation &&&\\

    \ebc{lllll}
    
    \rc{5}{0.93}{\textbf{Which AIET's final documentation would be more informative and better understood by your model's stakeholders?}}\\
    \hspace{0.5cm}\bc~Harms Modeling documentation & \bc~Model Cards documentation &&&\\

    \ebc{lllll|lllll}
    
    \rc{10}{0.93}{\textbf{How would you rate the AIETs, with 5 being the most positive and 1 the most negative?}}\\
    \rc{5}{0.45}{\hspace{0.5cm}Harms Modeling} & \rc{5}{0.45}{Model Cards}\\
    \hspace{0.5cm}\bc~1 & \bc~2 & \bc~3 & \bc~4 & \bc~5 & \bc~1 & \bc~2 & \bc~3 & \bc~4 & \bc~5 \\

    \ebc{lllll}
    
    \rc{5}{0.93}{\textbf{Before applying the AIETs/interviews, did you assess the ethical issues related to your model?}}\\
    \hspace{0.5cm}\bc~Yes & \bc~No &&&\\ 

    \ebc{lllll}
    
    \rc{5}{0.93}{\textbf{Before applying the AIETs/interviews, did you already know about any AIET?}}\\
    \hspace{0.5cm}\bc~Yes and I've used it & \bc~Yes, but I've never used it & \bc~No &&\\
    
    \rc{5}{0.93}{\textbf{If you answered “Yes” to the previous question, which AIET did you know? }}\\
    \rc{5}{0.93}{\hspace{0.5cm}[\textit{Open answer}]}\\

    \ebc{lllll}

    \rc{5}{0.93}{\textbf{Would you like to make any additional comments about the AIETs or your previous answers? }}\\
    \rc{5}{0.93}{\hspace{0.5cm}[\textit{Open answer}]}\\
    
    \end{tabular}\\\arrayrulecolor{black}\hline
    \end{tabular}
    \label{tab:forms1}
\end{table}

\begin{table}[!htb]
    \centering
    \renewcommand{\arraystretch}{1.29}
    \footnotesize
    \caption{Questionnaire to extract developers' perceptions about the analyzed AIETs [Part 2]}
    
    \newcommand{\rc}[3]{\multicolumn{#1}{>{\raggedright\arraybackslash}p{#2\textwidth}}{#3}}
    \newcommand{\bc}{$\bigcirc$}
    \newcommand{\ebc}[1]{\end{tabular}\\\hline\begin{tabular}{#1}}
    \newcommand{\opc}{\hspace{0.5cm}\bc~Harms Modeling & \bc~Model Cards & \bc~None \hspace{3cm} & \bc~Yes & \bc~No \\}
    
    \begin{tabular}{|l|}\hline\arrayrulecolor{black!15}
    
    \begin{tabular}{lll|ll}

    \rc{5}{0.93}{\textbf{In which AIET do you think that the following risk/harm was best addressed?} \textit{For each option, select the AIET and select if your model has the risk ('Yes' option) or not ('No' option)}
    }\\\hline
    
    \rc{5}{0.93}{\hspace{0.3cm}\textbf{Social stereotypes and unfair discrimination}}\\ \opc \hline
    
    \rc{5}{0.93}{\hspace{0.3cm}\textbf{Exclusionary norms}}\\ \opc \hline
    
    \rc{5}{0.93}{\hspace{0.3cm}\textbf{Toxic language}}\\ \opc \hline
    
    \rc{5}{0.93}{\hspace{0.3cm}\textbf{Lower performance by social group}}\\ \opc \hline
    
    \rc{5}{0.93}{\hspace{0.3cm}\textbf{Compromise privacy by leaking private information}}\\\opc \hline
    
    \rc{5}{0.93}{\hspace{0.3cm}\textbf{Compromise privacy by correctly inferring private information}}\\ \opc \hline
    
    \rc{5}{0.93}{\hspace{0.3cm}\textbf{Risks from leaking or correctly inferring sensitive information}}\\ \opc \hline
    
    \rc{5}{0.93}{\hspace{0.3cm}\textbf{Disseminating false or misleading information}}\\ \opc \hline
    
    \rc{5}{0.93}{\hspace{0.3cm}\textbf{Causing material harm by disseminating misinformation, \textit{e.g.}, in medicine or law}}\\ \opc \hline
    
    \rc{5}{0.93}{\hspace{0.3cm}\textbf{Nudging or advising users to perform unethical or illegal actions}}\\  \opc \hline
    
    \rc{5}{0.93}{\hspace{0.3cm}\textbf{Reducing the cost of disinformation campaigns}}\\  \opc \hline
    
    \rc{5}{0.93}{\hspace{0.3cm}\textbf{Facilitating fraud and impersonation scams}}\\  \opc \hline
    
    \rc{5}{0.93}{\hspace{0.3cm}\textbf{Assisting code generation for cyber attacks, weapons, or malicious use}}\\  \opc \hline
    
    \rc{5}{0.93}{\hspace{0.3cm}\textbf{Illegitimate surveillance and censorship}}\\ \opc \hline  
    
    \rc{5}{0.93}{\hspace{0.3cm}\textbf{Anthropomorphizing systems can lead to overreliance or unsafe use}}\\ \opc \hline
    
    \rc{5}{0.93}{\hspace{0.3cm}\textbf{Create avenues for exploiting user trust to obtain private information}}\\  \opc \hline
    
    \rc{5}{0.93}{\hspace{0.3cm}\textbf{Promoting harmful stereotypes by implying gender or ethnic identity}}\\  \opc \hline
    
    \rc{5}{0.93}{\hspace{0.3cm}\textbf{Environmental harms from operating language models}}\\  \opc \hline
    
    \rc{5}{0.93}{\hspace{0.3cm}\textbf{Increasing inequality and negative effects on job quality}}\\  \opc \hline
    
    \rc{5}{0.93}{\hspace{0.3cm}\textbf{Undermining creative economies}}\\  \opc \hline
    
    \rc{5}{0.93}{\hspace{0.3cm}\textbf{Disparate access to benefits due to hardware, software, skill constraints}}\\  \opc \hline
    
    \rc{5}{0.93}{\hspace{0.3cm}\textbf{Lack of representation of cultural and social aspects of the Brazilian population}}\\  \opc \hline
    
    \rc{5}{0.93}{\hspace{0.3cm}\textbf{Lack of representation of cultural and social aspects of the Portuguese-speaking population}}\\  \opc \hline
    
    \rc{5}{0.93}{\hspace{0.3cm}\textbf{Low performance in the Portuguese language}}\\  \opc \hline
    
    \end{tabular}\\\arrayrulecolor{black}\hline
    \end{tabular}
    \label{tab:forms2}
\end{table}

%%==================================%%
%%      Results and Discussion      %%
%%==================================%%
\section{Results and Discussion}
\label{sec:results}

This section presents the results and discusses the developers' perspectives and opinions on the AIETs applied in an interview format. Section~\ref{sec:res_capivara} presents the results obtained from the pilot interviews with the three main developers of \capivaraicon{1.2em}CAPIVARA model with the application of the AIETs Model Cards, ALTAI, FactSheets, and Harms Modeling. Section~\ref{sec:res_entrevistas} describes the aggregated results of the interviews with the developers of the four language models interviewed with the application of the AIETs Model Cards and Harms Modeling.

We organized both subsections with the following topics that encompass the questions we asked the developers through the AIETs final evaluation form:

\begin{description}
    \vspace{0.3cm}\item[\textbf{Evaluation of the AIETs using agree/disagree responses:}] The first set of questions asks for opinions on whether developers agree or disagree with certain affirmation about each analyzed AIET. For each question (Fig.~\ref{fig:CapivaraAgree}), the percentage of \cpl, \cep, \dep~and \dpl~answers for each AIET is shown. These percentages exclude the \neu~responses when a developer neither agreed nor disagreed with the affirmation (Fig.~\ref{fig:CA2}). In addition, for each question, the percentage of developers who agree with the affirmation is also highlighted in green. This figure includes the percentage of \cpl~and \cep~answers (Fig.~\ref{fig:CapivaraAgree}).
    
    \vspace{0.3cm}\item[\textbf{Classification of the AIETs using general questions:}] The second set of questions asks for general opinions about the analyzed AIETs. We asked developers to choose the AIET that best answered the question.
    
    \vspace{0.3cm}\item[\textbf{Ethical considerations identified using the AIETs:}] The third set of questions asks for opinions on the risks, harms, and ethical considerations identified using the AIETs. We asked developers which AIET best addressed or discussed the risks or harms of the model they developed. We presented 24 options of risks or harms that a language model can contain, where the developers had to answer as an option either the four analyzed AIETs or the option ``None''. The first 21 risks presented to the developers were extracted from the risks mapped by~\citet{weidinger2021}. As the models interviewed were developed with a focus on the Portuguese language (PT-BR), we added three possible specific and existing risks for these language models: ``Lack of representation of cultural and social aspects of the Brazilian population'', ``Lack of representation of cultural and social aspects of the Portuguese-speaking population'', and ``Low performance in the Portuguese language''.
    
    \vspace{0.3cm}\item[\textbf{Scores of the analyzed AIETs by the developers:}] Finally, we asked developers to provide a score for each analyzed AIET, following a 5-point scale, with 1 being the most negative perception of the AIET and 5 being the most positive perception of the AIET.
\end{description}

\subsection{Results of the CAPIVARA Model}
\label{sec:res_capivara}
According to the perspectives of the three developers of the \capivaraicon{1.2em}CAPIVARA language model, the results of this subsection show that:

\begin{description}
    \vspace{0.3cm}\item[\textbf{Model Cards}] was considered an easy-to-use AIET, effective in identifying the risks associated with language models and generating responsible documentation for these models. It was considered the AIET that generated the most informative documentation. The tool obtained an average score of 4.33 points, the developers' second-best-rated score.  
    
    \vspace{0.3cm}\item[\textbf{ALTAI}] was considered the least readable AIET, being the most complicated to use both for identifying the risks of language models and for creating responsible and informative documentation. The tool obtained an average score of 2 points, making it the worst rated by the developers.
    
    \vspace{0.3cm}\item[\textbf{FactSheets}] was an average AIET, being easy to use, and helpful for identifying language model risks and generating responsible model documentation. The tool obtained an average score of 3.67 points.
    
    \vspace{0.3cm}\item[\textbf{Harms Modeling}] was considered an easy-to-use AIET and the best tool for identifying language model risks, generating responsible and informative documentation about the models. The tool obtained an average score of 4.67 points, being the best rated by the developers.\vspace{0.2cm}
\end{description}

In the following, we present and discuss in detail these results following the structure presented in Section~\ref{sec:results}.

\subsubsection{Evaluation of the AIETs using Agree/Disagree Responses}
\label{sec:agree}

\begin{figure}[!htb]
\centering
    \subfigure[][\afirmacaoum]{\includegraphics[width=0.49\textwidth, clip]{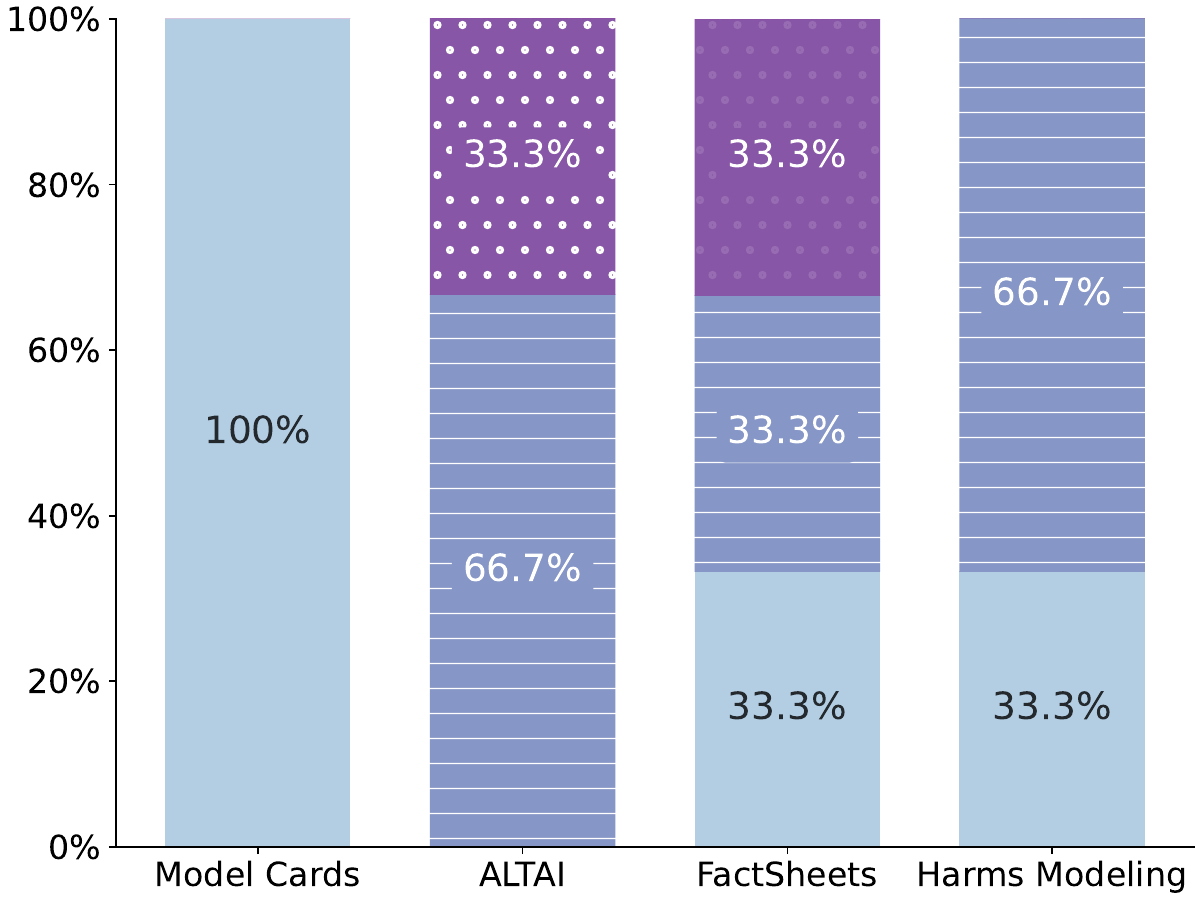}\label{fig:CA1}}\hspace{0.1cm}
    \subfigure[][\afirmacaodois]{\includegraphics[width=0.49\textwidth, clip]{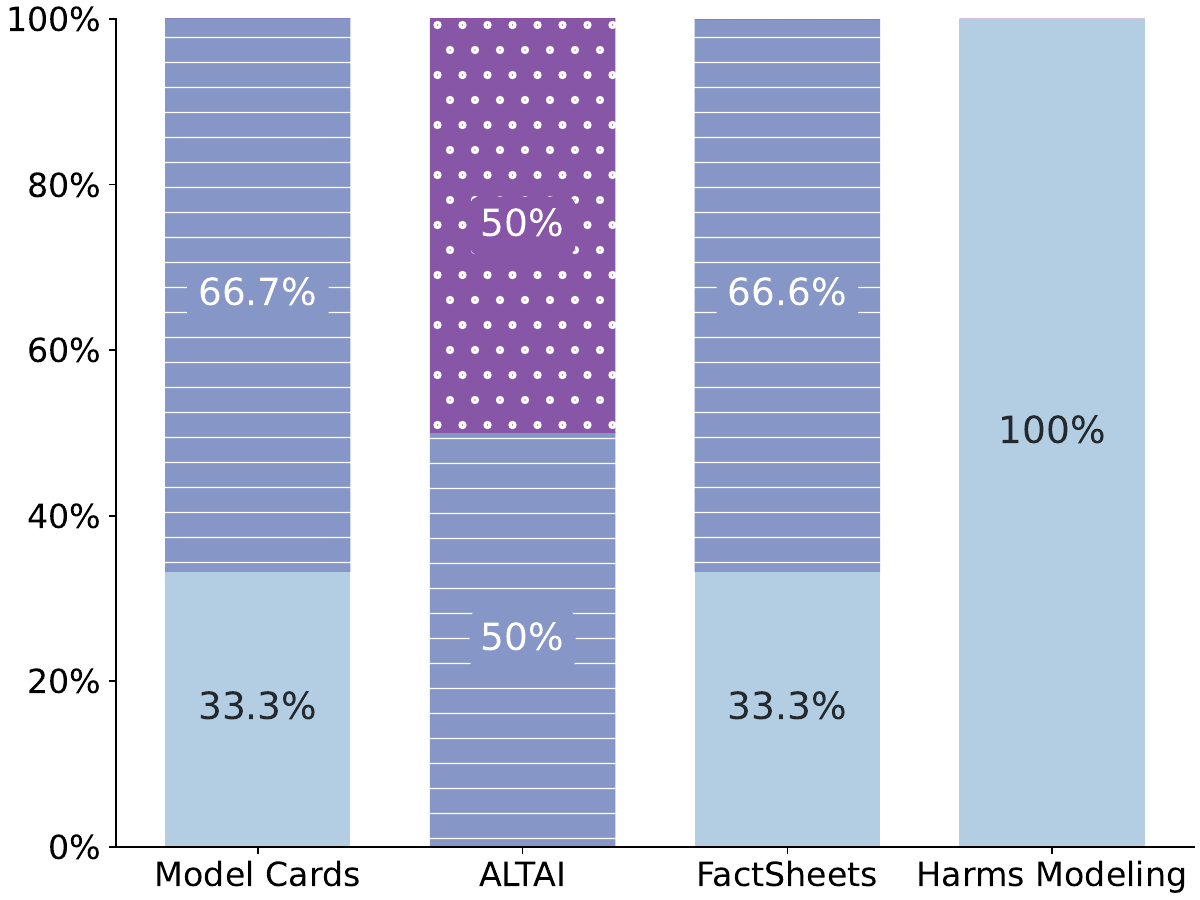}\label{fig:CA2}}\hspace{0.1cm}
    \subfigure[][\afirmacaotres]{\includegraphics[width=0.49\textwidth]{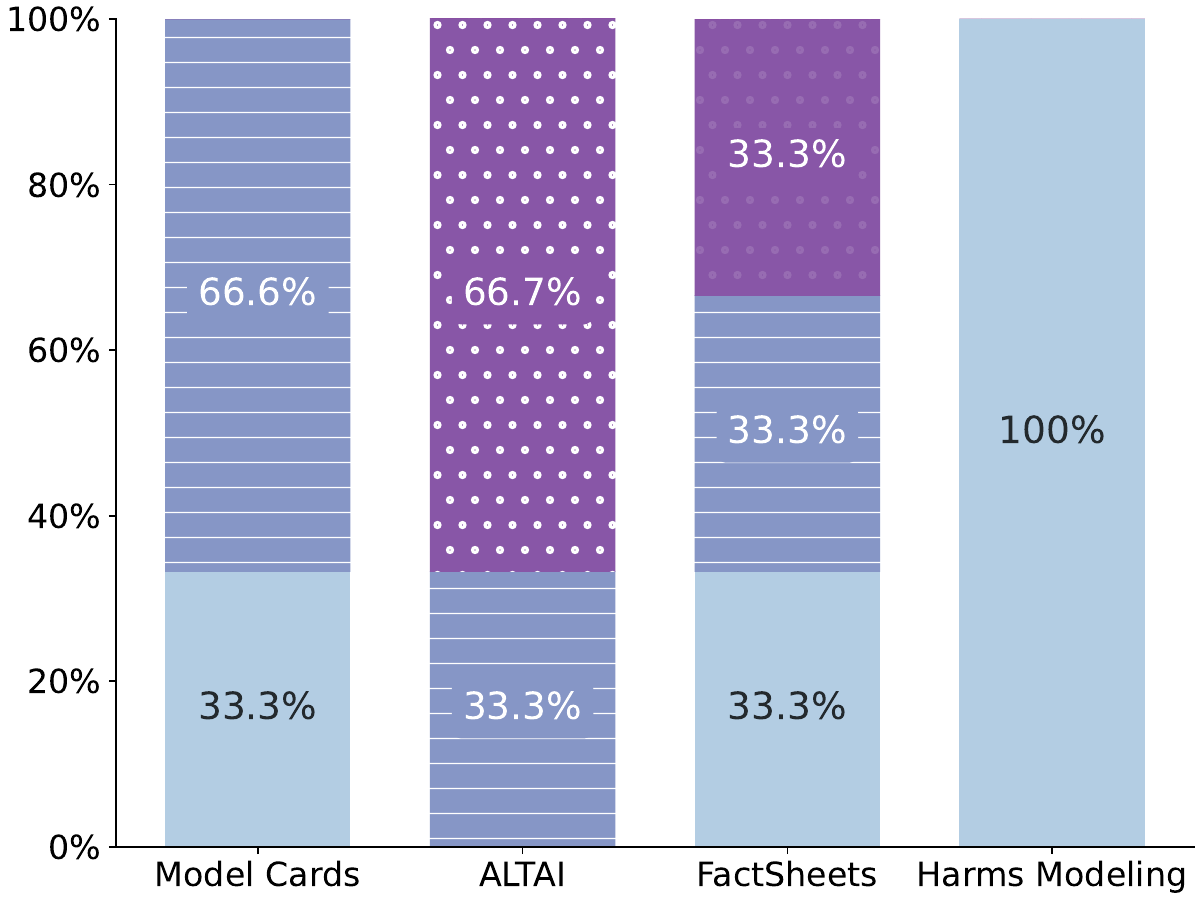}\label{fig:CA3}}\hspace{0.1cm}
    \subfigure[][\afirmacaoquatro]{\includegraphics[width=0.49\textwidth]{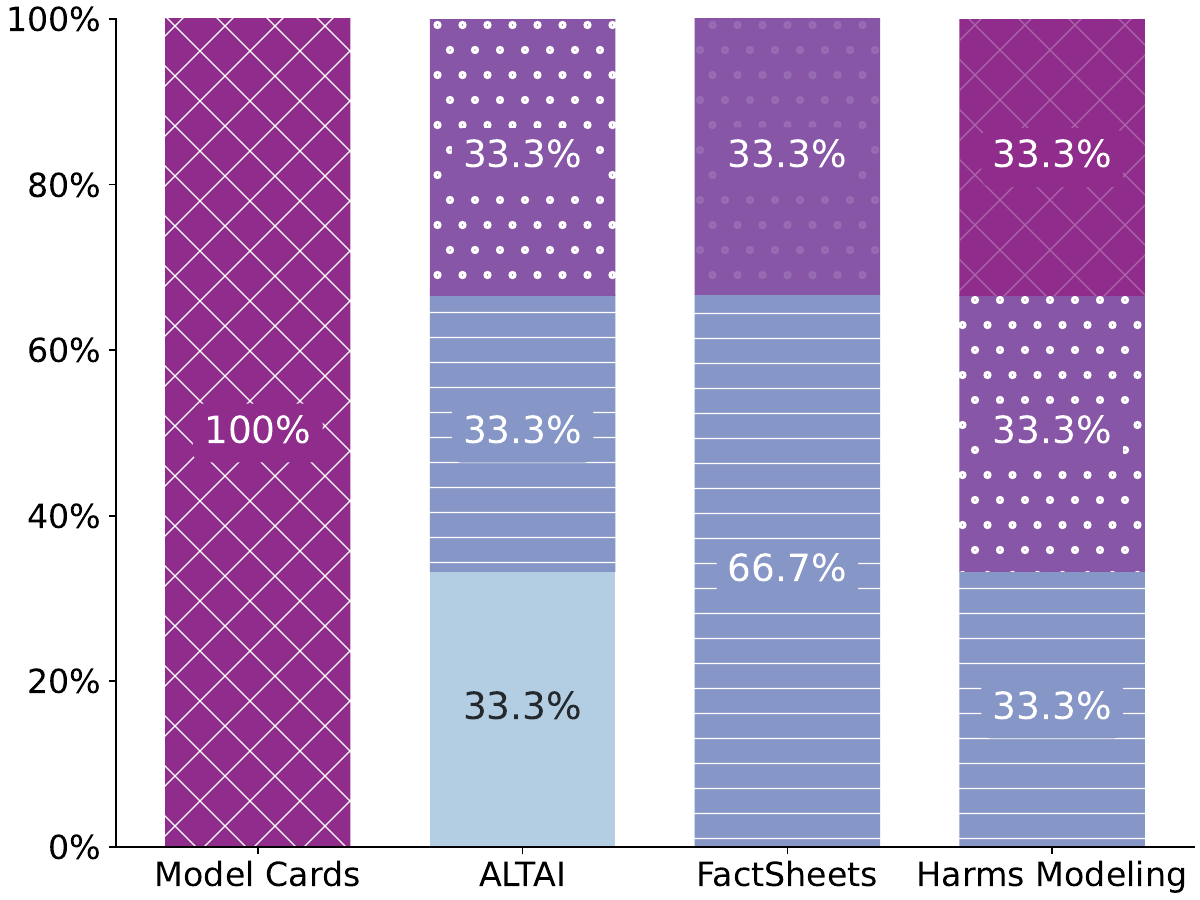}\label{fig:CA4}} 
    \caption[Evaluation of the AIETs from the perspective of the developers interviewed.]{Evaluation of the AIETs from the perspective of the developers interviewed. Here, the graphs do not include the ``neutral'' option. \includegraphics[width=0.3cm, trim={0.7cm 0.7cm 0.7cm 0.7cm}, clip]{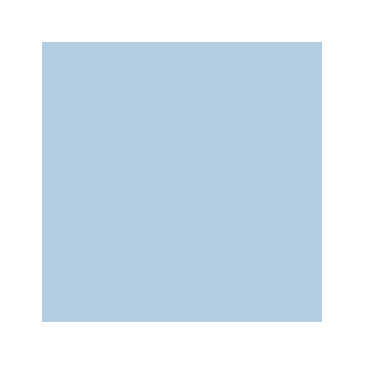}:~\cpl, \includegraphics[width=0.3cm, trim={0.7cm 0.7cm 0.7cm 0.7cm}, clip]{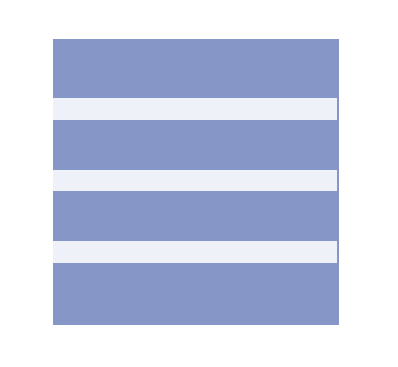}:~\cep, \includegraphics[width=0.3cm, trim={0.7cm 0.7cm 0.7cm 0.7cm}, clip]{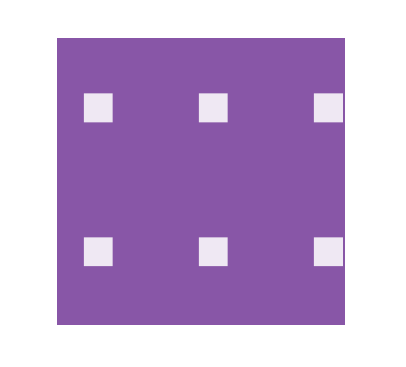}:~\dep, and \includegraphics[width=0.3cm, trim={0.7cm 0.7cm 0.7cm 0.7cm}, clip]{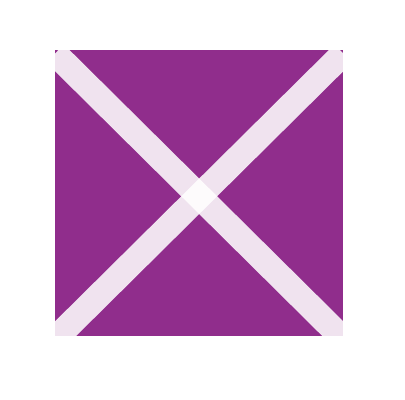}:~\dpl }
    \label{fig:CapivaraAgree}
\end{figure}

Fig.~\ref{fig:CapivaraAgree} presents the results of evaluating the AIETs stage using agree/disagree responses. In the following, we discuss each of the affirmations.

\vspace{0.4cm}
\textbf{\afirmacaoum} (Fig.~\ref{fig:CA1}). In this affirmation, Model Cards received 100\% of \cpl answers, showing the developers' agreement on the tool's ease of use, followed by Harms Modeling, with 100\% agreeing with the affirmation, the majority vote being \cep. The results make it possible to understand the developers' difficulty using ALTAI and FactSheets, respectively. An interesting contrast is that the AIET that took the longest to apply was Harms Modeling (around 4 hours), showing that the time taken to complete the AIET does not indicate difficulty in using it. During the application of the ALTAI, the developers commented on the tool's readability, with the questions being challenging to understand with terms they did not know. ALTAI offers a glossary in its documentation describing keywords. However, some developers did not understand some questions, even with the glossary. This result shows a difficulty in using the tool and a possible impediment to its use in the daily lives of developers. 

\vspace{0.4cm}
\textbf{\afirmacaodois} (Fig.~\ref{fig:CA2}). Harms Modeling received 100\% \cpl answers, showing agreement on the part of the developers as to the usability of the tool in identifying risks in a language model. In fact, Harms Modeling presents an extensive list of harms technology can generate. It offers a mechanism for evaluating these harms in terms of ``severity'', ``scale'', ``probability'' and ``frequency'', where for each option, the developers choose whether that harm is ``low'', ``moderate'' or ``high''. Harms Modeling brought dynamism to the interviews, where the developers could work together to analyze the harms in their model and indicate how harmful these risks could be to the stakeholders. An interesting point is the fact that, at random, Harms Modeling was the last AIET applied to the model. We therefore hypothesize that the order of the application interferes with how developers can evaluate the AIETs. Thus, it is possible that if Harms Modeling had been applied first in the model, this could have changed how Model Cards and FactSheets were answered, both with 100\% agreement with the affirmation, the majority vote being \cep. In this affirmation, ALTAI was the worst-rated tool, receiving 50\% agreement with the \cep option. Interestingly, at the same time as ALTAI was rated worst when asked ``Which AIET helped you the most to identify the ethical considerations of your model?'' (see Fig.~\ref{fig:CA5}), ALTAI was chosen by one out of three respondents, along with FactSheets and Harms Modeling, showing that how a question is asked can interfere with the answer given.

\vspace{0.4cm}
\textbf{\afirmacaotres} (Fig.~\ref{fig:CA3}). Harms Modeling received 100\% of \cpl answers, showing agreement on the part of the developers as to the quality of the tool in generating responsible documentation about the model developed, followed by Model Cards with 100\% of agreement with the affirmation, the majority vote being \cep. Both AIETs were the ones that generated documentation with the most information and which will possibly be better understood by the model's stakeholders (see Fig.~\ref{fig:CA5}). FactSheets was the third best-rated tool with 66.7\%, followed by ALTAI with 33.3\% of the votes. One contrast between the two AIETs is the documentation that introduces them. FactSheets has usage examples, while ALTAI does not. In addition, ALTAI presents ``yes''/``no'' questions, which makes it challenging to present a clear final documentation of the model. The final documentation for both FactSheets and ALTAI was accompanied by questions, unlike the proposal offered by Harms Modeling and Model Cards, making the final document more complicated to~read. 

\vspace{0.4cm}
\textbf{\afirmacaoquatro} (Fig.~\ref{fig:CA4}). ALTAI and FactSheets, with 66.7\% agreement, respectively, were considered the most complicated to use as a guide for identifying ethical considerations about the model, reinforcing the results achieved for the previous affirmation \afirmacaotres. Harms Modeling, with 33.3\% of the votes, was the third tool considered complicated to use as a guide, showing a contrast with the responses obtained for the affirmation \afirmacaodois, suggesting that although the tool can be helpful at the same time, it can be complicated. Model Cards received 100\% \dpl answers, showing total agreement among developers on the ease of using the tool as a guide for identifying the ethical considerations of the model. During the application, Model Cards was praised for offering examples of how to fulfill the requirements of the card, which indicates that the way the tool is presented changes the perception of developers when using it, and is in line with the results obtained in the affirmation \afirmacaoum~(Fig.~\ref{fig:CA1}).

\subsubsection{Classification of the AIETs using General Questions}
\label{sec:pg_capivara}
\begin{figure}[!htb]
    \centering
    \includegraphics[width=\textwidth, trim={0.5cm 0.5cm 0.5cm 0}, clip]{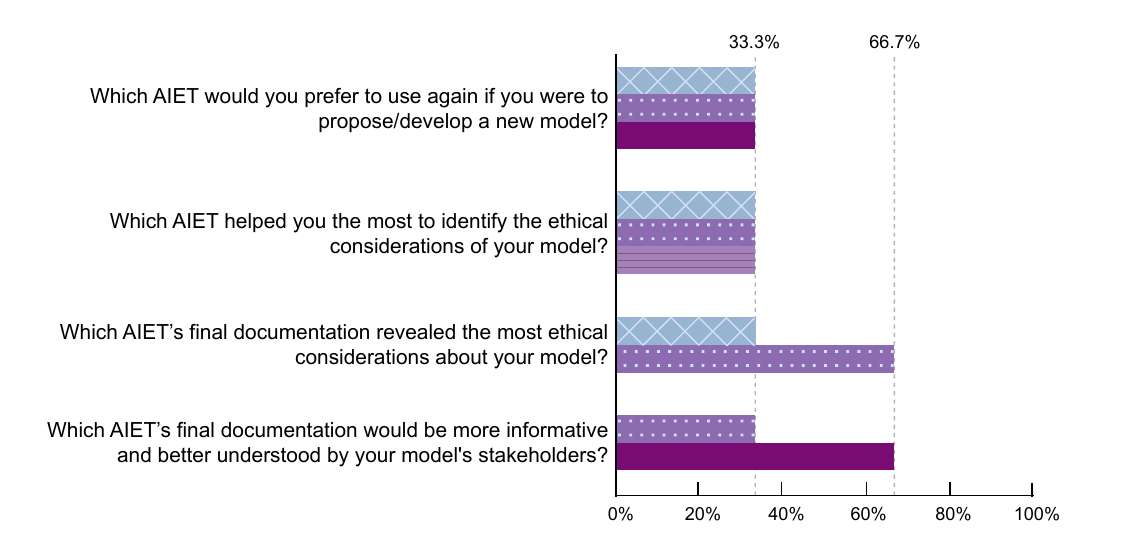}
    \caption[Classification of AIETs through general questions.]{Classification of AIETs through general questions. \includegraphics[width=0.3cm, trim={0.7cm 0.7cm 0.7cm 0.7cm}, clip]{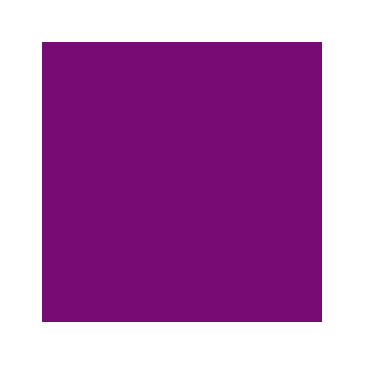}: Model Cards, \includegraphics[width=0.3cm, trim={0.7cm 0.7cm 0.7cm 0.7cm}, clip]{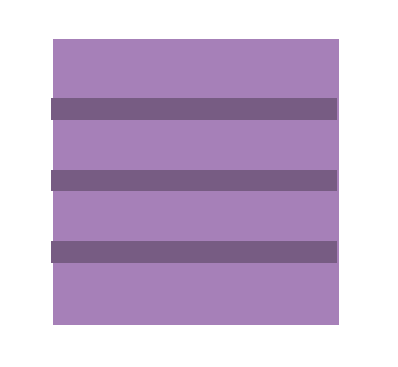}: ALTAI, \includegraphics[width=0.3cm, trim={0.7cm 0.7cm 0.7cm 0.7cm}, clip]{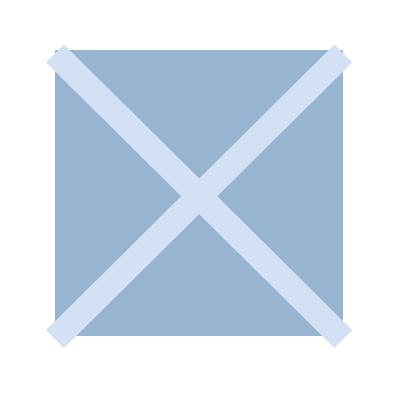}:~FactSheets, and \includegraphics[width=0.3cm, trim={0.7cm 0.7cm 0.7cm 0.7cm}, clip]{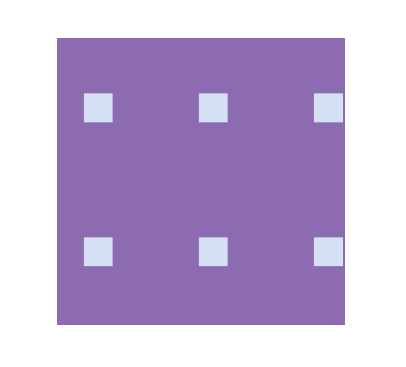}: Harms Modeling}
    \label{fig:CA5}
\end{figure}

Fig.~\ref{fig:CA5} shows the answers obtained from the four general questions asked at this stage. In the following, we discuss each of the questions.

\vspace{0.4cm}
\textbf{``Which AIET would you prefer to use again if you were to propose/develop a new model?''}. We asked developers whether they would prefer to use an AIET again if they were to implement a new AI model, regardless of the purpose of the model. The AIETs that were chosen, equally with 33.3\% of the votes, were FactSheets, Harms Modeling, and Model Cards, with no agreement among the developers. We also asked developers whether, although they have a preference for an AIET, they really would use this AIET in the future, and all answered ``yes'', showing an acceptance on the part of the developers to make use of AIETs to make the deployment of their models more ethical. However, at the same time, as the AIETs analyzed were all for the final stage of the AI life-cycle (when the model is already ready), the developers commented that it would be interesting to use intermediate AIETs alongside these AIETs for other stages of the AI life-cycle. According to the developers, changing the finished model and maintaining the same result the AI achieves is often not feasible, making it difficult to make the model ``more ethical''. Using an AIET earlier in the model development process makes it possible to anticipate and mitigate risks and negative impacts before the model is ready. Once the model is ready, it will unlikely be changed to include ethical adjustments.

\vspace{0.4cm}
\textbf{``Which AIET helped you the most to identify the ethical considerations of your model?''}. This question asked which AIET best helped the developers identify their model's ethical considerations. The AIETs that were chosen, equally with 33.3\% of the votes, were ALTAI, FactSheets, and Harms Modeling, with no agreement between the developers, showing that for the same model, different AIETs can serve as aids to help developers map the ethical considerations of a model. We also asked developers if they had already ethically evaluated the model they had developed, and 66.7\% answered ``yes'', but the affirmative respondents commented that they had only been concerned with two ethical considerations and that by using AIETs it was possible to discover other critical considerations about the model they had developed. 

\vspace{0.4cm}
\textbf{``Which AIET's final documentation revealed the most ethical considerations about your model?''}. This question asked which AIET's final documentation contained and revealed more ethical considerations about the model developed. The most voted AIET's final documentation was Harms Modeling with 66.7\% of the votes, followed by FactSheets with 33.3\%. This result is in line with the result obtained in Section~\ref{sec:agree} through the affirmation \afirmacaodois, where Harms Modeling received 100\% of ``agree'' votes for the statement. During the interviews, the developers praised how Harms Modeling presented several different harms. As it was the last AIET interviewed, the developers even commented that it was possible to identify ethical considerations with the help of Harms Modeling that had not yet been identified using any of the AIETs previously interviewed and that if Harms Modeling had been the first analyzed AIET, the results obtained in the other AIETs would probably have been different, for example, by raising more ethical considerations.

\vspace{0.4cm}
\textbf{``Which AIET's final documentation would be more informative and better understood by your model's stakeholders?''}. This question asked which AIET's final documentation could help developers disseminate responsible, transparent, and reliable documentation in a clear and informative way to their stakeholders, both direct and indirect. The most voted AIET's final documentation was Model Cards with 66.7\% of the votes, followed by Harms Modeling with 33.3\%. This result aligns with the result obtained in Section~\ref{sec:agree} through the affirmation \afirmacaotres, where Harms Modeling and Model Cards received 100\% of the votes in agreement with the affirmation. We also asked developers which AIET's final documentation they would like to share publicly with their stakeholders. Both Harms Modeling and Model Cards received 100\% of votes, followed by 33.3\% of votes for FactSheets final documentation. No developer chose ALTAI final documentation as an option, supporting the result obtained in Section~\ref{sec:agree} through the affirmation \afirmacaoum, where it was argued that the AIET was not so readable for developers, so at the same time, it would be complicated for such documentation to be shared to stakeholders in a clear and informative way. 

\subsubsection{Ethical Considerations Identified using the AIETs}
\label{sec:riscos_cap}

\begin{table}[!htb]
    \renewcommand{\arraystretch}{1.4}
    \footnotesize
    \caption{\capivaraicon{1.2em}CAPIVARA developers' choice of which AIET best addressed or discussed a risk to their language model. The cells show the percentage of developers who chose an AIET as the one that best addressed the risk under analysis. The table includes the option ``None'', meaning that any AIETs identified the analyzed risk, according to the developers' opinion. The first 21 risks in the table were extracted from the risks mapped by~\citet{weidinger2021}}
    \centering
    \arrayrulecolor{black!15}
    \begin{tabular}{>{\raggedright\arraybackslash}p{4.6cm}|>{\centering\arraybackslash}p{1.1cm}|>{\centering\arraybackslash}p{1.1cm}|>{\centering\arraybackslash}p{1.1cm}|>{\centering\arraybackslash}p{1.4cm}|>{\centering\arraybackslash}p{1.1cm}}\arrayrulecolor{black}
    \toprule
    \multirow{2}{*}{\textbf{Analyzed Risk}} & \textbf{Model Cards} & \multirow{2}{*}{\textbf{ALTAI}} & \textbf{Fact- Sheets} & \textbf{Harms Modeling} & \multirow{2}{*}{\textbf{None}}\\
    \midrule
    \arrayrulecolor{black!15}
    
    Social stereotypes and unfair discrimination &  \multirow{2}{*}{\scaleb{0}} & \multirow{2}{*}{\scaleb{33.3}} & \multirow{2}{*}{\scaleb{0}} & \multirow{2}{*}{\scalea{66.7}} & \multirow{2}{*}{\scaleb{0}} \\\hline
    
    Exclusionary norms & \scaleb{0} & \scaleb{0} & \scaleb{0} & \scalea{100} & \scaleb{0} \\\hline
    
    Toxic language & \scaleb{0} & \scaleb{0} & \scaleb{0} & \scalea{100} & \scaleb{0} \\\hline
    
    Lower performance by social group & \scaleb{33.3} & \scaleb{0} & \scaleb{33.3} & \scaleb{33.3} & \scaleb{0} \\\hline
    
    Compromise privacy by leaking private information & \multirow{2}{*}{\scaleb{0}} & \multirow{2}{*}{\scaleb{33.3}} & \multirow{2}{*}{\scaleb{0}} & \multirow{2}{*}{\scalea{66.7}} & \multirow{2}{*}{\scaleb{0}} \\\hline
    
    Compromise privacy by correctly inferring private information & \multirow{2}{*}{\scaleb{0}} & \multirow{2}{*}{\scaleb{33.3}} & \multirow{2}{*}{\scaleb{0}} & \multirow{2}{*}{\scalea{66.7}} & \multirow{2}{*}{\scaleb{0}} \\\hline
    
    Risks from leaking or correctly inferring sensitive information & \multirow{2}{*}{\scaleb{0}} & \multirow{2}{*}{\scaleb{0}} & \multirow{2}{*}{\scaleb{33.3}} & \multirow{2}{*}{\scalea{66.7}} & \multirow{2}{*}{\scaleb{0}} \\\hline
    
    Disseminating false or misleading information & \multirow{2}{*}{\scaleb{33.3}} & \multirow{2}{*}{\scaleb{0}} & \multirow{2}{*}{\scaleb{33.3}} & \multirow{2}{*}{\scaleb{33.3}} & \multirow{2}{*}{\scaleb{0}}\\\hline
    
    Causing material harm by disseminating misinformation, \textit{e.g.}, in medicine or law & \multirow{3}{*}{\scaleb{0}} & \multirow{3}{*}{\scaleb{0}} & \multirow{3}{*}{\scaleb{33.3}} & \multirow{3}{*}{\scalea{66.7}} & \multirow{3}{*}{\scaleb{0}} \\\hline
    
    Nudging or advising users to perform unethical or illegal actions & \multirow{2}{*}{\scaleb{0}} & \multirow{2}{*}{\scaleb{0}} & \multirow{2}{*}{\scaleb{33.3}} & \multirow{2}{*}{\scaleb{0}} &  \multirow{2}{*}{\scalea{66.7}} \\\hline
    
    Reducing the cost of disinformation campaigns & \multirow{2}{*}{\scaleb{33.3}} & \multirow{2}{*}{\scaleb{0}} & \multirow{2}{*}{\scaleb{0}} & \multirow{2}{*}{\scaleb{0}} & \multirow{2}{*}{\scalea{66.7}}   \\\hline
    
    Facilitating fraud and impersonation scams & \multirow{2}{*}{\scaleb{0}} & \multirow{2}{*}{\scaleb{0}} & \multirow{2}{*}{\scaleb{0}} & \multirow{2}{*}{\scaleb{33.3}} &  \multirow{2}{*}{\scalea{66.7}}  \\\hline
    
    Assisting code generation for cyber attacks, weapons, or malicious use & \multirow{2}{*}{\scaleb{33.3}} & \multirow{2}{*}{\scaleb{0}} & \multirow{2}{*}{\scaleb{0}} & \multirow{2}{*}{\scaleb{0}} & \multirow{2}{*}{\scalea{66.7}}  \\\hline
    
    Illegitimate surveillance and censorship & \multirow{2}{*}{\scaleb{0}} & \multirow{2}{*}{\scaleb{33.3}} & \multirow{2}{*}{\scaleb{0}} & \multirow{2}{*}{\scalea{66.7}}   & \multirow{2}{*}{\scaleb{0}} \\\hline
    
    Anthropomorphizing systems can lead to overreliance or unsafe use & \multirow{2}{*}{\scaleb{0}} & \multirow{2}{*}{\scaleb{33.3}} & \multirow{2}{*}{\scaleb{0}} & \multirow{2}{*}{\scalea{66.7}} & \multirow{2}{*}{\scaleb{0}}  \\\hline
    
    Create avenues for exploiting user trust to obtain private information & \multirow{2}{*}{\scaleb{0}} & \multirow{2}{*}{\scaleb{0}} & \multirow{2}{*}{\scaleb{0}} & \multirow{2}{*}{\scaleb{33.3}} &  \multirow{2}{*}{\scalea{66.7}}  \\\hline
    
    Promoting harmful stereotypes by implying gender or ethnic identity & \multirow{2}{*}{\scaleb{0}} & \multirow{2}{*}{\scaleb{0}} & \multirow{2}{*}{\scaleb{0}} & \multirow{2}{*}{\scalea{100}} &  \multirow{2}{*}{\scaleb{0}} \\\hline
    
    Environmental harms from operating language models & \multirow{2}{*}{\scaleb{0}} & \multirow{2}{*}{\scaleb{33.3}} & \multirow{2}{*}{\scaleb{0}} & \multirow{2}{*}{\scalea{66.7}} & \multirow{2}{*}{\scaleb{0}} \\\hline
    
    Increasing inequality and negative effects on job quality & \multirow{2}{*}{\scaleb{0}} & \multirow{2}{*}{\scaleb{0}} & \multirow{2}{*}{\scaleb{0}} & \multirow{2}{*}{\scalea{100}} & \multirow{2}{*}{\scaleb{0}}  \\\hline
    
    Undermining creative economies & \multirow{1}{*}{\scaleb{0}} & \multirow{1}{*}{\scaleb{0}} & \multirow{1}{*}{\scaleb{0}} & \multirow{1}{*}{\scalea{66.7}} & \multirow{1}{*}{\scaleb{33.3}}  \\\hline
    
    Disparate access to benefits due to hardware, software, skill constraints & \multirow{2}{*}{\scaleb{0}} & \multirow{2}{*}{\scaleb{0}} & \multirow{2}{*}{\scaleb{0}} & \multirow{2}{*}{\scalea{100}} &   \multirow{2}{*}{\scaleb{0}} \\\hline
    
    Lack of representation of cultural and social aspects of the Brazilian population & \multirow{3}{*}{\scaleb{0}} & \multirow{3}{*}{\scaleb{0}} & \multirow{3}{*}{\scaleb{0}} & \multirow{3}{*}{\scaleb{33.3}} &   \multirow{3}{*}{\scalea{66.7}} \\\hline
    
    Lack of representation of cultural and social aspects of the Portuguese-speaking population & \multirow{3}{*}{\scaleb{0}} & \multirow{3}{*}{\scaleb{0}} & \multirow{3}{*}{\scaleb{0}} & \multirow{3}{*}{\scaleb{33.3}} &   \multirow{3}{*}{\scalea{66.7}} \\\hline
    
    Low performance in the Portuguese language & \multirow{2}{*}{\scaleb{0}} & \multirow{2}{*}{\scaleb{0}} & \multirow{2}{*}{\scaleb{33.3}} & \multirow{2}{*}{\scaleb{33.3}} & \multirow{2}{*}{\scaleb{33.3}}  \\\arrayrulecolor{black}
    \bottomrule
    \end{tabular}
    \label{tab:riscos_cap}
\end{table}

In this stage, we asked developers to point out the ethical considerations identified using the analyzed AIETs and Table~\ref{tab:riscos_cap} shows the results achieved. The results suggest that Harms Modeling was the tool that best addressed or discussed the risks or harms of the model analyzed.

According to 66.7\% of the developers, no AIET has led to the discussion of harms related to malicious uses of the technology, such as encouraging unethical actions, reducing the cost of disinformation campaigns, facilitating fraud and impersonation scams, and generating code for malicious or unethical use.

In addition, the specific risks for the Portuguese language model, ``Lack of representation of cultural and social aspects of the Brazilian population'' and ``Lack of representation of cultural and social aspects of the Portuguese-speaking population'' according to 66.7\% of the developers, were not addressed by any AIET, showing that generalist AIETs may not map specific risks in an AI.

\subsubsection{Scores of the analyzed AIETs by the Developers}

Finally, we asked developers to provide a score for each analyzed AIET and Table~\ref{tab:CA6} shows the results obtained. The best-rated AIET was Harms Modeling, with an average of 4.67 points, followed by Model Cards, with 4.33~points. FactSheets scored 3.67 points, and ALTAI 2 points. 

\begin{table}[!htb]
    \renewcommand{\arraystretch}{1.1}
    \footnotesize
    \caption[Rating given to AIETs by the developers]{Rating given to AIETs by the developers. The last line shows the average rating for each AIET}
    \centering 
    \arrayrulecolor{black}
    \begin{tabular}{>{\arraybackslash}p{1.9cm}>{\centering\arraybackslash}p{2.1cm}>{\centering\arraybackslash}p{1.8cm}>{\centering\arraybackslash}p{2.0cm}>{\centering\arraybackslash}p{3cm}}
    \toprule
         & \textbf{Model Cards} & \textbf{ALTAI} & \textbf{FactSheets} &\textbf{Harms Modeling} \\\midrule
        Developer 1 & 4 & 3 & 3 & 5 \\
        Developer 2 & 4 & 2 & 3 & 5\\
        Developer 3 & 5 & 1 & 5 & 4\\\hline 
        \cellcolor{Gray!40}\textbf{Average} & \cellcolor{Gray!40}\textbf{4.33} & \cellcolor{Gray!40}\textbf{2} & \cellcolor{Gray!40}\textbf{3.67} & \cellcolor{Gray!40}\textbf{4.67}\\
        \bottomrule
    \end{tabular}
    \label{tab:CA6}
\end{table}

The results achieved are in line with the results presented and discussed earlier in this section, showing a preference on the part of developers to use the AIETs Harms Modeling and Model Cards as aids for identifying and disclosing ethical considerations about a developed model, with fewer difficulties in use and better clarity of the results achieved. Because of all these results, as mentioned above, the following evaluations of this paper are focused only on Harms Modeling and Model Cards.

\subsection{Results of the Interviews}
\label{sec:res_entrevistas}

As mentioned in Section~\ref{sec:sel_modelos_linguagem}, 11 developers participated in the interviews, forming four groups: 

\begin{itemize}
    \item  Group 1 (or language model 1) -- 2 participants; 
    \item  Group 2 (or language model 2) -- 2 participants; 
    \item  Group 3 (or language model 3) -- 3 participants; 
    \item  Group 4 (or language model 4) -- 4 participants. 
\end{itemize}

However, one participant from Group~4 refrained from answering the final form. Therefore, the results of this section will take into account the perspectives of 10 of the 11 interviewed developers. According to their perspective, the results of this subsection show that:

\begin{description}
    \vspace{0.3cm}\item[\textbf{Model Cards}] was considered an objective AIET with the most tangible documentation. As in Section~\ref{sec:res_capivara}, Model Cards was considered an easy-to-use AIET, effective in identifying the risks associated with language models and generating responsible documentation for these models. The tool obtained an average score of 4.1 points, being the best rated by the developers.
    
    \vspace{0.3cm}\item[\textbf{Harms Modeling}] was considered a more complete AIET. However, it is very general and broad, and does not consider specific aspects of language models. Nevertheless, Harms Modeling was still considered the best tool for identifying language model risks. The tool obtained an average score of 3.6 points.
    \vspace{0.3cm}
\end{description}

The results also indicate that the applied AIETs effectively map general ethical considerations about language models. However, it is important to note that they did not address the unique aspects of these models and failed to identify potential negative impacts for the Portuguese language. Additionally, the results suggest that it is still not common practice for developers to include an ethical analysis of their models before publishing them. Furthermore, there was a lack of agreement among developers when it came to identifying risks associated with their models individually. In the following section, we present and discuss these results in detail.

\subsubsection{Evaluation of the AIETs using Agree/Disagree Responses}
\label{sec:agree_entrevistas}

\begin{figure}[!htb]
\centering
    \subfigure[][\afirmacaoum]{\includegraphics[width=0.49\textwidth, clip, trim = 0 0 5cm 0]{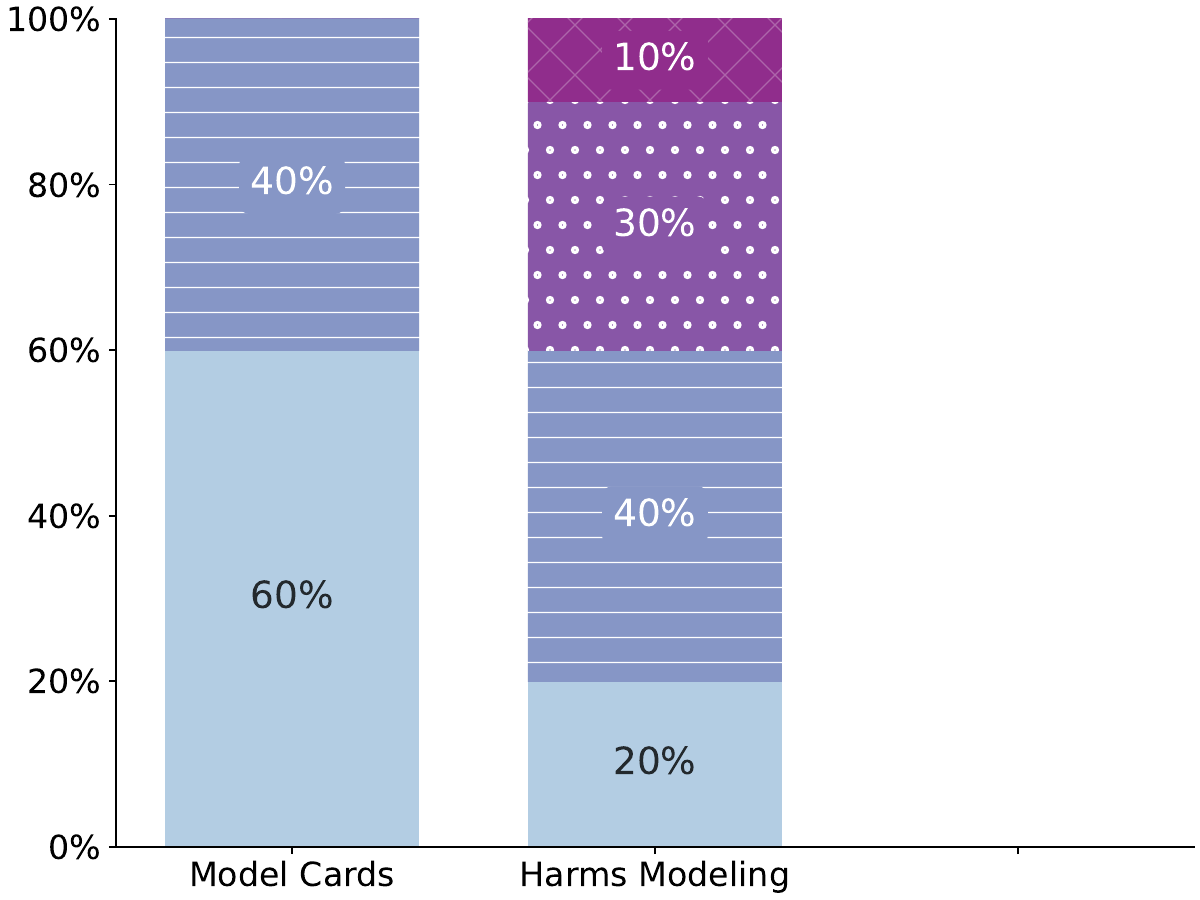}\label{fig:EA1}}\hspace{0.1cm}
    \subfigure[][\afirmacaodois]{\includegraphics[width=0.49\textwidth, clip, trim = 0 0 5cm 0]{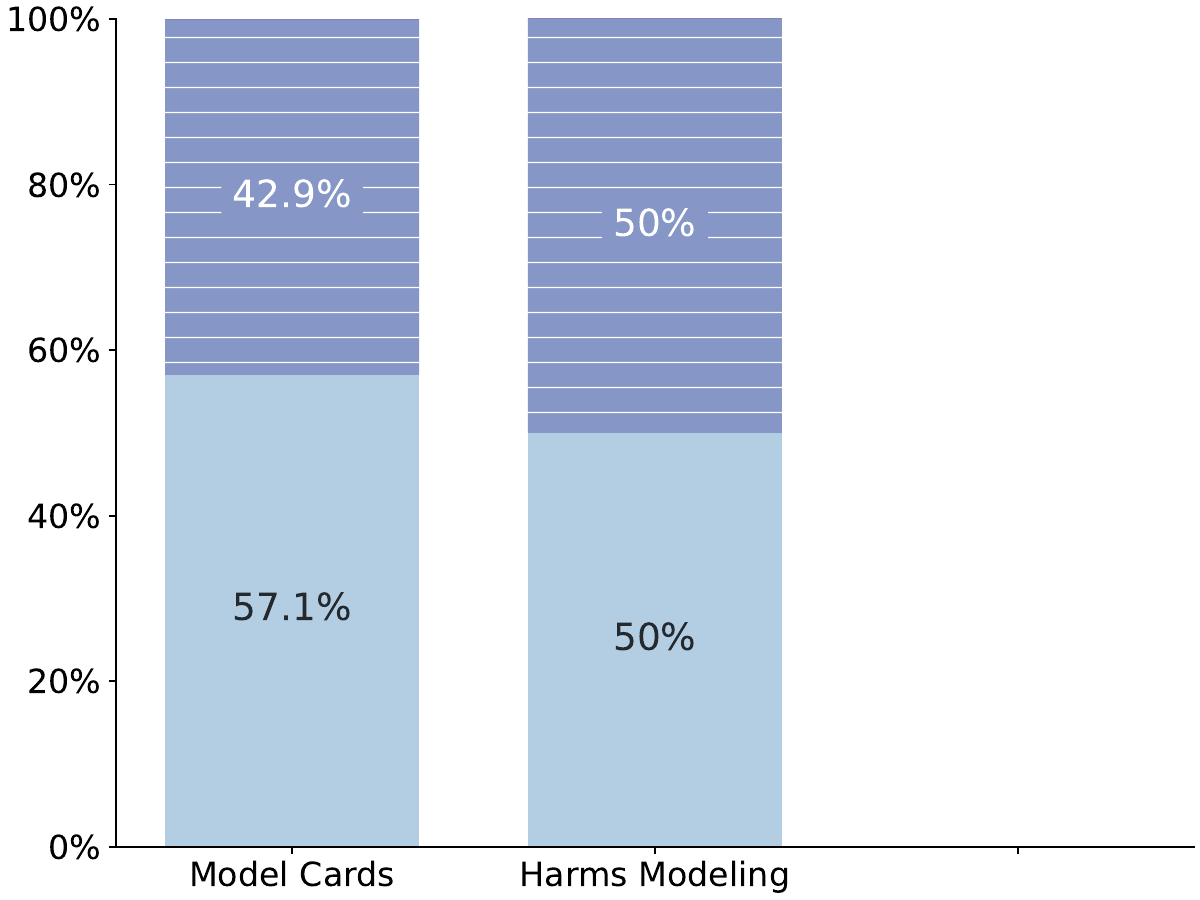}\label{fig:EA2}}\hspace{0.35cm}
    \subfigure[][\afirmacaotres]{\includegraphics[width=0.49\textwidth, clip, trim = 0 0 5cm 0]{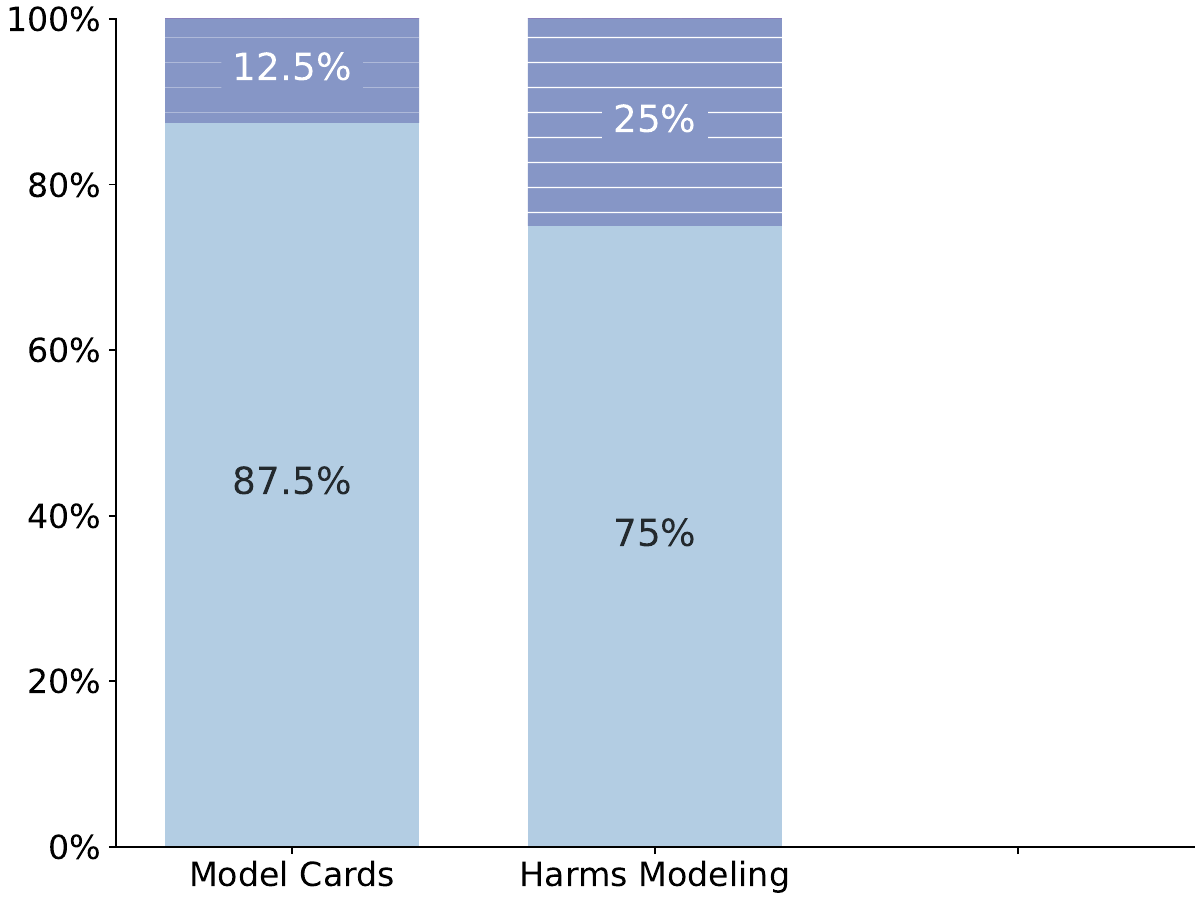}\label{fig:EA3}}\hspace{0.1cm}
    \subfigure[][\afirmacaoquatro]{\includegraphics[width=0.49\textwidth, clip, trim = 0 0 5cm 0]{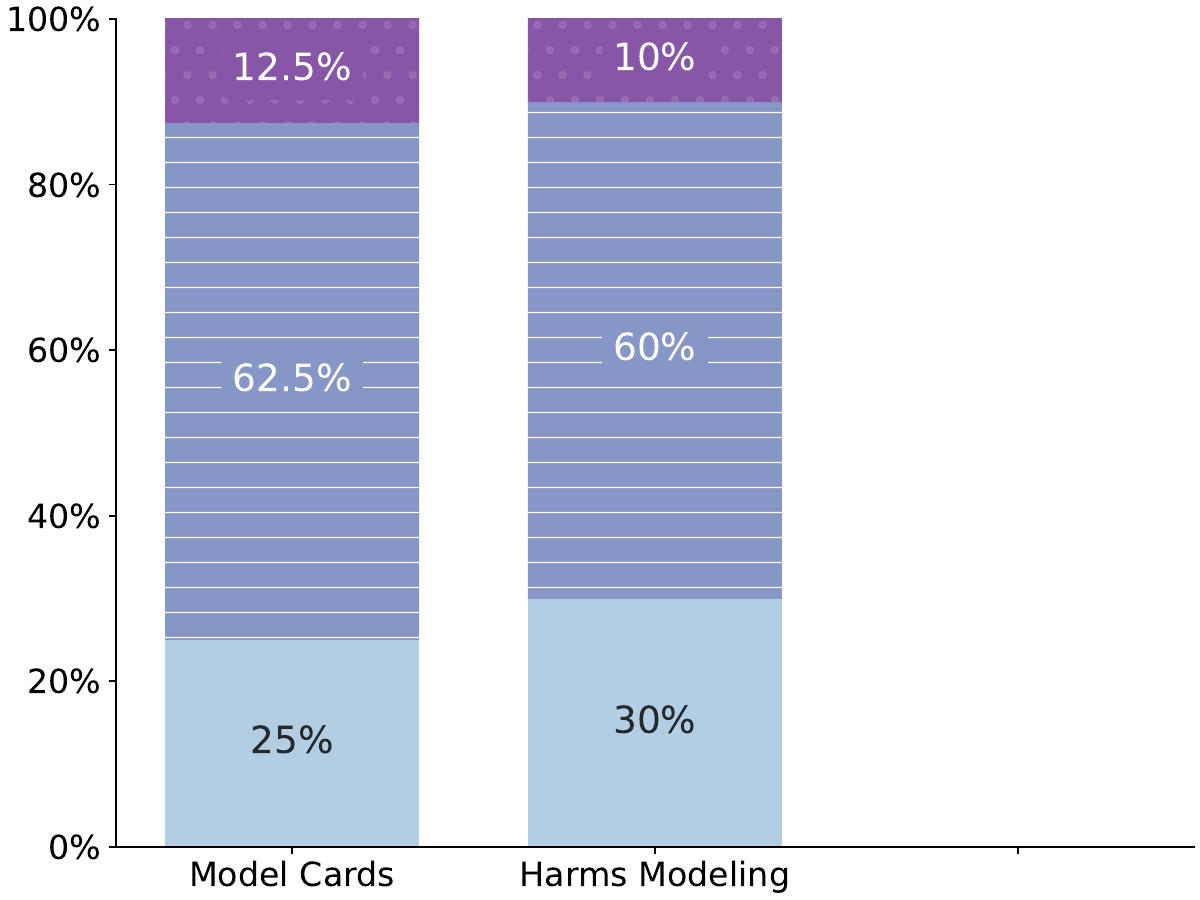}\label{fig:EA4}}\\
    \caption{Evaluation of the AIETs from the perspective of the developers interviewed. Here, the graphs do not include the ``neutral'' option. 
    %For each figure from top to bottom: Model Cards and Harms Modeling. 
    \includegraphics[width=0.3cm, trim={0.7cm 0.7cm 0.7cm 0.7cm}, clip]{Fig22.pdf}:~\cpl, \includegraphics[width=0.3cm, trim={0.7cm 0.7cm 0.7cm 0.7cm}, clip]{Fig23.pdf}:~\cep, \includegraphics[width=0.3cm, trim={0.7cm 0.7cm 0.7cm 0.7cm}, clip]{Fig24.pdf}:~\dep, and \includegraphics[width=0.3cm, trim={0.7cm 0.7cm 0.7cm 0.7cm}, clip]{Fig25.pdf}:~\dpl }
    \label{fig:EA}
\end{figure}

Fig.~\ref{fig:EA} presents the results of evaluating  the AIETs stage using agree/disagree responses. We discuss each of the affirmations as follows.

\textbf{\afirmacaoum} (Fig.~\ref{fig:EA1}). In this affirmation, Model Cards received 100\% of agreement, where the majority of votes were \cpl~demonstrating agreement on the part of the developers regarding the tool's ease of use. Harms Modeling, unlike Section~\ref{sec:agree}, received 60\% of agreement, where the majority of votes were \cep. During the interviews, some participants commented that Harms Modeling was generalist, too broad, and not ideal for language models, and that this made it difficult to answer the tool's questions related to the model in question.

\vspace{0.4cm}
\textbf{\afirmacaodois} (Fig.~\ref{fig:EA2}). In this affirmation, both Model Cards and Harms Modeling received 100\% agreement, showing that both AIETs can be used to help identify the ethical considerations of language models. However, it was possible to notice throughout the interviews that the analyzed AIETs do not have questions that would allow them to address more unique considerations of language models, such as considerations related to multilingual performance, accents, slang, and regional expressions. 

\vspace{0.4cm}
\textbf{\afirmacaotres} (Fig.~\ref{fig:EA3}).  In this affirmation, both Model Cards and Harms Modeling received 100\% agreement, where the majority of votes were \cpl~for both. This result shows that both AIETs can be used to help create responsible documentation of language models. However, it is important to note that this documentation was created by the research team of this paper, not by the developers. Harms Modeling does not explain how documentation should be created/generated from the tool's questions. Therefore, we followed a pattern in creating these documentations that we felt would be more informative, given the characteristics of the AIET. One comment we received from one of the participants was that he/she did not see how to create documentation from the questions he/she answered. In this way, the usability of Harms Modeling to generate good documentation will depend a lot on how that documentation is structured. In contrast, we can find examples of documentation to follow in the paper introducing Model Cards.

\vspace{0.4cm}
\textbf{\afirmacaoquatro} (Fig.~\ref{fig:EA4}). In this affirmation, Model Cards and Harms Modeling received 87.5\% and 90\% agreement, respectively, with the majority of votes being \cep~for both. This result significantly diverged from the findings reported in Section~\ref{sec:agree}, showing that in these interviews stage, the developers had more difficulties in both AIETs. This classification may be attributed to the fact that the \capivaraicon{1.2em}CAPIVARA model team used two other AIETs in addition to Model Cards (the first to be applied) and Harms Modeling (the last to be applied). This contributed to the team arriving better prepared in Harms Modeling, resulting in a good evaluation of the tool, while Model Cards was already a tool known to the interviewees, although never used. Another factor is that they are part of the same H.IAAC team as the researchers of this study, which introduces a high bias in the evaluation because they are in contact with the subject. In the general interviews, the other developers had no close contact with AIETs and had never used them---only 4 participants said they already knew about the existence of Model Cards. All these points contribute to the evaluation received by both AIETs, where the developers voted that it was complicated to draw up ethical considerations about language models using AIETs, even though they are useful for identifying the risks of these models---the result of the affirmation \afirmacaodois{} of the two (Fig.~\ref{fig:EA2}). 

\subsubsection{Classification of the AIETs through General Questions}
\label{sec:qg_entrevistas}

\begin{figure}[!htb]
    \centering
    \includegraphics[width=\textwidth, trim={0.5cm 0.5cm 0.5cm 0}, clip]{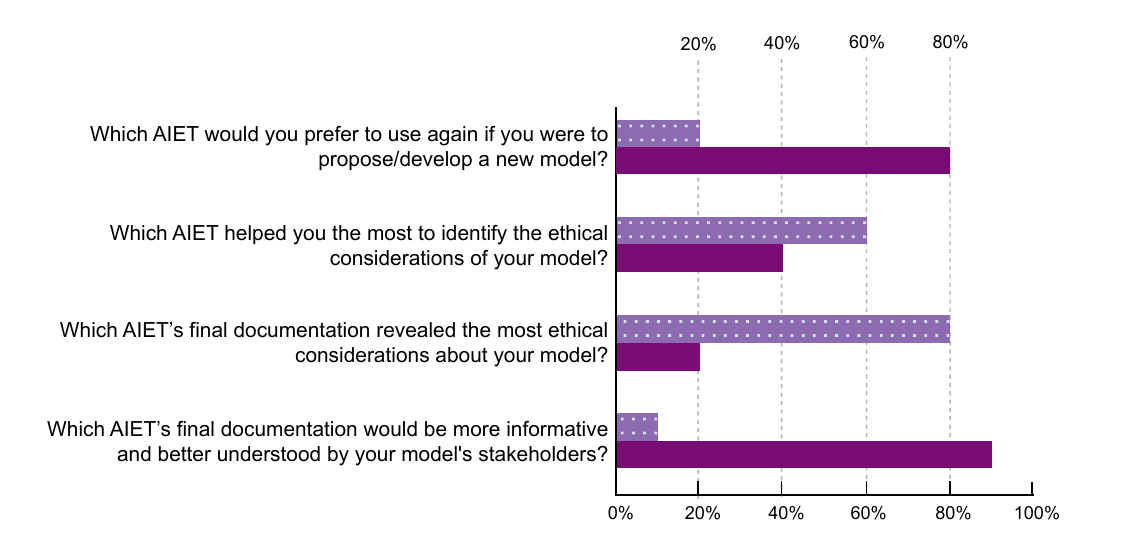}
    \caption[Classification of AIETs through general questions.]{Classification of AIETs through general questions. \includegraphics[width=0.3cm, trim={0.7cm 0.7cm 0.7cm 0.7cm}, clip]{Fig12.pdf}:~Model Cards and \includegraphics[width=0.3cm, trim={0.7cm 0.7cm 0.7cm 0.7cm}, clip]{Fig15.pdf}:~Harms Modeling}
    \label{fig:qgentrevistas}
\end{figure}

Fig.~\ref{fig:qgentrevistas} shows the answers obtained to the four general questions asked at this stage. In the following, we discuss each of the questions.

\vspace{0.4cm}
\textbf{``Which AIET would you prefer to use again if you were to propose/develop a new model?''}. 
We asked developers whether they would prefer to use an AIET again if they were to implement a new AI model, regardless of the purpose of the model. Model Cards received 80\% of the votes against Harms Modeling. We also asked developers whether, although they prefer an AIET, they would really use this AIET in the future. 50\% answered ``maybe'', 30\% ``yes'' and 20\% ``no''. This result shows that there is neither significant resistance nor adherence on the part of developers to use the AIETs again.

\vspace{0.4cm}
\textbf{``Which AIET helped you the most to identify the ethical considerations of your model?''}. This question asked which AIET best helped the developers identify their model's ethical considerations. Harms Modeling received 60\% of the developers' votes, indicating AIET's potential to help developers map the ethical considerations of a model. We also asked developers if they had already ethically evaluated the model they had developed, with or without AIET, and they all answered ``no'', although four had already met Model Cards. This result indicates that it is not yet natural for developers to add a step to ethically analyze their models before publishing them, showing the need for greater engagement and dissemination in the area of AI ethics.

\vspace{0.4cm}
\textbf{``Which AIET's final documentation revealed the most ethical considerations about your model?''}. This question asked which AIET's final documentation contained and revealed more ethical considerations about the model developed. The most voted AIET's final documentation was Harms Modeling with 80\% of the votes, in line with the result obtained in ``Which AIET helped you the most to identify the ethical considerations of your model?''. During application in the \capivaraicon{1.2em} CAPIVARA model, we raised the point that the order in which the AIETs were applied could generate different results from the others. However, this effect was not observed in the four groups interviewed, where the AIETs were applied in alternating order. The results were similar in all application orders. 

\vspace{0.4cm}
\textbf{``Which AIET's final documentation would be more informative and better understood by your model's stakeholders?''}. This question asked which AIET's final documentation could help developers disseminate responsible, transparent, and reliable documentation in a clear and informative way to their stakeholders, both direct and indirect. The most voted AIET's final documentation was Model Cards with 90\% of the votes against Harms Modeling. We also asked developers which AIET's final documentation they would like to share publicly with their stakeholders. The developers chose only the Model Cards documentation because it was clearer, more objective, and easier to read. Interestingly, in the statement ``Which AIET's final documentation revealed the most ethical considerations about your model?'' Harms Modeling scored the highest. This shows that even if a documentation reveals more ethical considerations about a model, it will not necessarily be more informative and better understood by stakeholders. 

\subsubsection{Ethical Considerations Identified using the AIETs}

In this stage, we asked developers to point out the ethical considerations identified using the analyzed AIETs and Table~\ref{tab:riscos_entrevistas} shows the results achieved. The results suggest that Harms Modeling was the tool that best addressed or discussed the risks or harms of the analyzed model. Unlike the results of Section~\ref{sec:riscos_cap}, the specific risks for Portuguese language models, ``Lack of representation of cultural and social aspects of the Brazilian population'' and ``Lack of representation of cultural and social aspects of the Portuguese-speaking population'', according to 70\% of the developers, were addressed either by Model Cards or by Harms Modeling.

At this stage,  we asked developers to indicate whether or not the language model presents the analyzed risk. The developers' opinions on the risks are shown in Table~\ref{tab:riscos_ent}. In addition, we assessed the disagreement of developers in the same group about the presence of risks in their language models. Table~\ref{tab:discordancia} shows these disagreement results. An important point about these results is the fact that all the developers in the same group took part in all the interviews together, in which they talked to each other about the ethical considerations raised by the AIETs about the analyzed language model, to reach an agreement on the final answer that would be given. However, even in this case, the results show disagreement about the impacts of the model among the developers themselves.

These results show how important it is for everyone involved in the process of creating and developing a language model to participate in the stages of identifying and documenting ethical considerations about the technology.

\begin{table}[!htb]
    \renewcommand{\arraystretch}{1.4}
    \footnotesize
    \caption{Developers' choice of which AIET best addressed or discussed a risk to their language model. The cells show the percentage of developers who chose an AIET as the one that best addressed the risk under analysis. The table includes the option ``None'', meaning that any AIETs identified the analyzed risk, according to the developers' opinion. The first 21 risks in the table were extracted from the risks mapped by~\citet{weidinger2021}}
    \centering
    \arrayrulecolor{black!15}
    \begin{tabular}{>{\raggedright\arraybackslash}p{7.3cm}|>{\centering\arraybackslash}p{1.3cm}|>{\centering\arraybackslash}p{1.4cm}|>{\centering\arraybackslash}p{1.3cm}} \arrayrulecolor{black}
    \toprule
    \multirow{2}{*}{\textbf{Analyzed Risk}} & \textbf{Model Cards} & \textbf{Harms Modeling} & \multirow{2}{*}{\textbf{None}}\\
    \midrule
    
    \arrayrulecolor{black!15}
    
    Social stereotypes and unfair discrimination  & \scaleb{0} & \scalea{90} & \scaleb{10} \\\hline
    
    Exclusionary norms   & \scaleb{0} & \scalea{80} & \scaleb{20} \\\hline
    
    Toxic language  & \scaleb{0} & \scalea{100} & \scaleb{0} \\\hline
    
    Lower performance by social group  & \scaleb{0} & \scalea{90} & \scaleb{10} \\\hline
    
    Compromise privacy by leaking private information   & \scaleb{10} & \scalea{80} & \scaleb{10} \\\hline
    
    Compromise privacy by correctly inferring private information  & \multirow{2}{*}{\scaleb{0}} & \multirow{2}{*}{\scalea{80}} & \multirow{2}{*}{\scaleb{20}} \\\hline
    
    Risks from leaking or correctly inferring sensitive information & \multirow{2}{*}{\scaleb{0}} & \multirow{2}{*}{\scalea{80}} & \multirow{2}{*}{\scaleb{20}} \\\hline
    
    Disseminating false or misleading information & \scaleb{0} & \scalea{90} & \scaleb{10} \\\hline
    
    Causing material harm by disseminating misinformation, \textit{e.g.}, in medicine or law  & \multirow{2}{*}{\scaleb{10}} & \multirow{2}{*}{\scalea{70}} & \multirow{2}{*}{\scaleb{20}} \\\hline
    
    Nudging or advising users to perform unethical or illegal actions  & \multirow{2}{*}{\scaleb{0}} & \multirow{2}{*}{\scalea{90}} & \multirow{2}{*}{\scaleb{10}} \\\hline
    
    Reducing the cost of disinformation campaigns  & \scaleb{0} & \scalea{90} & \scaleb{10} \\\hline
    
    Facilitating fraud and impersonation scams  & \scaleb{0} & \scalea{90} & \scaleb{10} \\\hline
    
    Assisting code generation for cyber attacks, weapons, or malicious use   & \multirow{2}{*}{\scaleb{0}} & \multirow{2}{*}{\scalea{70}} & \multirow{2}{*}{\scaleb{30}} \\\hline
    
    Illegitimate surveillance and censorship   & \scaleb{0} & \scalea{80} & \scaleb{20} \\\hline
    
    Anthropomorphizing systems can lead to overreliance or unsafe use   & \multirow{2}{*}{\scaleb{0}} & \multirow{2}{*}{\scalea{70}} & \multirow{2}{*}{\scaleb{30}} \\\hline
    
    Create avenues for exploiting user trust to obtain private information  & \multirow{2}{*}{\scaleb{0}} & \multirow{2}{*}{\scalea{70}} & \multirow{2}{*}{\scaleb{30}} \\\hline
    
    Promoting harmful stereotypes by implying gender or ethnic identity   & \multirow{2}{*}{\scaleb{0}} & \multirow{2}{*}{\scalea{90}} & \multirow{2}{*}{\scaleb{10}} \\\hline
    
    Environmental harms from operating language models & \scaleb{0} & \scalea{80} & \scaleb{20} \\\hline
    
    Increasing inequality and negative effects on job quality  & \scaleb{0} & \scalea{90} & \scaleb{10} \\\hline
    
    Undermining creative economies  & \scaleb{10} & \scalea{70} & \scaleb{20} \\\hline
    
    Disparate access to benefits due to hardware, software, skill constraints & \multirow{2}{*}{\scaleb{10}} & \multirow{2}{*}{\scalea{80}} & \multirow{2}{*}{\scaleb{10}} \\\hline
    
    Lack of representation of cultural and social aspects of the Brazilian population& \multirow{2}{*}{\scaleb{20}} & \multirow{2}{*}{\scaleb{50}} & \multirow{2}{*}{\scaleb{30}} \\\hline
    
    Lack of representation of cultural and social aspects of the Portuguese-speaking population & \multirow{2}{*}{\scaleb{40}} & \multirow{2}{*}{\scaleb{30}} & \multirow{2}{*}{\scaleb{30}} \\\hline
    
    Low performance in the Portuguese language  & \scaleb{40} & \scaleb{30} & \scaleb{30}\\\arrayrulecolor{black} 
    \bottomrule
    \end{tabular}
    \label{tab:riscos_entrevistas}
\end{table}

\begin{table}[!htb]
    \caption{Opinions of developers in the same group about the presence of risks in the analyzed model. This shows the percentage of developers in the same group who think the model has that risk. In other words, 100\% (darker color) means that all the developers in that group agree that the analyzed model has that risk. 0\% (lighter color) means that all the developers in that group agree that the analyzed model does not have that risk. The first 21 risks in the table were extracted from the risks mapped by~\citet{weidinger2021}}
    \renewcommand{\arraystretch}{1.4}
    \footnotesize
    \arrayrulecolor{black!35}
    \begin{tabular}{>{\raggedright\arraybackslash}p{6.1cm}|>{\centering\arraybackslash}p{1.2cm}|>{\centering\arraybackslash}p{1.2cm}|>{\centering\arraybackslash}p{1.2cm}|>{\centering\arraybackslash}p{1.2cm}}
    \arrayrulecolor{black} 
    \toprule
    \multirow{2}{*}{\textbf{Analyzed Risk}} & \textbf{Group~1\newline (2)} & \textbf{Group~2\newline (2)} & \textbf{Group~3\newline (3)} & \textbf{Group~4 \newline (3)}\\
    \midrule
    \arrayrulecolor{black!35}
       
    Social stereotypes and unfair discrimination  & \scaleb{50} & \scaleb{0} & \scalea{100} & \scalea{100} \\\hline
    
    Exclusionary norms  & \scaleb{50} & \scaleb{0} & \scaleb{33.3} & \scalea{66.7} \\\hline
    
    Toxic language   & \scaleb{50} & \scaleb{0} & \scalea{100} & \scalea{66.7}\\\hline
    
    Lower performance by social group  & \scaleb{50} & \scaleb{0} & \scalea{66.7} & \scalea{66.7}\\\hline
    
    Compromise privacy by leaking private information  & \multirow{2}{*}{\scaleb{0}} & \multirow{2}{*}{\scaleb{0}} & \multirow{2}{*}{\scalea{100}} & \multirow{2}{*}{\scalea{66.7}}\\\hline
    
    Compromise privacy by correctly inferring private information  & \multirow{2}{*}{\scaleb{50}} & \multirow{2}{*}{\scaleb{0}} & \multirow{2}{*}{\scalea{100}} & \multirow{2}{*}{\scalea{100}}\\\hline
    
    Risks from leaking or correctly inferring sensitive information  & \multirow{2}{*}{\scaleb{0}} & \multirow{2}{*}{\scaleb{0}} & \multirow{2}{*}{\scalea{100}} & \multirow{2}{*}{\scalea{100}}\\\hline
    
    Disseminating false or misleading information  & \scalea{100} & \scaleb{0} & \scalea{100} & \scalea{100} \\\hline
    
    Causing material harm by disseminating misinformation, \textit{e.g.}, in medicine or law  & \multirow{2}{*}{\scalea{100}} & \multirow{2}{*}{\scaleb{0}} & \multirow{2}{*}{\scalea{100}} & \multirow{2}{*}{\scalea{66.7}} \\\hline
    
    Nudging or advising users to perform unethical or illegal actions  & \multirow{2}{*}{\scaleb{0}} & \multirow{2}{*}{\scaleb{0}} & \multirow{2}{*}{\scalea{100}} & \multirow{2}{*}{\scalea{100}}\\\hline
    
    Reducing the cost of disinformation campaigns  & \scaleb{50} & \scaleb{0} & \scalea{100} & \scalea{66.7} \\\hline
    
    Facilitating fraud and impersonation scams & \scaleb{50} & \scaleb{0} & \scalea{100} & \scalea{66.7} \\\hline
    
    Assisting code generation for cyber attacks, weapons, or malicious use   & \multirow{2}{*}{\scaleb{0}} & \multirow{2}{*}{\scaleb{0}} & \multirow{2}{*}{\scaleb{0}} & \multirow{2}{*}{\scalea{66.7}} \\\hline
    
    Illegitimate surveillance and censorship  & \scaleb{0} & \scaleb{0} & \scalea{66.7} & \scalea{66.7} \\\hline
    
    Anthropomorphizing systems can lead to overreliance or unsafe use   & \multirow{2}{*}{\scaleb{0}} & \multirow{2}{*}{\scaleb{0}} &  \multirow{2}{*}{\scalea{100}} & \multirow{2}{*}{\scalea{66.7}} \\\hline
    
    Create avenues for exploiting user trust to obtain private information   & \multirow{2}{*}{\scaleb{0}} & \multirow{2}{*}{\scaleb{0}} & \multirow{2}{*}{\scalea{100}} & \multirow{2}{*}{\scalea{66.7}} \\\hline
    
    Promoting harmful stereotypes by implying gender or ethnic identity   & \multirow{2}{*}{\scaleb{50}} & \multirow{2}{*}{\scaleb{0}} & \multirow{2}{*}{\scalea{100}} & \multirow{2}{*}{\scalea{100}} \\\hline
    
    Environmental harms from operating language models  & \multirow{2}{*}{\scaleb{0}} & \multirow{2}{*}{\scaleb{0}} & \multirow{2}{*}{\scaleb{0}} & \multirow{2}{*}{\scalea{66.7}} \\\hline
    
    Increasing inequality and negative effects on job quality  & \multirow{2}{*}{\scalea{100}} & \multirow{2}{*}{\scaleb{0}} & \multirow{2}{*}{\scalea{100}} & \multirow{2}{*}{\scalea{66.7}} \\\hline
    
    Undermining creative economies  & \scaleb{0} & \scaleb{0} & \scalea{66.7} & \scalea{66.7} \\\hline
    
    Disparate access to benefits due to hardware, software, skill constraints & \multirow{2}{*}{\scaleb{0}} & \multirow{2}{*}{\scaleb{0}} & \multirow{2}{*}{\scalea{66.7}} & \multirow{2}{*}{\scalea{66.7}} \\\hline
    
    Lack of representation of cultural and social aspects of the Brazilian population & \multirow{2}{*}{\scaleb{50}} & \multirow{2}{*}{\scaleb{50}} &  \multirow{2}{*}{\scalea{66.7}} & \multirow{2}{*}{\scalea{66.7}} \\\hline
    
    Lack of representation of cultural and social aspects of the Portuguese-speaking population  & \multirow{2}{*}{\scaleb{50}} & \multirow{2}{*}{\scaleb{50}} & \multirow{2}{*}{\scalea{66.7}} & \multirow{2}{*}{\scalea{66.7}} \\\hline
    
    Low performance in the Portuguese language  & \scaleb{0} & \scaleb{0} & \scalea{66.7} & \scaleb{33.3}\\\arrayrulecolor{black} 
    \bottomrule
    \end{tabular}
    \label{tab:riscos_ent}
\end{table}

\begin{table}[!htb]
    \renewcommand{\arraystretch}{1.4}
    \footnotesize
    \caption{Disagreement between developers in the same group about the presence of risks in the analyzed model. The colored cells mean there was disagreement between the developers of the same language model regarding whether the model presents the analyzed risk. The first 21 risks in the table were extracted from the risks mapped by~\citet{weidinger2021}}
    \arrayrulecolor{black!35}
    \begin{tabular}{>{\raggedright\arraybackslash}p{6.1cm}|>{\centering\arraybackslash}p{1.2cm}|>{\centering\arraybackslash}p{1.2cm}|>{\centering\arraybackslash}p{1.2cm}|>{\centering\arraybackslash}p{1.2cm}}
    \arrayrulecolor{black} 
    \toprule
    \multirow{2}{*}{\textbf{Analyzed Risk}} & \textbf{Group~1\newline (2)} & \textbf{Group~2\newline (2)} & \textbf{Group~3\newline (3)} & \textbf{Group~4\newline (3)}\\
    \midrule
    \arrayrulecolor{black!35}
    
    Social stereotypes and unfair discrimination   & \scalecc{50} & \scalebb{0} & \scaleaa{100} & \scaleaa{100} \\\hline
    
    Exclusionary norms  & \scalecc{50} & \scalebb{0} & \scalebb{33.3} & \scaleaa{66.7} \\\hline
    
    Toxic language  & \scalecc{50} & \scalebb{0} & \scaleaa{100} & \scaleaa{66.7}\\\hline
    
    Lower performance by social group  & \scalecc{50} & \scalebb{0} & \scaleaa{66.7} & \scaleaa{66.7}\\\hline
    
    Compromise privacy by leaking private information  & \multirow{2}{*}{\scalebb{0}} & \multirow{2}{*}{\scalebb{0}} & \multirow{2}{*}{\scaleaa{100}} & \multirow{2}{*}{\scaleaa{66.7}}\\\hline
    
    Compromise privacy by correctly inferring private information   & \multirow{2}{*}{\scalecc{50}} & \multirow{2}{*}{\scalebb{0}} & \multirow{2}{*}{\scaleaa{100}} & \multirow{2}{*}{\scaleaa{100}}\\\hline
    
    Risks from leaking or correctly inferring sensitive information & \multirow{2}{*}{\scalebb{0}} & \multirow{2}{*}{\scalebb{0}} & \multirow{2}{*}{\scaleaa{100}} & \multirow{2}{*}{\scaleaa{100}}\\\hline
    
    Disseminating false or misleading information  & \scaleaa{100} & \scalebb{0} & \scaleaa{100} & \scaleaa{100} \\\hline
    
    Causing material harm by disseminating misinformation, \textit{e.g.}, in medicine or law  & \multirow{2}{*}{\scaleaa{100}} & \multirow{2}{*}{\scalebb{0}} & \multirow{2}{*}{\scaleaa{100}} & \multirow{2}{*}{\scaleaa{66.7}} \\\hline
    
    Nudging or advising users to perform unethical or illegal actions  & \multirow{2}{*}{\scalebb{0}} & \multirow{2}{*}{\scalebb{0}} & \multirow{2}{*}{\scaleaa{100}} & \multirow{2}{*}{\scaleaa{100}}\\\hline
    
    Reducing the cost of disinformation campaigns   & \scalecc{50} & \scalebb{0} & \scaleaa{100} & \scaleaa{66.7} \\\hline
    
    Facilitating fraud and impersonation scams  & \scalecc{50} & \scalebb{0} & \scaleaa{100} & \scaleaa{66.7} \\\hline
    
    Assisting code generation for cyber attacks, weapons, or malicious use   & \multirow{2}{*}{\scalebb{0}} & \multirow{2}{*}{\scalebb{0}} & \multirow{2}{*}{\scalebb{0}} & \multirow{2}{*}{\scaleaa{66.7}} \\\hline
    
    Illegitimate surveillance and censorship  & \scalebb{0} & \scalebb{0} & \scaleaa{66.7} & \scaleaa{66.7} \\\hline
    
    Anthropomorphizing systems can lead to overreliance or unsafe use  & \multirow{2}{*}{\scalebb{0}} & \multirow{2}{*}{\scalebb{0}} &  \multirow{2}{*}{\scaleaa{100}} & \multirow{2}{*}{\scaleaa{66.7}} \\\hline
    
    Create avenues for exploiting user trust to obtain private information  & \multirow{2}{*}{\scalebb{0}} & \multirow{2}{*}{\scalebb{0}} & \multirow{2}{*}{\scaleaa{100}} & \multirow{2}{*}{\scaleaa{66.7}} \\\hline
    
    Promoting harmful stereotypes by implying gender or ethnic identity  & \multirow{2}{*}{\scalecc{50}} & \multirow{2}{*}{\scalebb{0}} & \multirow{2}{*}{\scaleaa{100}} & \multirow{2}{*}{\scaleaa{100}} \\\hline
    
    Environmental harms from operating language models  & \multirow{2}{*}{\scalebb{0}} & \multirow{2}{*}{\scalebb{0}} & \multirow{2}{*}{\scalebb{0}} & \multirow{2}{*}{\scaleaa{66.7}} \\\hline
    
    Increasing inequality and negative effects on job quality  & \multirow{2}{*}{\scaleaa{100}} & \multirow{2}{*}{\scaleaa{100}} & \multirow{2}{*}{\scaleaa{100}} & \multirow{2}{*}{\scaleaa{66.7}} \\\hline
    
    Undermining creative economies  & \scalebb{0} & \scalebb{0} & \scaleaa{66.7} & \scaleaa{66.7} \\\hline
    
    Disparate access to benefits due to hardware, software, skill constraints & \multirow{2}{*}{\scalebb{0}} & \multirow{2}{*}{\scalebb{0}} & \multirow{2}{*}{\scaleaa{66.7}} & \multirow{2}{*}{\scaleaa{66.7}} \\\hline
    
    Lack of representation of cultural and social aspects of the Brazilian population & \multirow{2}{*}{\scalecc{50}} & \multirow{2}{*}{\scalecc{50}} &  \multirow{2}{*}{\scaleaa{66.7}} & \multirow{2}{*}{\scaleaa{66.7}} \\\hline
    
    Lack of representation of cultural and social aspects of the Portuguese-speaking population  & \multirow{2}{*}{\scalecc{50}} & \multirow{2}{*}{\scalecc{50}} & \multirow{2}{*}{\scaleaa{66.7}} & \multirow{2}{*}{\scaleaa{66.7}} \\\hline
    
    Low performance in the Portuguese language  & \scalebb{0} & \scalebb{0} & \scaleaa{66.7} & \scalebb{33.3}\\
    \arrayrulecolor{black} 
    \bottomrule
    \end{tabular}

\vspace{0.1cm}

\textbf{How to interpret this table:} This table presents two information. \textbf{Percentage} indicates the percentage of developers in the same group who think the model has that risk. 100\% means that all the developers in that group agree that the analyzed model has that risk. 0\% means that all the developers in that group agree that the analyzed model does not have that risk. \textbf{Colors} indicates a disagreement among developers within the same group about the presence of risk in the model. This disagreement (colored cells) occurs for values that differ from 0\% or 100\% (white cells). For example, in a group of two developers, 50\% means that one developer believes the model has that risk, while the other developer does not; this disagreement is reflected through the cell color.

    \label{tab:discordancia}
\end{table}

\subsubsection{Scores of the analyzed AIETs by the Developers}
\label{sec:pont_entrevistas}

\begin{table}[!htb]
    \renewcommand{\arraystretch}{1.1}
    \small
    \caption[Rating given to AIETs by the developers]{Rating given to AIETs by the developers. The last line shows the average rating for each AIET}
    \centering \rowcolors{12}{}{Gray!40}
    \arrayrulecolor{black}
    \begin{tabular}{>{\arraybackslash}p{2cm}>{\centering\arraybackslash}p{3cm}>{\centering\arraybackslash}p{3cm}}
    \toprule
         & \textbf{Model Cards} & \textbf{Harms Modeling} \\\midrule
        Developer 1 & 5 & 4 \\
        Developer 2 & 5 & 5 \\
        Developer 3 & 5 & 4 \\
        Developer 4 & 2 & 3 \\
        Developer 5 & 4 & 1 \\
        Developer 6 & 5 & 4 \\
        Developer 7 & 4 & 4 \\
        Developer 8 & 4 & 4 \\
        Developer 9 & 3 & 3 \\
        Developer 10 & 4 & 4 \\\hline 
        \textbf{Average} & \textbf{4.1} & \textbf{3.6} \\
        \bottomrule
    \end{tabular}
    \label{tab:pontuacaoentrevistas}
\end{table}

Finally, we asked developers to provide a score for each analyzed AIET and Table~\ref{tab:pontuacaoentrevistas} shows the results obtained. The best-rated AIET was Model Cards, with an average of 4.1 points. Harms Modeling  had an average score of 3.6 points. 
The results achieved are in line with those presented and discussed earlier in this section, showing that part of developers prefer to use the Model Cards. 
In the final evaluation form, we left a field open for developers to leave their comments on the AIETs if they wished. We received the following comments from three participants about their preference for Model Cards:

\begin{quote}
    \textit{``I found both AIETs very useful, but I preferred Model Cards because it is more objective; this makes it easier to develop and consume by other people who will read its content.''}.
\end{quote}

\begin{quote}
    \textit{``Harms Modeling, although more complete in terms of subject matter, always seemed very general to me, which ended up being an evaluation of the technology rather than the models themselves, as well as being very extensive. In this respect, Model Cards has my preference, as it is much more specific to the model itself, although it does not cover as many factors.''}.
\end{quote}

\begin{quote}
    \textit{``Overall, I came away with a feeling that Harms Modeling is a much more `nitpicky' tool in the sense that it is much more extensive and tries to cover as many considerations as possible, but this also means that in the case of language models, it covers many aspects that do not apply. At various times we have commented on topics that seemed very disconnected and I had to `struggle' to find something that fit.
    Model Cards seemed more succinct. Looking at the final documentation, I can say that the documentation resulting from Model Cards is much closer to what I would write in a model/article of my own (if I were writing it without the aid of any AIET), than the documentation resulting from Harms Modeling.''}.
\end{quote}

The comments summarize that Model Cards is an objective, specific tool with more tangible documentation. On the other hand, Harms Modeling is a more complete tool, but it is very general and does not consider specific aspects of language models in general.

%%===================================%%
%% Limitations and future directions %%
%%===================================%%
\section{Limitations and Future Directions}
\label{sec:limitations}

This paper presents the perspectives of language model developers about four distinct AIETs. However, we evaluated them using a pre-defined form with closed and multiple-choice questions. Relying the evaluation only on a quantitative questionnaire may not be the most suitable method for a comprehensive assessment of AIETs, as such evaluations encompass various nuances. In future work, it would be beneficial to enhance this analysis by incorporating more open-ended questions for the interviewees and gathering additional qualitative data from the interviews.

Currently, this research does not offer deeper insights into participants' reasoning and preferences for AIETs. This insight could lead to richer results, not only about which AIETs participants prefer, but for what reasons they prefer them. In addition, more in-depth analyses of the impacts of language models developed for the Portuguese language and how developers deal with these impacts have not been carried out. However, it would be essential to consider this aspect of the research, also taking into account the nuances of the conversations during the interviews and the logic for mapping the risks of the model. In future work, we intend to conduct a qualitative methodological analysis of the interviews.

In future research, it would be valuable to analyze the documentation produced by each AIET with individuals---other than the developers themselves---, who are intended to interpret the results. However, this analysis was not conducted in the current study, as a decision was made not to publicly disclose the documentation generated by each AIET about the models interviewed. The documentation produced in this study, particularly for the AIETs Harms Modeling and ALTAI, followed a specific format created by us (the research team of this study), as neither AIET included examples of the expected final documentation. Therefore, the findings of this study may be influenced by how we constructed these documentations. 

An important point is that many of the interviewees expressed a desire to receive feedback or a score regarding their model's potential ``damaging'' effects based on the assessment of the researchers involved in this study. However, due to the scope of this research and the team members' backgrounds, we did not address this aspect at this research stage.

The AIETs analyzed in this research are general and can be applied to various types of AIs. In future work, it will be essential to map the best and worst aspects of each AIET analyzed and propose a new AIET specifically tailored for language models. This is important as our interviews with developers highlighted several considerations that were not relevant to language models. Therefore, a new AIET must have the differential of filling the gap in the other AIETs, and, being specific to language models, it must have the means to identify their risks. Additionally, it is crucial to consider language, cultural, and country-specific aspects when assessing the language model's context. We hypothesize that each AIET has its own differential and that combining two or more could generate more significant results when evaluating a language model. Identifying these points is crucial when proposing a new AIET that can be used as a single tool with the best aspects of the others.

It is important to note that the results of this study currently only consider language models developed exclusively for the Portuguese language within the Brazilian context. As future work, it would be interesting to assess whether the results extend to (i) other Portuguese-speaking contexts, (ii) other linguistic contexts beyond Portuguese, and (iii) multilingual contexts. Such investigations could bring additional results to the research on the application of AIETs in language models, as we could understand whether the results obtained are general for language models or unique to models focused exclusively on Portuguese. 

Carrying out a study on an AIET requires significant time from both the researchers and the people who participate voluntarily. Consequently, it was not possible to analyze all the AIETs found in the AIET selection stage reported in Section~\ref{sec:AIETs_filtered}. In total, we conducted 35 interview sessions, each lasting roughly one hour. Additionally, documenting each conversation required about two hours. Overall, each participant group dedicated approximately 2 to 3 months to the study.

This study primarily analyzes AIETs to assess their usability and effectiveness in aiding language model developers in identifying, mapping, and documenting the ethical issues associated with their models. It is important to highlight that the scope of this study is deliberately limited to these specific functions. However, it should be noted that AIETs can serve additional purposes, such as evaluating datasets, assessing early stages of AI development, and conducting quantitative analyses of models like fairness metrics, which were not explored in this~study.

As far as we know, there is no method in the literature for comparing AIETs~\citep{qiang2023}, and we hope to provoke the community's interest in finding ways of standardizing AIET evaluations. A method for comparing AIETs would allow for better adoption by researchers, developers, companies, and organizations since the AI literature  lacks papers with exclusive sections for ethical considerations, and such AIETs could be a guide for the elaboration of this section. Having a tool to help these people identify the ethical considerations of what is being proposed could make the future of AI more transparent, responsible, and reliable. However, we understand that an AIET alone will not lead us down this path and will require from the AI (ethics) research community efforts. 

%%==================================%%
%%             Conclusion           %%
%%==================================%%
\section{Conclusion}
\label{sec:conclusion}

This paper introduced a methodology to carry out a systematically filter for selecting and evaluating AIETs in language models. To this end, we conducted a case study where we applied four AIETs (Model Cards, ALTAI, FactSheets, and Harms Modeling) through interviews with developers of language models developed with a focus on Portuguese (PT-BR models). We performed the interviews using the items of each AIET analyzed as a script. The total time to interview the developers was around two hours for Model Cards, three hours for ALTAI, two hours for FactSheets, and four hours for Harms Modeling. The developers answered the AIETs verbally during the interviews, and we conducted the interviews in Portuguese. 

We evaluated the AIETs from the developers' perspective, using a form each developer answered. The results suggest that the AIETs applied are effective for use as a guide in drawing up general ethical considerations about language models. However, we note that they are general and do not address unique aspects of these models, such as idiomatic expressions. In addition, the AIETs did not help to identify potential negative impacts of models for the Portuguese language, such as the lack of representation of social and cultural aspects of Portuguese-speaking populations in the Brazilian context. 

According to the developers' perspective, the final AIET documentation presented the potential negative impacts contained in the language models and is responsible and informative document for the stakeholders. However, the results show that documentation presenting the risks in detail differs from being informative and easy to understand. From the analyzed AIETs, the developers pointed Harms Modeling as the most helpful AIET in identifying ethical considerations about their model and pointed Model Cards as the AIET with the best usability and that generates the most responsible documentation for disclosing ethical considerations to their stakeholders. Therefore, using multiple AIETs when raising ethical considerations about a model would be interesting. 

The developers interviewed so far said they had never ethically evaluated their models. This result indicates that it is not yet natural for developers to add a step to ethically analyze their models during the development stage before publishing them, showing the need for greater engagement and dissemination in the area of AI ethics. During the interviews, all the developers of the same model participated together in the interviews regarding the language model they had developed, where they discussed and raised thoughts together. Even so, when it came to individually pointing out the risks and negative impacts of the developed model, there was disagreement about the risks contained in the model. This shows us how important it is that the ethical examination of a technology is carried out jointly by everyone involved in its life cycle.

The results showed that AIETs met their objectives of helping to map ethical issues in AI. However, we noted that relying solely on AIETs is insufficient; developers must also possess a grounding in AI ethics to effectively consider the implications of their technologies. In addition, some AIETs were exhausting to complete due to the complexity of use and readability of the content.

%%==================================%%
%%           Ethics Statement       %%
%%==================================%%
\section*{Ethics Statement}
\label{sec:ethicstt}

This research evaluates AIETs through a case study involving the developers of a language model. The developers participated in the study through remote interviews. We applied the AIETs to the developed language model through an orally answered questionnaire. This study was approved by the Research Ethics Committee of the Universidade Estadual de Campinas under CAAE number 74643023.8.0000.5404. To carry out this work, we followed all the guidelines and protocols required by the Committee. We are committed to not disclosing AIETs' final documentation about the model. We offered the participants the option to disclose the documentation to their stakeholders by their means. All results published in this paper have been obtained and disclosed with the consent of the participants. 

This research was carried out exclusively by computer science researchers. As such, this paper carries all our biases in the field, influencing how each AIET was interpreted, how the interviews were conducted, how the AIETs' final documentations were drawn up, how the results were analyzed, and how the paper was written. In the future, we intend to involve the participation of a multidisciplinary group of researchers so that ethical considerations about the work could be raised at all stages. In addition, the contribution of a multidisciplinary group would promote deeper discussions about the importance, use, and quality of existing AIETs. 

This paper may generate the adverse and unintended impact of one AIET being perceived better than another, which was different from this paper's intention. It was not our aim to provide a rank of the analyzed AIETs or to discourage using a specific AIET because it has a low score from the developers' perspective---following the criteria adopted in this paper. Here, we showed a case study where the interviewed developers of specific language models classified the AIETs. The result may differ if analyzed with another language models---or even another type of AI---and with other developers. Since there are no methods for comparing AIETs in the literature, we followed the criteria described in this paper. For this reason, we urge caution when interpreting these results.

%%==================================%%
%%            Declarations          %%
%%==================================%%

\section*{Declarations}
%Some journals require declarations to be submitted in a standardised format. Please check the Instructions for Authors of the journal to which you are submitting to see if you need to complete this section. If yes, your manuscript must contain the following sections under the heading `Declarations':

\begin{itemize}
\item \textbf{Funding:} This project was supported by the Brazilian Ministry of Science, Technology and Innovations, with resources from Law nº 8,248, of October 23, 1991, within the scope of PPI-SOFTEX, coordinated by Softex and published Arquitetura Cognitiva (Phase 3), DOU 01245.003479/2024-10. J.Silva was partially financed by the Coordination for the Improvement of Higher Education Personnel (CAPES)---Finance Code 001. J.Silva (2024/23118-1), D.Moreira (2023/05939-5), G.Santos (2024/07969-1), and S.Avila (2023/12086-9, 2023/12865-8, 2020/09838-0, 2013/08293-7) are also partially funded by FAPESP. S.Avila (316489/2023-9) and H.Pedrini (304836/2022-2) are also partially funded by CNPq.

\item \textbf{Conflict of interest:} The authors declare no competing interests.
\item \textbf{Ethics approval and consent to participate:} This study was approved by the Research Ethics Committee of the Universidade Estadual de Campinas under CAAE number 74643023.8.0000.5404. To carry out this work, we followed all the guidelines and protocols required by the Committee.  
\item \textbf{Consent for publication:} All the participants signed an informed consent form accepting their participation in the study and authorizing it to be carried out. 
\item \textbf{Data availability:} None of the data collected in the interviews will be made available, nor will the AIETs' final documentation about the model (sensitive data). 
\item \textbf{Material availability:} `Not applicable'.
\item \textbf{Code availability:} `Not applicable'.
\item \textbf{Additional Acknowledgments:} The authors would like to thank all the members (and former members) of the Natural Language Processing research group at the Artificial Intelligence and Cognitive Architectures Hub (H.IAAC). The authors also thank all participants in this study. 
\item \textbf{Author contribution:} J.Silva was responsible for writing the manuscript, conducting all steps of this research and the interviews with the developers. G.Santos, D.Moreira, and A.Ferreira participated in the initial interviews with the \text{CAPIVARA} model, helping to adapt the interview script. H.Maia contributed to discussing about this research at all stages and with the Research Ethics Committee project. S.Avila and H.Pedrini are the advisors of this research, providing guidance on all tasks and contributing to the writing process. All authors reviewed the manuscript and provided critical feedback to enhance its quality. All authors have approved the final version of this manuscript and agree to be responsible for all aspects of the work, ensuring its accuracy and integrity. 
\end{itemize}

\noindent
%If any of the sections are not relevant to your manuscript, please include the heading and write `Not applicable' for that section. 

\bibliography{sn-bibliography}

\begin{appendices}

%%=============================================%%
%% For submissions to Nature Portfolio Journals %%
%% please use the heading ``Extended Data''.   %%
%%=============================================%%

%%=============================================================%%
%% Sample for another appendix section			       %%
%%=============================================================%%

%% \section{Example of another appendix section}\label{secA2}%
%% Appendices may be used for helpful, supporting or essential material that would otherwise 
%% clutter, break up or be distracting to the text. Appendices can consist of sections, figures, 
%% tables and equations etc.

\newpage
\setcounter{table}{0}
\renewcommand{\thetable}{A\arabic{table}}
\setcounter{figure}{0}
\renewcommand{\thefigure}{A\arabic{figure}}

\section{AIETs list -- until July 2023}
\label{sec:aietslist}
{\footnotesize 
\begin{enumerate}
\item A confidence-based approach for balancing fairness and accuracy~\citep{fish_confidence_2016}
\item A generic framework for privacy preserving deep learning~\citep{ryffel2018generic}
\item A Matrix for Selecting Responsible AI Frameworks~\citep{narayanan2023matrix}
\item A Playbook for Ethical Technology Governance~\citep{playboook}
\item A responsible AI framework: pipeline contextualisation~\citep{vyhmeister2023responsible}
\item A Right to Reasonable Inferences: Re-Thinking Data Protection Law in the Age of Big Data and AI~\citep{wachter2019right}
\item A seven-layer model with checklists for standardising fairness assessment throughout the AI lifecycle~\citep{agarwal2024seven}
\item A survey of methods for explaining black box models~\citep{Guidotti}
\item A Survey of Value Sensitive Design Methods~\citep{HCI-015}
\item A systematic methodology for privacy impact assessments: a design science based approach~\citep{Oetzel01032014}
\item A toolkit of dilemmas: Beyond debiasing and fairness formulas for responsible AI/ML~\citep{10227133}
\item A Toolkit to Enable the Design of Trustworthy AI~\citep{Schmager}
\item A3i the Trust in AI Framework~\citep{ai3}
\item Adversarial Robustness - Theory and Practice~\citep{advrob}
\item Accenture Fairness Evaluation Tool~\citep{accfair}
\item Accountable Algorithms~\citep{kroll2017accountable}
\item Advbox: a toolbox to generate adversarial examples that fool neural networks~\citep{goodman2020advboxtoolboxgenerateadversarial}
\item Adversarial Robustness Toolbox v1.0.0~\citep{nicolae2019adversarialrobustnesstoolboxv100}	
\item Aequitas: A Bias and Fairness Audit Toolkit~\citep{saleiro2019aequitas}
\item Agile Ethics for AI (HAI)~\citep{hai}
\item AI Audit: A Card Game to Reflect on Everyday AI Systems~\citep{Ali_Kumar_Breazeal_2024}
\item AI and Big Data: A blueprint for a human rights, social and ethical impact assessment~\citep{MANTELERO2018754}
\item AI Blindspot: A discovery process for spotting unconscious biases and structural inequalities in AI systems~\citep{Blindspots}
\item AI Commons~\citep{AICommons}
\item AI Digital Tool Product Lifecycle Governance Framework through Ethics and Compliance by Design~\citep{10195137}
\item AI Explainability 360 Toolkit~\citep{10.1145/3430984.3430987}
\item AI Fairness 360 toolkit~\citep{aif360-oct-2018}
\item AI Incident Database~\citep{aiincident}
\item AI Privacy Toolkit~\citep{GOLDSTEEN2023101352}
\item AI Procurement in a Box~\citep{aipb}
\item AI Usage Cards: Responsibly Reporting AI-Generated Content~\citep{10266234}
\item AI-RFX Procurement Framework~\citep{airfx}
\item Algorithm tips resources and leads for investigating algorithms in society~\citep{AlgorithmTips}
\item Algorithmic Accountability and Public Reason~\citep{binns2018algorithmic}
\item Algorithmic Accountability Policy Toolkit~\citep{aapt}
\item Algorithmic Accountability: Journalistic investigation of computation power structures~\citep{Diakopoulos04052015}
\item Algorithmic Impact Assessment (AIA)~\citep{canadaaist}
\item Algorithmic Impact Assessments: A Practical Framework for Public Agency Accountability~\citep{reisman2018algorithmic}
\item Algorithmic Transparency via Quantitative Input Influence: Theory and Experiments with Learning Systems~\citep{7546525}
\item Alibi Explain: Algorithms for Explaining Machine Learning Models~\citep{JMLR:v22:21-0017}
\item Amazon SageMaker Clarify: Machine Learning Bias Detection and Explainability in the Cloud~\citep{10.1145/3447548.3467177}
\item An Ethical Framework for Evaluating Experimental Technology~\citep{van2016ethical}
\item An Ethical Toolkit for Engineering/Design Practice~\citep{etp}
\item Assessing Radiology Research on Artificial Intelligence: A Brief Guide for Authors, Reviewers, and Readers—From the Radiology Editorial Board~\citep{bluemke2020assessing}
\item Assessment List for Trustworthy Artificial Intelligence (ALTAI) for self-assessment~\citep{altai}
\item Audit-AI~\citep{auditai}
\item Auditing Algorithms @ Northeastern~\citep{auditinga}
\item Auditing Algorithms: Research Methods for Detecting Discrimination on Internet Platforms~\citep{sandvig2014auditing}
\item Auditing large language models: a three-layered approach~\citep{mokander2024auditing}
\item Augmented Datasheets for Speech Datasets and Ethical Decision-Making~\citep{10.1145/3593013.3594049}
\item Beyond Accuracy: Behavioral Testing of NLP models with CheckList~\citep{ribeiro-etal-2020-beyond}
\item BOLD: Dataset and Metrics for Measuring Biases in Open-Ended Language Generation\citep{dhamala2021bold}
\item Building Data AI Ethics Committees~\citep{accbuilding}
\item Captum~\citep{captum}
\item Cards for Humanity~\citep{cardshuman}
\item Certifying and Removing Disparate Impact~\citep{10.1145/2783258.2783311}
\item Charting the Sociotechnical Gap in Explainable AI: A Framework to Address the Gap in XAI~\citep{10.1145/3579467}
\item CleverHans~\citep{papernot2018cleverhans}
\item Closing the AI Accountability Gap: Defining an End-toEnd Framework for Internal Algorithmic Auditing~\citep{raji2020closing}
\item Co-Designing Checklists to Understand Organizational Challenges and Opportunities around Fairness in AI~\citep{10.1145/3313831.3376445}
\item Community Jury~\citep{communityjury}
\item Consequence Scanning~\citep{doteveryone}
\item Contrastive Explanation Method~\citep{dhurandhar2018explanations}
\item Corporate Digital Responsibility~\citep{LOBSCHAT2021875}
\item Counterfactual Explanations without Opening the Black Box: Automated Decisions and the GDPR~\citep{wachter2018counterfactualexplanationsopeningblack}
\item Counterfactual Fairness~\citep{NIPS2017_a486cd07}
\item Cyber Resilience Review (CRR)~\citep{cyber}
\item Data Cards: Purposeful and Transparent Dataset Documentation for Responsible AI~\citep{10.1145/3531146.3533231}
\item Data Ethics Canvas: user guide~\citep{odidata}
\item Data Ethics Framework~\citep{def_uk}
\item Data Management and Use: Governance in the 21st Century~\citep{british2017data}
\item Data Minimisation: A Language-Based Approach~\citep{10.1007/978-3-319-58469-0_30}
\item Data Statements for Natural Language Processing: Toward Mitigating System Bias and Enabling Better Science~\citep{bender2018data}
\item Data Statements: From Technical Concept to Community Practice~\citep{10.1145/3594737}
\item Datasheets for Datasets~\citep{gebru2021}
\item DEDA: De Ethische Data Assistent~\citep{deda}
\item Deep Inside Convolutional Networks: Visualising Image Classification Models and Saliency Maps~\citep{simonyan2014deepinsideconvolutionalnetworks}
\item Deep Learning for Case-Based Reasoning Through Prototypes: A Neural Network That Explains Its Predictions~\citep{li2018deep}
\item DeepExplain~\citep{deepexplain}
\item DeepLIFT~\citep{deeplift}
\item DEON - An ethics checklist for data scientists~\citep{deon}
\item Design Ethically Toolkit~\citep{det}
\item Designing for Motivation, Engagement and Wellbeing in Digital Experience~\citep{10.3389/fpsyg.2018.00797}
\item Development of privacy design patterns based on privacy principles and UML~\citep{8022748}
\item DIANES: A DEI Audit Toolkit for News Sources~\citep{10.1145/3477495.3531660}
\item DiCE: Diverse Counterfactual Explanations for ML~\citep{mothilal2020dice}
\item Digital Impact Toolkit~\citep{dtoolkit}
\item ELI5~\citep{ELI5}
\item Ellpha~\citep{Ellpha}
\item Empowering AI Leadership~\citep{eai}
\item Empowering AI Leadership: AI C-Suite Toolkit~\citep{eail}
\item Enigma: Decentralized Computation Platform with Guaranteed Privacy~\citep{8333139}
\item Equity Evaluation Corpus (EEC)~\citep{kiritchenko2018examining}
\item Ethical framework for Artificial Intelligence and Digital technologies~\citep{ASHOK2022102433}
\item Ethical Requirements Stack: A framework for implementing ethical requirements of AI in software engineering practices~\citep{10.1145/3593434.3593489}
\item Ethically Aligned Design IEEE~\citep{shahriari2017ieee}
\item Ethics \& Algorithms Toolkit~\citep{eatoolkit}
\item Ethics and Privacy in AI and Big Data: Implementing Responsible Research and Innovation~\citep{8395078}
\item Ethics Cards~\citep{ethicskit}
\item Ethics for designers -- the toolkit~\citep{ethicsfordesigners}
\item Ethics Kit~\citep{ethicskittool}
\item EthicsNet~\citep{ethicsnet}
\item Explainability fact sheets: a framework for systematic assessment of explainable approaches~\citep{10.1145/3351095.3372870}
\item Explainable AI: Driving business value through greater understanding~\citep{oxborough2018explainable}
\item Explanations based on the Missing: Towards Contrastive Explanations with Pertinent Negatives~\citep{dhurandhar2018explanations}
\item FactSheets: Increasing trust in AI services through supplier's declarations of conformity~\citep{arnold2019factsheets}
\item Fair, Transparent, and Accountable Algorithmic Decision-making Processes~\citep{lepri2018fair}
\item Fairlearn: A toolkit for assessing and improving fairness in AI~\citep{bird2020fairlearn}
\item Fairness Constraints: Mechanisms for Fair Classification~\citep{pmlr-v54-zafar17a}
\item Fairness in Classification~\citep{finc}
\item Fairness in Design: A Framework for Facilitating Ethical Artificial Intelligence Designs~\citep{10091496}
\item Fairness Score and process standardization: framework for fairness certification in artificial intelligence systems~\citep{agarwal2023fairness}
\item Foolbox: A Python toolbox to benchmark the robustness of machine learning models~\citep{rauber2018foolboxpythontoolboxbenchmark}
\item Foresight into AI Ethics (FAIE)~\citep{orif}
\item Framework for Evaluating Ethics in AI~\citep{10099747}
\item Futures Wheel~\citep{futureweeel}
\item Generative adversarial networks (GANs) for synthetic dataset generation with binary classes~\citep{joshi2019generative}
\item Glass-Box: Explaining AI Decisions With Counterfactual Statements Through Conversation With a Voice-enabled Virtual Assistant~\citep{ijcai2018p865}
\item Guide to the UK General Data Protection Regulation (UK GDPR)~\citep{officerguide}
\item H20.ai Machine Learning Interpretability Resources~\citep{h20}
\item Harms Modeling~\citep{harmsmodeling}
\item HAX Workbook~\citep{mhax}
\item Hazy~\citep{hazi}
\item How to stimulate effective public engagement on the ethics of artificial intelligence~\citep{adams2019stimulate}
\item Human Decisions and Machine Predictions~\citep{10.1093/qje/qjx032}
\item IBM Watson OpenScale~\citep{ibmw}
\item ICO Anonymisation: managing data protection risk. Code of Practice~\citep{graham2012anonymisation}
\item ICO Guide to the General Data Protection Regulation~\citep{gicoa}
\item IDEO’s AI Ethics Cards~\citep{ideo2019}
\item IEEE Draft Model Process for Addressing Ethical Concerns During System Design P7000/D3~\citep{olszewska2021ieee}
\item IEEE Recommended Practice for Assessing the Impact of Autonomous and Intelligent Systems on Human WellBeing Std 7010~\citep{ieeerecome}
\item IEEE Ethics Certification Program for Autonomous and Intelligent Systems (ECPAIS)~\citep{ieeesa}
\item Improving Social Responsibility of Artificial Intelligence by Using ISO 26000~\citep{Zhao_2018}
\item Interactive Model Cards: A Human-Centered Approach to Model Documentation~\citep{10.1145/3531146.3533108}
\item InterpretML: A Unified Framework for Machine Learning Interpretability~\citep{nori2019interpretml}
\item Judgment Call the Game: Using Value Sensitive Design and Design Fiction to Surface Ethical Concerns Related to Technology~\citep{ballard2019judgmentcall}
\item Learning Adversarially Fair and Transferable Representations~\citep{pmlr-v80-madras18a}
\item Learning Important Features Through Propagating Activation Differences~\citep{pmlr-v70-shrikumar17a}
\item Learning Representation for Counterfactual Inference~\citep{pmlr-v48-johansson16}
\item LIME -- ``Why Should I Trust You?'': Explaining the Predictions of Any Classifier~\citep{10.1145/2939672.2939778}
\item Man is to Computer Programmer as Woman is to Homemaker? Debiasing Word Embeddings~\citep{bolukbasi2016mancomputerprogrammerwoman}
\item Metrics for explainable AI: Challenges and prospects~\citep{hoffman2019metricsexplainableaichallenges}
\item Microsoft’s framework for building AI systems responsibly~\citep{mfres}
\item Model Cards for Model Reporting~\citep{mitchell2019}
\item Model Ethical Data Impact Assessment~\citep{modelea}
\item Model-Agnostic Private Learning via Stability~\citep{bassily2018modelagnosticprivatelearningstability}
\item Moral Machine~\citep{moralmachine}
\item Navigating Data-Centric Artificial Intelligence With DC-Check: Advances, Challenges, and Opportunities~\citep{Seedat_2024}
\item Neural Network Exchange Format (NNEF)~\citep{nnef}
\item New Economy Impact Model~\citep{neim}
\item OECD Framework for the Classification of AI systems~\citep{oecd}
\item On Pixel-Wise Explanations for Non-Linear Classifier Decisions by layer-wise relevance propagation~\citep{10.1371/journal.pone.0130140}
\item ONNX Open Neural Network Exchange~\citep{onne}
\item OpenMined~\citep{openmi}
\item OpenML~\citep{openml}
\item Peeking Inside the Black Box: Visualizing Statistical Learning With Plots of Individual Conditional Expectation~\citep{Goldstein02012015}
\item People + AI Research Guidebook~\citep{apir}
\item Personal AI Privacy Watchdog Could Help You Regain Control of Your Data~\citep{orcutt2017personal}
\item PMML: An Open Standard for Sharing Models~\citep{guazzelli2009pmml}
\item Practical verification of decision-making in agent-based autonomous systems~\citep{dennis2016practical}
\item Principles for Accountable Algorithms and a Social Impact Statement for Algorithms~\citep{principlesfor}
\item Privacy by Design: essential for organizational accountability and strong business practices~\citep{cavoukian2010privacy}
\item Privacy Principles: Towards a common privacy audit methodology~\citep{10.1007/978-3-319-22906-5_17}
\item PROBAST: A Tool to Assess the Risk of Bias and Applicability of Prediction Model Studies~\citep{wolff2019probast}
\item PROV-ML -- Provenance Data in the Machine Learning Lifecycle in Computational Science and Engineering~\citep{8943505}
\item Questioning the assumptions behind fairness solutions~\citep{overdorf2018questioningassumptionsfairnesssolutions}
\item Responsible AI Licenses (RAIL)~\citep{rail}
\item Responsible AI Toolbox~\citep{responsiblemicrosoft}
\item Responsible AI Toolkit~\citep{tensorflowresponsible}
\item Responsible and Inclusive Technology Framework: A Formative Framework to Promote Societal Considerations in Information Technology Contexts~\citep{sandoval2023responsibleinclusivetechnologyframework}
\item Responsible natural language processing: A principlist framework for social benefits~\citep{BEHERA2023122306}
\item Risks, Harms and Benefits Assessment Tool~\citep{rsbtool}
\item SAFETYKIT: First Aid for Measuring Safety in Open-domain
Conversational Systems~\citep{dinan2022safetykit}
\item Scalable Private Learning with PATE~\citep{papernot2018scalableprivatelearningpate}
\item SHAP - A Unified Approach to Interpreting Model Predictions~\citep{shap2017}
\item Should I disclose my dataset? Caveats between reproducibility and individual data rights~\citep{m-benatti-etal-2022-disclose}
\item System Cards, a new resource for understanding how AI systems work~\citep{green2021system}
\item Ten simple rules for responsible big data research~\citep{10.1371/journal.pcbi.1005399}
\item Tensorflow's Fairness Evaluation and Visualization Toolkit~\citep{tensorflowfairness}
\item TensorFlow Privacy~\citep{tensorflowprivacy}
\item Text Characterization Toolkit (TCT)~\citep{simig-etal-2022-text}
\item The “big red button” is too late: an alternative model for the ethical evaluation of AI systems~\citep{arnold2018big}
\item The Algorithmic Equity Toolkit (AEKit)~\citep{aekit}
\item The Dataset Nutrition Label: A Framework To Drive Higher Data Quality Standards~\citep{holland2018}
\item The Dataset Nutrition Label (2nd Gen): Leveraging Context to Mitigate Harms in Artificial Intelligence~\citep{chmielinski2022datasetnutritionlabel2nd}
\item The Field Guide to Human-Centred Design~\citep{designkit}
\item The ICO and artificial intelligence: the role of fairness in the GDPR framework~\citep{BUTTERWORTH2018257}
\item The LinkedIn Fairness Toolkit (LiFT)~\citep{vasudevan20lift}
\item The Machine Learning Reproducibility Checklist~\citep{mlchecklist}
\item The Model Card Authoring Toolkit: Toward Community-centered, Deliberation-driven AI Design~\citep{10.1145/3531146.3533110}
\item The ONS Methodology working paper on synthetic data~\citep{ons}
\item The Scored Society: Due process for automated predictions~\citep{citron2014scored}
\item The Selective Labels Problem: Evaluating Algorithmic Predictions in the Presence of Unobservables~\citep{10.1145/3097983.3098066}
\item The Turing Way~\citep{turing2021turing}
\item The Whole Tale~\citep{wholetale}
\item They shall be fair, transparent, and robust: auditing learning analytics systems~\citep{simbeck2024they}
\item Three naïve Bayes approaches for discrimination-free classification~\citep{calders2010three}
\item Toward situated interventions for algorithmic equity: lessons from the field~\citep{10.1145/3351095.3372874}
\item Towards a Principled Approach for Engineering Privacy by Design~\citep{10.1007/978-3-319-67280-9_9}
\item Towards better understanding of gradient-based attribution methods for Deep Neural Networks~\citep{ancona2018better}
\item TuringBox: An Experimental Platform for the Evaluation of AI Systems~\citep{ijcai2018p851}
\item Uber Differential Privacy~\citep{uber}
\item UnBias Fairness Toolkit~\citep{lane_2018_2667808}
\item Understanding artificial intelligence ethics and safety: A guide for the responsible design and implementation of AI systems in the public sector~\citep{leslie2019}
\item Value-based Engineering for Ethics by Design~\citep{spiekermann2020valuebasedengineeringethicsdesign}
\item Weights \& Biases~\citep{wb}
\item Wellcome Data -- A new method for ethical data science~\citep{wellcome}
\item What-If Tool~\citep{whatiftool}
\item When Worlds Collide: Integrating Different Counterfactual Assumptions in Fairness~\citep{NIPS2017_1271a702}
\item White Paper on Data Ethics in Public Procurement of AI-based Services and Solutions~\citep{hasselbalch2020white}
\item ``Why Should I Trust You?'': Explaining the Predictions of Any Classifier~\citep{10.1145/2939672.2939778}
\item XAI - An eXplainability toolbox for machine learning~\citep{EthicalMLxai}	
\item Zeno: An Interactive Framework for Behavioral Evaluation of Machine Learning~\citep{10.1145/3544548.3581268}
\end{enumerate}
}

\section{AIETs list update -- from August 2023 until March 2025}
\label{sec:aietslistupdate}
{\footnotesize 
\begin{enumerate}
\setcounter{enumi}{213}
\item A Capability Approach to Ethical Development and Internal Auditing of AI Technology~\citep{graves_capability_2024}
\item A Conceptual Framework for Solving Ethical Issues in Generative Artificial Intelligence~\citep{zlateva_conceptual_2024}
\item A Human Rights-Based Approach to Artificial Intelligence in Healthcare: A Proposal for a Patients' Rights Impact Assessment Tool~\citep{van_kolfschooten_human_2024}
\item A Lifecycle Approach for Artificial Intelligence Ethics in Energy Systems~\citep{el-haber_lifecycle_2024}
\item A new framework for ethical artificial intelligence: keeping HRD in the loop~\citep{wang_new_2024}
\item A Questionnaire of Artificial Intelligence Use Motives: A Contribution to Investigating the Connection between AI and Motivation~\citep{yurt_questionnaire_2024}
\item A semi-automated software model to support AI ethics compliance assessment of an AI system guided by ethical principles of AI~\citep{cappelli_semi-automated_2024}
\item A Soft Constraint-Based Framework for Ethical Reasoning~\citep{hosobe_soft_2024}
\item A tangible toolkit to uncover clinician's ethical values about AI clinical decision support systems~\citep{eindhoven_tangible_2024}
\item A theoretical framework to guide AI ethical decision making~\citep{ferrell_theoretical_2024}
\item ACESOR: a critical engagement in systems of oppression AI assessment tool~\citep{mcfadden_acesor_2024}
\item AI Design: A Responsible AI Framework for Impact Assessment Reports~\citep{bogucka_ai_2024}
\item AI Ethical Framework: A Government-Centric Tool Using Generative AI~\citep{kone2024ai}
\item AI Ethics: Integrating Transparency Fairness and Privacy in AI Development~\citep{radanliev_ai_2025}
\item An Explainable AI Tool for Operational Risks Evaluation of AI Systems for SMEs~\citep{han_explainable_2023}
\item Artificial Intelligence for next generation cybersecurity: The AI4CYBER framework~\citep{iturbe_artificial_2023}
\item Artificial intelligence in governance: recent trends, risks, challenges, innovative frameworks and future directions~\citep{ghosh_artificial_2025}
\item Assessment of Artificial Intelligence Credibility in Evidence-Based Healthcare Management with “AERUS” Innovative Tool~\citep{sallam_assessment_2024}
\item Bridging the Gap between Theory and Practice: Towards Responsible AI Evaluation~\citep{gollner_bridging_2023}
\item Bridging Theory and Practice: A Tool for Translating Ethical AI Requirements into Ethical User Stories~\citep{rossi_de_borba_bridging_2024}
\item Card-Based Approach to Engage Exploring Ethics in AI for Data Visualization~\citep{wang_card-based_2024}
\item Checklist for Artificial Intelligence in Medical Imaging (CLAIM)~\citep{tejani_checklist_2024}
\item Creative Explainable AI Tools to Understand Algorithmic Decision-Making~\citep{bhat_creative_2024}
\item Crossing the principle–practice gap in AI ethics with ethical problem-solving~\citep{correa_crossing_2024}
\item Data Ethics Emergency Drill: A Toolbox for Discussing Responsible AI for Industry Teams~\citep{hanschke_data_2024}
\item Data-driven framework for evaluating digitization and artificial intelligence risk: a comprehensive analysis~\citep{badawy_data-driven_2023}
\item Datasheets for Machine Learning Sensors~\citep{stewart_datasheets_2023}
\item Decisional value scores: A new family of metrics for ethical AI-ML~\citep{waters_decisional_2024}
\item E-LENS: User Requirements-Oriented AI Ethics Assurance~\citep{zhou_e-lens_2025}
\item ECS: an interactive tool for data quality assurance~\citep{sieberichs_ecs_2024}
\item ED-AI Lit: An Interdisciplinary Framework for AI Literacy in Education~\citep{allen_ed-ai_2023}
\item Embedded Ethics for Responsible Artificial Intelligence Systems (EE-RAIS) in disaster management: a conceptual model and its deployment~\citep{afroogh_embedded_2024}
\item Ensuring fundamental rights compliance and trustworthiness of law enforcement AI systems: the ALIGNER Fundamental Rights Impact Assessment~\citep{casaburo_ensuring_2024}
\item Ethical practices of artificial intelligence: a management framework for responsible AI deployment in businesses~\citep{tripathi_ethical_2025}
\item Ethical, Legal and Social Aspects (ELSA) for AI: an assessment tool for Agri-food~\citep{van_hilten_ethical_2024}
\item Ethics by design for artifcial intelligence~\citep{brey_ethics_2023}
\item Evaluation of AI Solutions in Health Care Organizations — The OPTICA Tool~\citep{dagan_evaluation_2024}
\item Evolving AI Risk Management: A Maturity Model based on the NIST AI Risk Management Framework~\citep{dotan_evolving_2024}
\item Explainable Fairness in Regulatory Algorithmic Auditing~\citep{oneil_explainable_2024}
\item FAIR Enough: Develop and Assess a FAIR-Compliant Dataset for Large Language Model Training?~\citep{raza_fair_2024}
\item Fairframe: a fairness framework for bias detection and mitigation in news~\citep{sallami_fairframe_2024}
\item FairPIVARA: Reducing and Assessing Biases in CLIP-Based Multimodal Models~\citep{moreira_fairpivara_2024}
\item FAMEWS: a Fairness Auditing tool for Medical Early-Warning Systems~\citep{hoche_famews_2024}
\item Framework for Bias Detection in Machine Learning Models: A Fairness Approach~\citep{rosado_gomez_framework_2024}
\item Generative artificial intelligence and ethical considerations in health care: a scoping review and ethics checklist~\citep{ning_generative_2024}
\item Governance of Artificial Intelligence – A Framework Towards Ethical AI Applications~\citep{lachenmaier_governance_2023}
\item How to design an AI ethics board~\citep{schuett_how_2024}
\item Introducing the ethical-epistemic matrix: a principle-based tool for evaluating artificial intelligence in medicine~\citep{adams_introducing_2024}
\item MLCA: A Tool for Machine Learning Life Cycle Assessment~\citep{morand_mlca_2024}
\item Multi-Value Alignment for Ml/AI Development Choices A Four-step Process Including Approaches to Trade-offs~\citep{jethwani_multi-value_2025}
\item POLARIS: A framework to guide the development of Trustworthy AI systems~\citep{baldassarre_polaris_2024}
\item POTDAI: A Tool to Evaluate the Perceived Operational Trust Degree in Artificial Intelligence Systems~\citep{martin-moncunill_potdai_2024}
\item Responsible AI Question Bank: A Comprehensive Tool for AI Risk Assessment~\citep{une_lee_responsible_2024}
\item Risk Assessment of Large Language Models Beyond Apocalyptic Visions~\citep{maathuis_risk_2024}
\item RuleGLM: An Ethics Evaluation Framework with Knowledge Vector Space~\citep{li_ruleglm_2024}
\item Safe LoRA: The Silver Lining of Reducing Safety Risks when Finetuning Large Language Models~\citep{hsu_safe_2024}
\item SafeInfer: Context Adaptive Decoding Time Safety Alignment for Large Language Models~\citep{banerjee_safeinfer_2024}
\item Scaling AI responsibly: Leveraging MLOps for sustainable machine learning deployments~\citep{kodakandla2024scaling}
\item SORRY-Bench: Systematically Evaluating Large Language Model Safety Refusal~\citep{xie_sorry-bench_2024}
\item Stream: social data and knowledge collective intelligence platform for TRaining Ethical AI Models~\citep{wang_stream_2024}
\item tachAId—An interactive tool supporting the design of human-centered AI solutions~\citep{bauroth_tachaidinteractive_2024}
\item TAI-PRM: trustworthy AI—project risk management framework towards Industry 5.0~\citep{vyhmeister_tai-prm_2024}
\item The Artificial Intelligence Assessment Scale (AIAS): A Framework for Ethical Integration of Generative AI in Educational Assessment~\citep{perkins_artificial_2024}
\item The BIG Argument for AI Safety Cases~\citep{habli_big_2025}
\item The Principle-at-Risk Analysis (PaRA): Operationalising Digital Ethics by Bridging Principles and Operations of a Digital Ethics Advisory~\citep{nemat_principle-at-risk_2023}
\item Toolkit for specification, validation and verification of social, legal, ethical, empathetic and cultural requirements for autonomous agents~\citep{getir_yaman_toolkit_2024}
\item Toward a framework for risk mitigation of potential misuse of artificial intelligence in biomedical research~\citep{trotsyuk_toward_2024}
\item Towards a framework for local interrogation of AI ethics: A case study on text generators, academic integrity, and composing with ChatGPT~\citep{vetter_towards_2024}
\item Towards a Human-Centric AI Trustworthiness Risk Management Framework~\citep{kioskli_towards_2024}
\item Towards a Responsible AI Metrics Catalogue: A Collection of Metrics for AI Accountability~\citep{xia_towards_2024}
\item Towards Responsible AI Music: an Investigation of Trustworthy Features for Creative Systems~\citep{berardinis_towards_2025}
\item TRIPOD+AI statement: updated guidance for reporting clinical prediction models that use regression or machine learning methods~\citep{collins_tripodai_2024}
\item Trustworthy Artificial Intelligence: Design of AI Governance Framework~\citep{sharma_trustworthy_2023}
\item What Would You do? An Ethical AI Quiz~\citep{teo_what_2023}
\item XAutoML: A Visual Analytics Tool for Understanding and Validating Automated Machine Learning~\citep{zoller_xautoml_2023}
\end{enumerate}}

\end{appendices}

%%===========================================================================================%%
%% If you are submitting to one of the Nature Portfolio journals, using the eJP submission   %%
%% system, please include the references within the manuscript file itself. You may do this  %%
%% by copying the reference list from your .bbl file, paste it into the main manuscript .tex %%
%% file, and delete the associated \verb+\bibliography+ commands.                            %%
%%===========================================================================================%%

\end{document}